\documentclass[a4paper,11pt]{article}

\usepackage{hyperref}
\usepackage{graphicx}
\usepackage{amssymb}
\usepackage{amsfonts}
\usepackage{amsbsy}
\usepackage{amsmath}
\usepackage{longtable}
\usepackage{a4wide}
\usepackage{soul}
\usepackage{color}
\usepackage{multirow}

\newcommand{\nm}{\phantom{-}}

\begin{document}

\thispagestyle{empty}
\begin{flushright}
TUW-10-07\\
IPMU10-0090
\end{flushright}
\vspace{1cm}
\begin{center}
{\LARGE Global $SO(10)$ F-theory GUTs}
\end{center}
\vspace{8mm}
\begin{center}
{\large Ching-Ming Chen${}^{\ast}$\footnote{e-mail: cmchen AT hep.itp.tuwien.ac.at}, Johanna Knapp${}^{\dagger}$\footnote{e-mail: johanna.knapp AT ipmu.jp}, Maximilian Kreuzer${}^{\ast}$\footnote{e-mail: kreuzer AT hep.itp.tuwien.ac.at}\\
and Christoph Mayrhofer${}^{\ast}$\footnote{e-mail: cmayrh AT hep.itp.tuwien.ac.at }}
\end{center}
\vspace{3mm}
\begin{center}
{\it $^{\ast}$Institut f\"ur theoretische Physik \\
TU Vienna\\
Wiedner Hauptstrasse 8-10\\
1040 Vienna\\
Austria }
\end{center}
\begin{center}
{\it $^{\dagger}$ Institute for the Physics and Mathematics of the Universe (IPMU)\\
The University of Tokyo \\
5-1-5 Kashiwanoha\\
Kashiwa 277-8583\\
Japan
}
\end{center}
\vspace{15mm}
\begin{abstract}
\noindent Making use of toric geometry we construct a class of
global F-theory GUT models. The base manifolds are blowups of Fano
threefolds and the Calabi-Yau fourfold is a complete intersection
of two hypersurfaces. We identify possible GUT divisors and
construct $SO(10)$ models on them using the spectral cover
construction. We use a split spectral cover to generate chiral
matter on the ${\bf 10}$ curves in order to get more degrees of
freedom in phenomenology. We use abelian flux to break $SO(10)$ to
$SU(5)\times U(1)$ which is interpreted as a flipped $SU(5)$
model.  With the GUT Higgses in the $SU(5)\times U(1)$ model it is
possible to further break the gauge symmetry to the Standard
Model. We present several phenomenologically attractive examples
in detail.
\end{abstract}
\newpage
\setcounter{tocdepth}{1}
\tableofcontents
\setcounter{footnote}{0}

\section{Introduction}
Triggered by~\cite{Donagi:2008ca,Beasley:2008dc,Beasley:2008kw}
F-theory has recently received new attention as a natural setup
for constructing realistic GUT models within string theory. One of
the key properties of the recent F-theory GUT constructions is
that it is possible to decouple gravity from the gauge theory and
to consider the GUT model locally on the seven brane geometry.
After the embedding of a local model into a global geometry the
issue of the decoupling limit becomes more delicate and its existence will be a central guideline in our search of
global models. Some attractive features of the local models are
that they are straightforward to construct and manage to solve
many problems that plague GUT models, such as proton decay,
GUT-breaking and doublet--triplet splitting, the $\mu$-term problem,
and flavor hierarchy elegantly from geometry and with little fine
tuning. Given these features it is desirable to see whether it is
possible to embed the local models consistently into a global
setup of F-theory compactifications on compact Calabi-Yau
fourfolds. One first step in this direction was done by so called
semi-local
models~\cite{Hayashi:2008ba,Hayashi:2009ge,Andreas:2009uf,Donagi:2009ra,Marsano:2009gv} where conditions for the embedding of
local models into a Calabi-Yau fourfold have been discussed, and
some global features such as fluxes and monodromies have been
introduced into the local setup. Global F-theory GUT models have
been first constructed
in~\cite{Marsano:2009ym,Blumenhagen:2009yv,Marsano:2009wr,Grimm:2009yu}. The authors
of~\cite{Blumenhagen:2009yv,Grimm:2009yu} have made use of toric
geometry to construct and examine Calabi-Yau fourfolds. It turned
out that for F-theory GUTs it is natural not to look for
fourfolds that are hypersurfaces in some toric ambient space but
to go to the sometimes more involved case of complete
intersections. Concretely, one looks at complete intersections of
two hypersurfaces in a six-dimensional ambient space. One of
these equations describes a three dimensional base manifold in a
four dimensional projection of the sixfold. The other equation
describes the elliptic fibration. The aim of this publication is
to systematically construct a class of global models within this
framework, to discuss their properties and eventually explicitly
construct semi-realistic examples. Since $SU(5)$ F-theory GUTs
have already received a lot of attention in the literature, our
discussion will focus on $SO(10)$ models. Note however that the geometries we will construct are also perfectly suitable for $SU(5)$ F-theory models.

Let us now give an overview on how we are going to approach the
discussion of global models. The first step is to make use of
toric geometry to construct a complex three-dimensional base
manifold $B$ which is a hypersurface in a toric ambient space. As
a starting point we will consider Fano threefolds with one
K\"ahler class. It has been argued in~\cite{Cordova:2009fg} that
Fanos are not good base manifolds for F-theory GUTs because they
do not allow for a decoupling limit. To remedy this problem we
will blow up curves and points inside the Fano threefolds. As a
consequence the resulting manifold will in general no longer be
Fano. After the construction of a suitable base manifold we go on
to search for possible candidates of GUT divisors inside the base.
We impose two requirements on the divisor which are important for F-theory GUTs. Firstly, we will look for divisors which are del Pezzo and
secondly there should exist a mathematical or at least a physical
decoupling limit~\cite{Cordova:2009fg,Grimm:2009yu}. The former
condition means that the GUT divisor can be shrunk to zero size
while the volume of the base manifold remains finite. The latter
condition means that we can keep the volume of the GUT divisor
finite while the volume of the base manifold becomes infinitely
large. Having found a base manifold with a divisor satisfying
these elementary requirements we torically construct a Calabi-Yau
fourfold which is an elliptic fibration over the base manifold and is characterized in toric geometry by reflexive polyhedra. We
note in passing that this toric construction is actually not
possible for every base manifold. Having an explicit construction
of the Calabi-Yau fourfolds enables us to directly calculate the
Euler number of the fourfold. We have compared our results with a
formula for the Euler number proposed in~\cite{Blumenhagen:2009yv}
and report on a mismatch we find for several examples. We propose
possible resolutions for this discrepancy in the Conclusions
section.

We then go on to construct $SO(10)$ models in our geometries
making use of the spectral cover
construction~\cite{hitchin,donagi95,Friedman:1997yq,Donagi:1997pb}.
Originally the spectral cover has been introduced to describe
vector bundles in the heterotic string, which has a connection to
F-theory via the heterotic/F-theory
duality~\cite{Friedman:1997yq}. However, it has been realized
in~\cite{Hayashi:2009ge,Donagi:2009ra} that the spectral cover can
also be used in F-theory models without a heterotic dual to
locally describe fluxes near the GUT brane. Starting with an
elliptic fibration described by a Weierstrass model near the GUT
cycle, we extract the terms responsible for the breaking of  a
$E_8$ singularity down to a $SO(10)$ singularity. Via the spectral
cover picture, these are related to some data of a bundle with
structure group  in the complement of the GUT group in $E_8$ which
breaks the excess symmetry of $E_8$. For the case of $SO(10)$
models one must look at a $SU(4)$ spectral cover.  The $SU(4)$
vector bundle $V$ represents the $\bf 16$, while the bundle
$\wedge^2 V$ represents the $\bf 10$ of $SO(10)$. Thus we can have
a minimal $SO(10)$ GUT model with the $\bf 16\,16\,10$ Yukawa
coupling. However, it has been shown that the net chirality on the
${\bf 10}$ curve vanishes because it forms a double
curve~\cite{Hayashi:2008ba}, which would leave us without a
suitable Higgs candidate. In addition, there are not enough
degrees of freedom to adjust the number of generations on the
matter curves, so the model would not look very realistic. A
solution to these problems is to introduce a split spectral cover
\cite{Blumenhagen:2006wj, Marsano:2009gv}. Furthermore, having
constructed a $SO(10)$ GUT we have yet to show that we can obtain
the Standard Model after gauge breaking. Direct breaking from
$SO(10)$ to the Standard Model gauge group requires to turn on
non-abelian fluxes~\cite{Beasley:2008kw}. Up to now this mechanism
has not been explicitly studied. We will make an intermediate step
and first break to a model with a $SU(5)$ gauge group by a $U(1)$
flux $F_X$. Since we have then exhausted the option to break the
gauge symmetry by turning on flux, the further gauge breaking will
rely on different breaking mechanisms in the GUT model. The
natural choice is to break from $SO(10)$ to an $SU(5)$ GUT.
However since from the local model construction we have no
chirality for the adjoint representation~\cite{Beasley:2008kw},
the GUT Higgs $\bf 24$ which is needed for breaking the $SU(5)$ is
absent. Therefore we consider a flipped $SU(5)$ model
\cite{Barr:1981qv, Derendinger:1983aj, Antoniadis:1987dx}, (for
recent discussion in F-theory, see \cite{Jiang:2009zza,
Chen:2009me, King:2010mq, Chen:2010tp}) where the gauge breaking
to the Standard Model can be achieved by the ${\bf 10}_H$ and
$\overline{\bf 10}_H$ superheavy Higgs fields.  We will illustrate
our discussion by several examples.

This paper is organized as follows. In Section~\ref{sec-golbaltoric} we will describe the toric construction of the geometries of the global models. Section~\ref{sec-spectralcover} is devoted to the spectral cover construction. After a brief outline of the construction we will specialize to $SU(4)$ spectral covers and $S(U(3)\times U(1))$ covers which are relevant for $SO(10)$ GUTs. In Section~\ref{sec-examples} we will give explicit examples of some models. Section~\ref{sec-pheno} is devoted to the phenomenology of the models we have constructed. Conclusions and directions for further research will be given in Section~\ref{sec-conclusion}. In Appendix~\ref{app-models} we collect the data of the base manifolds which contain possible GUT divisors. In Appendix~\ref{fourfold-data} we present the toric data of the Calabi-Yau fourfolds of the models we have worked out in Section~\ref{sec-examples}.\\\\
{\bf Acknowledgments:} We gratefully acknowledge Nils-Ole Walliser for writing useful code extensions for PALP and pointing out a mistake we had in our del Pezzo calculation. Furthermore we would like to thank Taizan Watari for valuable comments on the manuscript. C-MC would like to thank Y.-C. Chung for useful discussions. CM would like to thank A. Braun and A. Collinucci for useful discussions. JK would like to acknowledge discussions with Emanuel Scheidegger. The work of JK was supported by World Premier International Research Center Initiative (WPI Initiative), MEXT, Japan. The work of CM and C-MC was supported in part by the Austrian Research Funds FWF under grant numbers P21239, and I192.\\
%%%%%%%%%%%%%%%%%%%%%%%%%%%%%%%%%%%%%%%%%%%%%%%%%%%%%%%%%%%%%%%%%%%%%%%
\section{Global toric constructions}
\label{sec-golbaltoric}
In this section we will discuss toric constructions of both the base manifold and the Calabi-Yau fourfold which is an elliptic fibration over the base.
%%%%%%%%%%%%%%%%%%%%%%%%%%%%%%%%%%%%%%%%%%%%%%%%%%%%%%%%%%%%%%%%%%%%%%%%%%%%
\subsection{General idea}
In~\cite{Blumenhagen:2009yv} a prescription for constructing F-theory models as a complete intersection of two hypersurfaces has been given. The hypersurface constraints have the following form:
\begin{equation}
\label{cicydef}
P_{B}(y_i,w)=0\qquad\qquad P_W(x,y,z,y_i,w)=0
\end{equation}
The first constraint describes the threefold base, the second equation the elliptic fibration encoded in the Weierstrass model, which is conveniently written in its Tate form:
\begin{equation}
\label{tate}
P_W=x^3-y^2+xyz a_1+x^2z^2a_2+yz^3a_3+xz^4a_4+z^6a_6,
\end{equation}
where the $a_n(y_i,w)$ are sections of $K_B^{-n}$. The GUT divisor $S$ is specified by $w=0$.\\
For the base manifold~\cite{Blumenhagen:2009yv} suggests the following construction. The starting point is a Fano threefold. A simple choice are Fano threefolds which have one K\"ahler class. These are $\mathbb{P}^3$ and quadric, cubic and quartic hypersurfaces in $\mathbb{P}^4$. The next step  is to generate a suitable four cycle where we will wrap the GUT brane. For local models it was argued~\cite{Donagi:2008ca,Beasley:2008dc,Beasley:2008kw} that, in order to have a well defined local model, the GUT divisor should be a del Pezzo surface. Furthermore, it has been argued in~\cite{Cordova:2009fg} that Fano base manifolds do not allow for a decoupling limit in a global F-theory GUT. Following~\cite{Blumenhagen:2009yv,Grimm:2009yu} we will blow up points and curves inside the Fano threefolds until the requirements for a GUT model are satisfied. This is discussed in Section~\ref{sec-toric-base}. Most of the data that are relevant for the construction of the GUT model are already encoded in the base manifold. However, in order to discuss global issues like the cancellation of D3 tadpoles it is necessary to have an explicit construction of the fourfold. Furthermore not every base manifold will allow for a toric elliptically fibered Calabi-Yau fourfold whose geometry is encoded in terms of reflexive lattice polytopes. We will discuss this in Section~\ref{sec-toric-fourfold}.
%%%%%%%%%%%%%%%%%%%%%%%%%%%%%%%%%%%%%%%%%%%%%%%%%%%%%%%%%%%%%%%%%%%%%%%%%%%%
\subsection{Toric construction of the base manifold}
\label{sec-toric-base}
We will now discuss how to obtain a suitable base manifold for a GUT model by blowing up points and curves inside a Fano threefold. Such constructions were first introduced into the F-theory literature in~\cite{Blumenhagen:2009yv}. Fano threefolds have been classified in~\cite{morimukai1,morimukai2}. Since we would like to construct our base manifold as a hypersurface in a toric ambient space it is suggestive to start with the following Fanos which are hypersurfaces in $\mathbb{P}^4$:
\begin{equation}
\label{fano}
\mathbb{P}^4[d]=\{P_d(y_1,\ldots,y_5)=0|[y_i:\ldots:y_5]\in\mathbb{P}^4\}\qquad d=2,3,4,
\end{equation}
where $P_d(y_1,\ldots,y_5)$ stands for a polynomial of degree $d$ in the homogeneous coordinates $y_i,\ldots,y_5$.

It has been argued in~\cite{Cordova:2009fg} that Fanos are not suitable as base manifolds for an F-theory GUT model due to the non-existence of a decoupling limit. For our examples this is trivial to see since the hypersurfaces in~(\ref{fano}) only have a single K\"ahler modulus, and it is therefore impossible to keep the volume of a GUT divisor finite while the size of the base becomes infinitely large, or conversely shrink the GUT divisor to zero size while the volume of the base remains non-zero. For a decoupling limit we need at least two K\"ahler moduli. We can increase the number of K\"ahler parameters by blowing up points and curves in the toric ambient space. Such a procedure reduces the number of complex structure moduli and increases the number of K\"ahler parameters.

The toric data to describe the ambient space and the blowups is encoded in weight vectors~\cite{2008arXiv0809.1188K} which give the homogeneous weights of the coordinates describing the toric variety, or in other words, the weights are the $U(1)$--charges of the gauged linear linear sigma model which describes the toric space. For a prescription of how to obtain the toric data from the weight matrices we refer the reader to one of the many books and reviews on toric geometry. The weight matrices are the input data for the package PALP \cite{Kreuzer:2002uu} and extensions thereof which can compute all the relevant data we need for the calculations described in this section.

The weight vector for $\mathbb{P}^4$ is given in table~\ref{tab-p4ambient}, along with some further data: $d$ stands for the degree of the hypersurface equation as in~(\ref{fano}), and the next to last entry gives the sum of the weights.
\begin{table}
\begin{center}
\begin{tabular}{c|ccccc|c|c}
&$y_1$&$y_2$&$y_3$&$y_4$&$y_5$&$\sum$&$B$\\
\hline
$w_1$&$1$&$1$&$1$&$1$&$1$&$5$&$d$\\
\end{tabular}
\end{center}
\caption{Weight vector and degree of the hypersurface equation of $\mathbb{P}^4[d]$.}\label{tab-p4ambient}
\end{table}
We can now blow up curves and points inside the Fano threefold by adding further weight vectors. Blowing up a curve means adding a weight vector as in table~\ref{tab-p41curve}.
 \begin{table}
\begin{center}
\begin{tabular}{c|cccccc|c|c}
&$y_1$&$y_2$&$y_3$&$y_4$&$y_5$&$y_6$&$\sum$&$B$\\
\hline
$w_1$&$1$&$1$&$1$&$1$&$1$&$0$&$5$&$d$\\
$w_2$&$0$&$0$&$0$&$1$&$1$&$1$&$3$&\\
\end{tabular}
\end{center}
\caption{Blowup of the first curve.}\label{tab-p41curve}
\end{table}
Note that due to the symmetry of the ambient space the choice of the weight vector is unique in the sense that we will get the same base equation up to permutations of variables.
In table~\ref{tab-p41curve} we have not specified the hypersurface degree the base should have in the new weight vector. For $d=(4,2)$ we recover an example given in~\cite{Blumenhagen:2009yv}. For a point blowup we have to add a weight vector as in table~\ref{tab-p41point}.
\begin{table}
\begin{center}
\begin{tabular}{c|cccccc|c|c}
&$y_1$&$y_2$&$y_3$&$y_4$&$y_5$&$y_6$&$\sum$&$B$\\
\hline
$w_1$&$1$&$1$&$1$&$1$&$1$&$0$&$5$&$d$\\
$w_2$&$1$&$0$&$0$&$0$&$0$&$1$&$2$&\\
\end{tabular}
\end{center}
\caption{Blowup of one point.}\label{tab-p41point}
\end{table}
For most interesting models it will not be enough to blow up just one curve or one point. We have systematically searched for models which come from up to three blowups of curves and points inside the Fano threefolds~(\ref{fano}).

After the blowups, the resulting base manifolds will in general no longer be Fano. One can ask if these manifolds fit into some mathematically well-defined class of generalizations of Fano threefolds. One possible generalization are the almost Fano manifolds. For a definition see e.g.~\cite{hubsch}. An almost Fano threefold is an algebraic threefold $\mathcal{F}$ that has a non-trivial anti-canonical bundle with at least one non-zero section at every point in $\mathcal{F}$. The Hodge diamond of an (almost) Fano threefold is:
\begin{equation}
\begin{array}{ccccccc}
&&&1&&&\\
&&0&&0&&\\
&0&&b_{1,1}&&0&\\
0&&b_{2,1}&&b_{2,1}&&0\\
&0&&b_{1,1}&&0&\\
&&0&&0&&\\
&&&1&&&
\end{array}
\end{equation}
The Euler number is $\chi=2(b_{1,1}+1-b_{2,1})$. Furthermore, all almost Fano threefolds satisfy~\cite{hubsch}:
\begin{equation}
\int_{\mathcal{F}}c_1c_2=24
\end{equation}
We discuss the results of our construction of base manifolds, together with a detailed discussion of the restrictions we made, in Appendix~\ref{app-models}, where we in particular give details about the base manifolds which satisfy the 'almost Fano' condition. For the models we have constructed we observe that the 'almost Fano' property seems to be a necessary but not sufficient condition for the fourfold to be described by a reflexive polytope.
It would be interesting to see if observed connection holds more generally and whether it can be put on mathematically solid ground. We plan to further investigate this relation in the future. Section~\ref{sec-examples} will be devoted to the construction of $SO(10)$ F-theory GUTs from examples among this class.\\

Having constructed a suitable base manifold, the next step is to find suitable candidates for GUT branes among its divisors. In our search of models we only made two restrictions which are important for the construction of a global GUT model: the GUT divisor should be del Pezzo and there should be a decoupling limit.

We look for candidates for del Pezzo surfaces by identifying divisors which have the same topology as a del Pezzo. Suppose the base manifold has hyperplane class $B$ and is embedded in a toric ambient space with divisors $D_i$. The total Chern class of a particular divisor $S$ in $B$ is:
\begin{equation}
c(S)=\frac{\prod_i(1+D_i)}{(1+B)(1+S)}
\end{equation}
A necessary condition for a divisor $S$ to be $dP_n$ is that it must have the following topological data:
\begin{equation}
\int_Sc_1(S)^2=9-n\qquad \int_Sc_2(S)=n+3\qquad\Rightarrow\qquad \chi_h=\int_S\mathrm{Td}(S)=1,
\end{equation}
where $\chi_h$ is the holomorphic Euler characteristic.
Since del Pezzos are Fano twofolds, we have a further necessary condition. The integrals of $c_1(S)$ over all torically induced curves on $S$ have to be positive:
\begin{equation}
\label{poscurve}
D_i\cap S\cap c_1(S) > 0\qquad D_i\neq S \qquad \forall D_i\cap S\neq\emptyset\,.
\end{equation}
Finding divisors $S$ that have the above properties provides strong evidence that $S$ is indeed a del Pezzo. Whenever we speak about del Pezzo divisors in this presentation we mean divisors which satisfy the conditions above. To be absolutely certain that the divisor is del Pezzo one should explicitly construct it. We did this for the examples we work out in detail in Section~\ref{sec-examples}.

For the model to have a physical decoupling limit, one must be able to tune the K\"ahler moduli in such a way that the volume of $S$ remains finite while the volume of the base manifold goes to infinity. It is to be contrasted with the mathematical decoupling limit where the GUT divisor shrinks to $0$ while the volume of the base remains finite. According to~\cite{Grimm:2009yu} these two decoupling limits may be governed by different vectors in the K\"ahler cone. In order for the volumes to be positive we have to find a basis $K_i$ of the K\"ahler cone such that the K\"ahler form $J$ has the form $J=\sum_ir_iK_i$ with $r_i>0$. The volumes of the base $B$ and the GUT divisor $S$ are then given by:
\begin{equation}
\label{gutvol}
\mathrm{Vol}(B)=J^3\qquad \mathrm{Vol}(S)=S\cdot J^2
\end{equation}
A sufficient condition for the existence of the physical decoupling limit is that the volume of the divisor $S$ does not depend on one or more of the moduli $r_i$ whereas the volume of the base does depend on them. By sending these K\"ahler moduli to infinity we can achieve infinite volume of the base while keeping the volume of the GUT divisor finite. The mathematical decoupling limit is slightly more difficult to check. There it is necessary to set sufficiently many K\"ahler parameters to zero such that $ \mathrm{Vol}(S)=0$ while still having terms in $\mathrm{Vol}(B)$ which are independent of the parameters we have set to zero. In order to perform this analysis we need the triple intersection numbers of the divisors, restricted to the base $B$, and a basis of the K\"ahler cone\footnote{We constructed the K\"ahler cone for the toric ambient space. We did not take into account that different phases of the toric ambient space (triangulations) can lead to equivalent phases on the hypersurface, which would yield a larger K\"ahler cone.}. This data can be extracted from the weight matrices and the divisor specifying the hypersurface $B$. The necessary calculations can be done with help of an extended version of the package PALP \cite{Kreuzer:2002uu}. In Appendix~\ref{app-models} we list, within our class of examples, the del Pezzo divisors that satisfy at least one of the decoupling conditions.\\\\
Having identified a 'nice' del Pezzo inside the base manifold we can now specify the gauge theory we would like to have. For $SO(10)$ models, the $a_i$s of  the Weierstrass equation~(\ref{tate}) have to degenerate in a certain way~\cite{Bershadsky:1996nh}. According to Kodaira's classification of singularities of elliptic curves \cite{kodaira} and Tate's algorithm~\cite{tate}, they have to have the following form\footnote{For an $SU(5)$ model the structure would be:
\[
a_1=b_5\quad a_2=b_4w^1\quad a_3=b_3w^2\quad a_4=b_2w^3\quad a_6=b_0w^5
\]},
\begin{equation}
a_1=b_5w^1\quad a_2=b_4w^1\quad a_3=b_3w^2\quad a_4=b_2w^3\quad a_6=b_0w^5\,,
\end{equation}
where the $b_i$s are sections of some appropriate line bundle over $B$ that have at least one term independent of $w$. In this description matter curves and Yukawa couplings are located at the subsets where the discriminant of~(\ref{tate}) has an order one and an order two enhancement, respectively. The order refers here to the vanishing power of the discriminant in the vicinity of $S$. The matter curves for the $SO(10)$ models are at
\begin{equation}
b_3=0\quad \textrm{{\bf 10} matter} \qquad b_4=0\quad \textrm{{\bf 16} matter}\,,
\end{equation}
and the Yukawa couplings are at
\begin{equation}
b_3=0 \cap b_4=0\quad \textrm{$E_7$ Yukawas} \qquad b_2^2-4 b_0 b_4=0 \cap b_3=0 \quad \textrm{$SO(14)$ Yukawas}\,.
\end{equation}
See Section~\ref{SO10-models} for a more detailed discussion. Now we can compute the genus (or respectively, the Euler number) of the generic matter curves and the number of their intersections (Yukawa couplings).  The genera of the matter curves can be computed via their first Chern classes. The number of Yukawa couplings is given by the triple intersection of the divisors of the curves with the GUT brane. In practice this is done by expressing $c_1$ of the curves and the $b_i$ in terms of toric divisors and making use of their triple intersections.
%%%%%%%%%%%%%%%%%%%%%%%%%%%%%%%%%%%%%%%%%%%%%%%%%%%%%%%%%%%%%%%%%%%%%%%%%%%%%%
\subsection{Fourfold}
\label{sec-toric-fourfold}
Having constructed a threefold base $B$ we obtain a Calabi-Yau fourfold by fibering a torus over it. Our aim is to construct a complete intersection~(\ref{cicydef}) of two hypersurfaces in a six dimensional toric manifold which describes this situation. One equation should define the base manifold  in a four dimensional projection of the toric sixfold. The other one, given in~(\ref{tate}), is the Weierstrass equation which gives us the torus fiber. Since the Weierstrass equation~(\ref{tate}) has to be a well defined equation and the $a_i$s are sections of certain line bundles over the base, also the fiber coordinates $x$ and $y$ have to transform non-trivially over $B$. To obtain a Calabi-Yau fourfold it turns out that $x$ and $y$ have to be sections of $K^{-2}_B$ and $K^{-3}_B$, respectively.\\

A complete intersection Calabi-Yau of codimension $r$ in a toric variety
$\mathbb P_\Sigma$ is described by $r$ equations $f_i=0$ where the $f_i$
are sections of line bundles whose support polytopes \cite{Kreuzer:2006ax}
are the Newton polytopes $\Delta_i$ of $f_i$ and $\Sigma$ is the fan\footnote{The fan $\Sigma$ is not to be confused with the matter curves which we will also denote by $\Sigma$ in the following sections.} over
the faces of a lattice polytope $\Delta^\circ$. By the adjunction formula
we obtain a Calabi-Yau if the Minkowski sum $\Delta=\Delta_1+\ldots+\Delta_r$
is dual to $\Delta^\circ$ in the sense that the inequality
$\langle \Delta,\Delta^\circ\rangle\ge-1$ is saturated. In particular,
$\Delta$ and $\Delta^\circ$ is a reflexive
pair of lattice polytopes. We will restrict our attention to the case
where $\Delta^\circ$ admits a nef partition into a convex hull of
lattice polytopes $\nabla_{i\le r}$ \cite{Batyrev:1994pg}, whose combinatorics is
summarized by
\begin{eqnarray}                                        \label{cicy-nef}
        \Delta=\Delta_1+\ldots+\Delta_r&& \Delta^{\circ}
        =\langle \nabla_1,\ldots,\nabla_r\rangle_{\mathrm{conv}}\nonumber\\
                &(\nabla_n,\Delta_m)\geq-\delta_{nm}&\\
        \nabla^{\circ}=\langle \Delta_1,\ldots,\Delta_r\rangle_{\mathrm{conv}}
        &&\nabla=\nabla_1+\ldots+\nabla_r\nonumber
\end{eqnarray}

The Batyrev-Borisov mirror construction \cite{Batyrev:1994pg} exchanges $\Delta$'s with
$\nabla$'s so that the mirror manifold is described by equations whose
Newton polytopes are $\nabla_i$. While we have no need for mirror symmetry
in the present context,
the advantage of this class of complete intersections is that combinatorial
formulas for the ``string theoretic'' Hodge numbers have been derived
\cite{Batyrev:1995ca,Kreuzer:2001fu} and implemented in PALP \cite{Kreuzer:2002uu}.

Calabi-Yau 4-folds in toric varieties in general may have terminal
singularities, which are inherited from the ambient space. In order
to resolve as many singularities as possible the first step is to
choose a maximal projective crepant resolution \cite{Batyrev:1994hm}, where crepant
means that we do not change the canonical class. In combinatorial terms
this means that we choose a maximal coherent triangulation of $\Delta^\circ$
and take for $\Sigma$ the fan over the faces of that triangulation.
Since $\mathbb P_\Sigma=\bigcup U_\sigma$ is covered by the affine patches
$U_\sigma$ for $\sigma\in\Sigma$ every singularity is located in some
$U_\sigma$ and the maximal dimension of a component of the singular locus
is equal to the minimal dimension of a singular cone. For
reflexive polytopes the lattice distance of the interior point to every facet
of $\Delta^\circ$ is one so that $U_\sigma$ is non-singular if and only if
$\sigma$ is the cone over a simplex of lattice volume 1. Since all triangles
of a maximal triangulation of a polygon have minimal volume, the
highest-dimensional singularities of $\mathbb P_\Sigma$ come from
4-dimensional cones and therefore have codimension 4. For a Calabi-Yau
$k$-fold $X$ of codimension $r$ the solution set of the $r$ equations
generically avoids singular sets of dimension smaller than $r$ so that $X$
is generically smooth for $k\le 3$.
Depending on the choice of
triangulation, terminal singularities of dimension
$k-4$ are possible, however, for $k>3$,
and no triangulation might exist for which $X$ is smooth. Therefore,
independently of the codimension, smoothness of a CY 4-fold needs to be
checked by computing the volumes of the 4-dimensional cones
$\sigma\in\Sigma^{(4)}$.
Note that string theoretic Hodge numbers are well-defined
and independent of the triangulation even in the singular case \cite{Batyrev:1995ca}.

While fibration structures depend on the intersection ring and therefore
on the (computationally expensive) resolution of singularities,
toric fibrations, which are inherited from morphisms of the ambient space,
have the advantage of being visible as reflexive plane sections
$\Delta^\circ_f$
of $\Delta^\circ$ \cite{Kreuzer:2000qv,Kreuzer:1997zg,Avram:1996pj}. Generically the codimension of the
fiber is equal to the codimension $r$ of the Calabi-Yau because it is
described by the same (number of) equations, while the base is a toric
variety $\mathbb P_{\Sigma^B}$ whose rays are the images of $\Sigma^{(1)}$
under projection along the fiber polytope $\Delta^\circ_f$.
It may happen, however,
that the complete fiber polytope is contained
in one of the $\nabla_i$ of the nef partition \cite{Klemm:2004km}.
Then the remaining equations $f_j=0$ do not contain the coordinates of the
fiber polytope and constrain the base, while the generic fiber becomes
a hypersurface. Since we want a fibration in Weierstrass form we are
interested in precisely this situation with $\Delta_f$ being the
Weierstrass triangle.

A transparent way to describe the polytopes is in terms of their
coordinate-independent weight matrices \cite{2008arXiv0809.1188K} (i.e. charges of
the GLSM, called ``combined weight systems'' in \cite{Kreuzer:2002uu}).
The fibration is then visible, up to permutations of the columns,
as a degree 6 weight vector $(1,\,2,\,3,\,0,\,\ldots\,,\,0)$ in a block form with
the weight matrix of the base and the choice of the remaining integers in
the columns supporting the Weierstrass fiber define the fibration. It is,
of course, a very selective condition that the resulting 6-dimensional
polytope is reflexive and admits a nef-partition of the required form.
In Appendix~\ref{app-models} we will indicate which of the base manifolds
satisfy these constraints.\\

For our purposes we now specialize to the case of a complete intersection of two hypersurfaces, i.e.~$r=2$ in~(\ref{cicy-nef}). The hypersurface constraints can be reconstructed from the toric data as follows:
\begin{equation}
f_m=\sum_{w_k\in\Delta_m}c_k^{m}\prod_{n=1}^{2}\prod_{\nu_i\in\nabla_n}x_i^{\langle\nu_i,w_k\rangle+\delta_{mn}}\qquad m,n=1,2\,,
\end{equation}
where the $c_k^{m}$ are complex structure parameters. So far we have only specified a generic elliptic fibration over the base manifold. In order to define a specific GUT model we also have to specify the GUT group. We can realize this torically by dropping all the monomials from $(\Delta_1,\Delta_2)$ which do not have the GUT group specific vanishing degrees as determined by the Tate classification. This procedure amounts to choosing a very non-generic complex structure. This may entail that the Calabi-Yau may not miss the singularities of the toric ambient space. A detailed discussion of these issues is beyond the scope of this article. Note however that the discrepancy we find for several examples between the Euler numbers computed for the fourfolds using toric methods and those calculated a by formula given in ~\cite{Blumenhagen:2009yv} may be due to additional singularities away from the GUT surface. We plan to return to these issues in future work.

The Tate form~(\ref{tate}) implies that the $a_n$ appear in the monomials which contain $z^n$. We can isolate these monomials by identifying the vertex $\nu_z$ in $(\nabla_1,\nabla_2)$ that corresponds to the $z$--coordinate.  All the monomials that contain $z^r$ are then in the following set:
\begin{equation}
A_r=\{w_k\in\Delta_m\::\:\langle\nu_z,w_k\rangle-1=r\}\qquad \nu_z\in\nabla_m,
\end{equation}
where $\Delta_m$ is the dual of $\nabla_m$, which denotes the polytope containing the $z$--vertex. The polynomials $a_r$ are then given by the following expressions:
\begin{equation}
a_r=\sum_{w_k\in A_r}c_k^{m}\prod_{n=1}^2\prod_{\nu_i\in\nabla_n}y_i^{\langle \nu_i,w_k\rangle+\delta_{mn}}\vert_{x=y=z=1}
\end{equation}
Since we are only interested in the coefficients $a_n$ in the Tate form, we have set those $y_i$ which correspond to $(x,y,z)$ to $1$.

Next we have to select the subset of the $a_n$ which is compatible with the GUT group. This amounts to fixing a particular gauge group on the GUT brane $S$ defined by $w=0$. The order of vanishing of the $a_n$ is dictated by the power of $w$ in the monomial. The Tate classification then implies that $(a_1,a_2,a_3,a_4,a_6)$ can be factored as $(b_5 w^{k_1},b_4 w^{k_2},b_3 w^{k_3},b_2 w^{k_4},b_0 w^{k_6})$ for some positive integers $k_i$. Since the sections $b_i$ can still depend on $w$, we only drop monomials whose $w$--powers are smaller than indicated by the Tate classification.

Using this procedure we get polyhedra $(\tilde{\Delta}_1,\tilde{\Delta}_2)$ which contain fewer points than $({\Delta_1},{\Delta_2})$. The duals $(\tilde{\nabla}_1,\tilde{\nabla}_2)$ will then contain more points. The crepant resolution of this new fourfold probably resolves all the GUT singularities. However again the issue with terminal singularities may arise. In this way we have explicitly constructed the toric Calabi-Yau fourfold which encodes the Euler number for the D3 brane tadpole cancellation condition to a particular F-theory GUT model.

Note that removing points from a polytope is a quite delicate procedure that may destroy certain features such as reflexivity. In our Model 4 discussed in Section~\ref{sec-examples} and Appendix~\ref{fourfold-data} reflexivity is lost for one of the $dP_5$s after imposing the $SO(10)$ Weierstrass model.
%%%%%%%%%%%%%%%%%%%%%%%%%%%%%%%%%%%%%%%%%%%%%%%%%%%%%%%%%%%%%%%%%%%%%%%
\section{Spectral cover}
\label{sec-spectralcover}
Originally the spectral cover construction has been introduced in the context of heterotic string theory as a means to describe stable bundles on elliptically fibered threefolds~\cite{Friedman:1997yq,Donagi:1997pb,Donagi:2004ia,Blumenhagen:2006wj}. Via heterotic/F-theory duality it is natural to use this construction also in F-theory models with heterotic duals. In fact  this structure is intrinsic to eight dimensional supersymmetric Yang-Mills theory. Therefore it was realized that the spectral cover construction does not only apply to models with a heterotic dual, see~\cite{Hayashi:2008ba} and~\cite{Donagi:2009ra}. In type II language its data can be interpreted as B-branes in an auxiliary non-compact Calabi-Yau threefold $X$. The authors of~\cite{Blumenhagen:2009yv,Grimm:2009yu} found that the spectral cover also seems to have some validity beyond the local limit. In~\cite{Blumenhagen:2009yv} a formula based on the spectral cover for the Euler number of the fourfold has been given which has been shown to match with the direct calculation of the Euler number using toric geometry. In the following sections we will put this formula to the test in several examples.

There are
two more equivalent ways to specify an eight dimensional supersymmetric gauge theory: ALE fibration over the GUT surface and $G$-flux and the  Higgs bundle picture. We will not elaborate on them here but it may be useful to switch between these pictures. For further details see~\cite{Donagi:2009ra,Marsano:2009gv}.

%%%%%%%%%%%%%%%%%%%%%%%%%%%%%%%%%%%%%%%%%%%%%%%%%%%%%%%%%%%%%%%
%%%%%%%%%%%%%%%%%%%%%%%%%%%%%%%%%%%%%%%%%%%%%%%%%%%%%%%%%%%%%%%
\subsection{SO(10) models}
\label{SO10-models}
At the end of Section~\ref{sec-toric-base} we have already considered $SO(10)$ singularities (gauge groups) along a surface $S$ ($w=0$). In this section we will give more details about this. The form of the degeneration of the elliptic fiber of the $CY_4$ on $S$ was given by:
\begin{equation}
y^2 = x^3 + b_5w\, xy + b_4 w\, x^2 + b_3 w^2\,y  + b_2 w^3\,x + b_0 w^5,\,\label{eq:local-tate}
\end{equation}
where now we are looking at the patch $z\ne 0$. By completing the square we can rewrite the above equation into the Weierstrass form,
\begin{equation}
\tilde{y}^2 = \tilde{x}^3 + f\tilde{x} + g\,,
\end{equation}
where $f$ and $g$ are sections of $K^{-4}_{B}$ and $K^{-6}_{B}$, respectively, with the following expansion in $w$:
\begin{eqnarray}
f &=& \sum f_m w^m=-\frac{b_4^2}{3}\,w^2 + (b_2 + \frac{b_3\,b_5}{2} - \frac{b_4\,b_5^2}{6})\,w^3 - \frac{b_5^4}{48}\,w^4\,,\\
g &=& \sum g_n w^n=\frac{2\,b_4^3}{27}\,w^3 +\frac{1}{36}\,(9\,b_3^2 - 12\,b_2\,b_4 - 6\,b_3\,b_4\,b_5 + 2\,b_4^2\,b_5^2)\,w^4\nonumber\\
  & &  +\,(b_0 - \frac{6\,b_2\,b_5^2}{72} - 3\,b_3\,b_5^3 + b_4\,b_5^4)\,w^5 + \frac{b_5^6}{864}\,w^6\,.
\end{eqnarray}
For the discriminant, $\Delta=4f^3+27g^2$, we obtain,
\begin{eqnarray}
\Delta &=& b_3^2\,b_4^3\,w^7+\frac{1}{16}\,(27\,b_3^4 + 16\,b_4^2\,(-b_2^2 + 4\,b_0\,b_4)-4\,b_3\,b_4\,b_5\,(9\,b_3^2 + 4\,b_2\,b_4)\nonumber\\
       & & +8\,b_3^2\,b_4\,(-9\,b_2 + b_4\,b_5^2))\,w^8 +\mathcal{O}(w^9)\,.
\end{eqnarray}
Therefore we can now really identify the singularity enhancements stated in Section~\ref{sec-toric-base}:
\begin{itemize}
\item First order enhancement to an $E_6$ singularity along the curve $b_4=0$---$\bf 16$ matter curve.
\item First order enhancement to an $SO(12)$ singularity along the curve $b_3=0$---$\bf 10$ matter curve.
\item Second order enhancement to an $E_7$ singularity over the intersection points of $b_4=0$ and $b_3=0$---$\bf 16\,16\,10$ Yukawa coupling points.
\item Second order enhancement to an $SO(14)$ singularity over the intersection points of $b_3=0$ and $b_2^2-4 b_0 b_4=0$---$\bf 10\,10\,16$ Yukawa coupling points.
\end{itemize}
The reason for the labeling of the curves and Yukawa couplings comes from group theory and the fact that the singularities of the fibration are directly related to the gauge theory on $S$~\cite{Bershadsky:1996nh,Katz:1996xe}.
For an $E_8$ singularity we would have $E_8$ SYM on $S$, and hence, matter in the adjoint representation of $E_8$. If we now deform the singularity to $SO(10)$ we obtain the following breaking pattern:
\begin{eqnarray}
E_8 & \supset & %SO(16) \supset
SU(4)\times SO(10)\nonumber\\
\textbf{248} %& \rightarrow & \textbf{120}\,+\,\textbf{128} \\
& \rightarrow & (\textbf{15},\textbf{1})\,+\,(\textbf{1},\textbf{45})\,+\,(\textbf{6},\textbf{10})\,
+\,(\textbf{4},\textbf{16})\,+\,(\overline{\textbf{4}},\overline{\textbf{16}})\,.
\end{eqnarray}
The \textbf{45} is the adjoint representation of the $SO(10)$ gauge theory on $S$. The $\textbf{16}$, $\overline{\textbf{16}}$ and the fundamental representation arise at the curves of higher singularity.
To see that we have a \textbf{10} matter curve we can use the IIB picture.
In type IIB we obtain a $SO(12)$ gauge group if we place six D-branes on top of the orientifold plane. If we now tilt one of the branes we reduce the  gauge theory on the O-plane to $SO(10)$ and get matter in the fundamental representation of $SO(10)$ along the intersection of the tilted brane and the O-plane. For the \textbf{16} matter curve we have to use group theory since one does not have a IIB analogue at hand. For the adjoint of $E_6$  we observe the following breaking pattern:
\begin{eqnarray}
E_6 & \supset  &
U(1)\times SO(10)\nonumber\\
\textbf{78} %& \rightarrow & \textbf{120}\,+\,\textbf{128} \\
& \rightarrow &
\textbf{1}_0\,+\,\textbf{45}_0\,+\,\textbf{16}_{-3}\,+\,\overline{\textbf{16}}_{3}\,,
\end{eqnarray}
where \textbf{45} is again the adjoint representation of the $SO(10)$ gauge theory on $S$. Hence, \textbf{16} matter arises at the $E_6$ enhancement. To see the labeling of the Yukawas, one has to look at the cubic coupling of the adjoints of $E_7$ and $SO(14)$, respectively, and its breaking to $SO(10)$.

Now we go from the global picture to the local picture in the vicinity of $S$. We do this by extracting an ALE-fibration from~(\ref{eq:local-tate}). First we 'localize' the $b_i$s by removing the $w$-dependent part such that they become sections of bundles on $S$. From now on we will denote by $b_i$ the 'localized' $b_i$s which are sections of the following bundles:
\begin{equation}
 b_i\in \Gamma(K_S^{-(6-i)}\otimes\mathcal{N}_{S|B})\qquad i=0,\ldots,4\,,
\end{equation}
and $b_5\in \Gamma(K_S^{-1})$. Then like in~\cite{Donagi:2009ra}, we assign the scaling dimensions $(\frac{1}{3},\frac{1}{2},\frac{1}{5})$ to $(x,y,w)$ in~(\ref{eq:local-tate}) such that the terms of order one give the $E_8$ singularity. If we drop all terms of scaling dimension greater than one we are left with
\begin{equation}
 y^2 = x^3 + b_4 w\, x^2 + b_3 w^2\,y  + b_2 w^3\,x + b_0 w^5\,,
\end{equation}
where the terms with dimension less than one are relevant deformations of the $E_8$ singularity. Hence we see that $b_5$ does not appear in the local construction. This we perceived already in our analysis above where it did not show up in the singularity considerations on $S$.
Since the intersection matrix of the two-cycles introduced by the partial resolutions of the $E_8$ singularity by relevant deformations is minus the Cartan matrix of $A_3$, we can equivalently describe these resolutions by an $A_3$ ($SU_4$) singularity that is completely resolved. Thus, we obtain an $\widehat{SU_4}$ ALE-fibration:
\begin{equation}
 y^2=x^2+b_0 s^4 +b_2 s^2+b_3 s+b_4\,,\label{eq:ALE-su4}
\end{equation}
where $s$ is a coordinate on $K_S$. As explained in~\cite{Donagi:2009ra}
the ALE fibration and the Higgs bundle are  equivalent  descriptions for the eight dimensional
gauge theory on $S$. The information about the Higgs bundle is encoded in the spectral cover.
Since this is a more appropriate description for our purpose, we elaborate on it in the next section.

%%%%%%%%%%%%%%%%%%%%%%%%%%%%%%%%%%%%%
\subsection{$SU(4)$ spectral cover}

The spectral cover was introduced in heterotic string theory to encode
the breaking of the $E_8$ gauge symmetry to some gauge
group $H$. It characterizes
all the information about the bundle with structure group $G$
($\subset E_8$) that is the commutant of $H$ in $E_8$. In our
case, we have an $SO(10)$ gauge group on $S$, and thus $G$ is $SU(4)$.
Since we also arrived at the $SU(4)$ in our above reasoning we
have convincing evidence that the spectral cover is applicable, locally in the vicinity of the GUT divisor $S$, even for F-theory models without a heterotic dual.

The starting point of the construction is the local Calabi-Yau threefold $X$ = ($K_S\,\rightarrow\,S$)~\cite{Donagi:2009ra}. It is convenient to compactify this non-compact CY to
\begin{equation}
 \bar X=\mathbb{P}( K_S\oplus\mathcal{O}_S) \quad\textmd{with}\quad\pi:\,\bar X\rightarrow S \,,
\end{equation}
where $[u:v]$ are the homogeneous coordinates of the fiber. $\bar X$ is a compact space but no longer Calabi-Yau. The divisor classes of the sections $\sigma$ and $\sigma_c$ are $\sigma\sim u$ and $\sigma_c\sim v$ in $\bar X$, respectively, where $\sigma_c$ denotes the section at infinity. As in the non-compact case $S$ is given by the zero section, $\sigma=0$. Since $\sigma_c\cdot\sigma=0$ we have:
\begin{equation}
 \sigma \cdot \sigma=-\sigma\cdot c_1(S)\,.
\end{equation}
The first Chern class of $\bar X$ is $c_1(\bar X)=2 \sigma_c=2(\sigma+c_1(S))$. The $SU(4)$ spectral cover is now given by the hypersurface,
\begin{equation}
 b_0 s^4+b_2 s^2+b_3 s + b_4 =0\,,\label{SU4cover}
\end{equation}
inside $\bar X$ where $s$ is $u/v\sim c_1(S)$ and the $b_i$s are the sections from equation~(\ref{eq:ALE-su4}). Together with the projection $\pi$ of $\bar X$ this induces a fourfold cover of $S$. For later convenience we denote the divisor $b_0=0$ on $S$ by $\eta$. This gives us the following relations between the divisors on $S$:
\begin{equation}
 b_i\sim \eta - i\,c_1(S)\sim (6-i)c_1(S) -t\qquad i=0,\ldots,4\quad,\label{eq:classlocsec}
\end{equation}
where $-t$ is a section of $\mathcal{N}_{S|B}$. From equation~(\ref{SU4cover}) and~(\ref{eq:classlocsec}) we obtain
\begin{equation}
 [C_V]=4\sigma+p_4^*\eta
\end{equation}
for the divisor class of the spectral cover, where $p_4$: $C_V\rightarrow S$ denotes the projection from the cover to the GUT surface and $V$ denotes the fundamental representation of $G$.

As mentioned in the beginning of this section, from a type II perspective the spectral cover can be interpreted as an auxiliary flavor brane. In type II string theory (fundamental) matter arises at the intersection of two branes. Hence, the matter curve(s) in $\bar X$ are given by the intersection of $C_V$ with $\sigma=0$. The corresponding curve $\Sigma_{V}$ ($\Sigma_{\bf 16}$) on $S$ can be calculated by
\begin{equation}
[C_{V}]\cdot\sigma=(4\sigma+\pi^{\ast}\eta)\cdot\sigma=\sigma\cdot\pi^{\ast}(\eta-4c_1),
\end{equation}
which implies that $\Sigma_{\bf 16}\sim\eta-4c_1$. This is in accord with our singularity analysis above because $b_4\sim\eta-4\,c_1$. Since $b_4$ is the product of the four roots of~(\ref{SU4cover}), denoted by $\lambda_i$ in the following, $\Sigma_{\bf 16}$ is determined by $\lambda_i=0$. Na\"ively one would guess that one obtains four distinct curves from the four solutions $\lambda_i=0$, but this is of course not true. This happens only in the case where the polynomial $b_4$ factors, as we will see in Section~\ref{spec-cov-split}. The reason for this is that the $\lambda_i$s are
multivalued functions. So monodromies connect the $\lambda_i$s, and only when $b_4$ splits also the monodromy group splits.

As mentioned above the $E_8$ symmetry is broken by a VEV
of the two cycles in the ALE fibration on $S$.  Locally the  $G$ flux of the ALE fibration over $S$ can be encoded
in terms of the spectral line bundle $\mathcal{L}$ along the
spectral cover $C_V$. Since $C_V$ is a four cover
of $S$, after pushing forward $\mathcal{L}$ one can define a
rank four vector bundle $V=p_{4\ast}\mathcal{L}$. The first Chern class of $V$ has to vanish because we have an
$SU(4)$ cover, which implies:
\begin{equation}
0= c_1(p_{4\ast} \mathcal{L})=\pi_{\ast}(\mathcal{L}-\frac{1}{2}r),
\end{equation}
where   $r$ is the ramification divisor
$r=\pi^{\ast}c_1(S)-c_1(C_V)$. If we assume:
\begin{equation}
c_1(\mathcal{L})= \frac{1}{2}r + \gamma_u,
\end{equation}
then $\pi_{\ast}\gamma_u =0$. $\gamma_u$ is the universal
flux and for the $SU(4)$ cover $C_V$ we get:
\begin{equation}
\gamma_u = C_V\cdot(4\sigma-\pi^{\ast}(\eta-4c_1))= 4 \Sigma_{V} -
\pi^{\ast}(\eta-4c_1(S)). \label{universalflux}
\end{equation}
The net chirality for fermions on the curve $\Sigma_{\bf 16}$ is
given by \cite{Curio:1998vu,Diaconescu:1998kg,Hayashi:2008ba}:
\begin{equation}
n_{\bf 16}-n_{\overline{\bf 16}} = \int_{\Sigma_V} \gamma =
-\eta\cdot (\eta-4c_1(S))\,.
\end{equation}

Up to now we have only discussed the {\bf 16} matter curve. The
$\bf 10$ curve can be obtained from a $\wedge^2 V$ bundle.  In
terms of the solutions of the spectral cover (\ref{SU4cover}) the
roots associated to $\wedge^2 V$ are given by $\lambda_i+
\lambda_j=0$. Thus the spectral surface $C_{\wedge^2 V}$ can be
represented as:
\begin{eqnarray}
C_{\wedge^2 V}:~~ b_0^2\prod_{i<j} (s+\lambda_i +\lambda_j)=b_0^2
s^6+ 2b_0 b_2 s^4 + ( b_2^2 -4b_0 b_4)s^2  - b_3^2 . \label{CV^2}
\end{eqnarray}
Therefore $b_3=0\cap s=0$ defines a $\bf 10$ curve. Using the
formula given in \cite{Donagi:2004ia}, or from (\ref{CV^2}), it
would be natural to write $C_{\wedge^2 V} = 6\sigma +
2\pi^{\ast}\eta$. We denote the curve supporting the $\bf 10$
representation by $\Sigma_{\bf 10}=C_{\wedge^2 V}\cdot \sigma$.
However usually $C_{\wedge^2 V}$ is singular, therefore we would
like to construct a resolved cover via $C_V$. Consider the intersection
of $C_V$ and its image under an involution $\tau$. Then one can
make the following decomposition \cite{Donagi:2004ia}:
\begin{eqnarray}
\tau C_V\cap C_V = C_V\cdot \sigma + C_V\cdot 3\sigma_{c} +D.
\end{eqnarray}
Here $\sigma$ is the zero section and $3\sigma_c$ is the
trisection intersecting the fiber at the three non-trivial points.
$D$ is contained in $X$. It is the double cover of the support
curve $\Sigma_{\wedge^2 V}$ with the map $\pi_D: D \rightarrow
\Sigma_{\wedge^2 V}$ and it can be written as $D =
C_V\cap(C_V-\sigma -3 \sigma_{c})$.

Since the spectral surface $C_{\wedge^2 V}$ forms a double curve,
we need to resolve this double-curve singularity. After blowing up
the singularities we can define a degree two cover
$\tilde{\Sigma}_{\wedge^2 V}$ with a projection $\tilde{\pi}_D: D
\rightarrow \tilde{\Sigma}_{\wedge^2 V}$ \cite{Hayashi:2008ba}.
The $\bf 10$ curve is then obtained from $D\cdot \sigma$
which implies the class of $\Sigma_{\bf 10}$
is~\cite{Hayashi:2008ba}:
\begin{equation}
\Sigma_{10} = -3c_1 + \eta.
\end{equation}

After further resolving the codimension-2 singularities along
$C_{\wedge^2 V}$ we obtain the following expression for the chirality for
$\Sigma_{\bf 10}$ \cite{Hayashi:2008ba}:
\begin{eqnarray}
\chi(\wedge^2 V) &=& \int_{\tilde \Sigma_{\wedge^2 V}} \tilde
\pi_{D\ast} \gamma = \int_{\Sigma_{\wedge^2 V}} \pi_{D\ast} \gamma
=0.
\end{eqnarray}
Thus, the net chirality of the $\bf 10$ curve always vanishes for the $SU(4)$
cover, which is not very
promising for the $SO(10)$ GUT model construction, because of the
absence of the $\bf 10$ Higgs.  Even if we turn on a flux in
the bulk to break the $SO(10)$ gauge group to $SU(5)$,
the structure is not abundant enough to realize a realistic model.
Therefore in what follows we will consider splitting the spectral
cover to enrich this construction.

%%%%%%%%%%%%%%%%%%%%%%%%%%%%%%%%%%%%%%%
\subsection{Spectral cover splitting} \label{spec-cov-split}
Trying to build $SO(10)$ GUT models from $SU(4)$ cover we
encounter the problem that the chirality of the $\bf 10$ curve is
generically zero. A possible way to solve this problem is to
factorize the spectral curve in order to obtain chiral matter on
each of the individual curves.  A convenient choice is the $(1,3)$
factorization. This means that we split the cover group $SU(4)$
into $S[U(3)\times U(1)]$\footnote{The reason we do not write
$SU(3)\times U(1)$ is that we only demanded $c_1(V)=0$ for the
$SU(4)$ vector bundle before the cover is split. The $U(1)$ can be split off
as a gauge symmetry if one extends the spectral cover
factorization to a global restriction of the Tate model
\cite{Grimm:2010ez}. In this case, $E_8\rightarrow SO(10)\times
[SU(3)\times U(1)$] and the gauge group of the GUT model can be
taken as $SO(10)\times U(1)$.}. The total group structure of the
$E_8$ breaking is:
\begin{eqnarray}
E_8 & \supset & SU(4)_\bot \times SO(10) \supset S[U(3)\times U(1)] \times SO(10)\nonumber\\
{\bf 248} & \rightarrow & (\textbf{15},\textbf{1})\,+\,(\textbf{1},\textbf{45})\,+\,(\textbf{6},\textbf{10})\,+\,(\textbf{4},\textbf{16})\,+\,(\overline{\textbf{4}},\overline{\textbf{16}})\nonumber\\
& \rightarrow & (\textbf{8},\textbf{1})_0\,+\,(\textbf{3},\textbf{1})_{-4}\,+\,(\overline{\textbf{3}},\textbf{1})_{4}\,+\,(\textbf{1},\textbf{1})_0\, +\,(\textbf{1},\textbf{45})_0 \,+\,(\textbf{3},\textbf{10})_{2}\\& &+\,(\bar{\textbf{3}},{\textbf{10}})_{-2}\,+\,(\textbf{1},\textbf{16})_{3}\,+\,(\textbf{3},\textbf{16})_{-1}\,+\,(\textbf{1},\overline{\textbf{16}})_{-3}\,
+\,(\bar{\textbf{3}},\overline{\textbf{16}})_{1}\nonumber
\end{eqnarray}
The cover $C_V$ can be factorized as $C_V \rightarrow C^{(1)} +
C^{(3)}$. With the homogeneous coordinates $[u:v]$ of ${\mathbb{P}}_1$
the polynomial (\ref{SU4cover}) turns out to be:
\begin{eqnarray}
b_0u^4+b_1u^3v+b_2u^2v^2+b_3uv^3+b_4v^4=(a_0 u^3 + a_1 u^2v +a_2
uv^2 + a_3 v^3)(d_0 u +d_1v)=0\,,
\end{eqnarray}
where
\begin{eqnarray}
b_0 = a_0 d_0,~~ b_1 = a_1 d_0 + a_0 d_1=0,~~ b_2 = a_2 d_0 + a_1
d_1,~~ b_3 = a_3 d_0 + a_2 d_1,~~ b_4 = a_3 d_1.
\end{eqnarray}
Since the factorization is not unique, we make the following assumption for the corresponding class:
\begin{equation}
[d_0]= c_1+\xi,~~ [d_1]= \xi, ~~ [a_i]=\eta -(i+1)c_1 -\xi.
\end{equation}
Therefore we can write the class of the split cover as:
\begin{equation}
C_V=C^{(1)} + C^{(3)}=(\sigma+\pi^{\ast}(c_1+\xi))+
(3\sigma+\pi^{\ast}(\eta -c_1-\xi))
\end{equation}

As mentioned above the $\bf 10$ curve is obtained from the
bundle $\wedge^2 V$, i.e.~from $C_V\cap \tau
C_V$ with the involution $\tau: v\rightarrow -v$.  After splitting
the cover we get:
\begin{eqnarray}
C_V\cap \tau C_V\rightarrow C^{(1),(1)} + C^{(1),(3)}+C^{(3),(1)}
+C^{(3),(3)}.
\end{eqnarray}
Following the monodromy group analysis of the roots $\lambda_i$ in
\cite{Donagi:2008kj, Donagi:2009ra, Marsano:2009gv,
Marsano:2009wr}, we find that $C^{(1),(1)}$ for $\bf 10$ cannot be
realized on the surface.  Therefore there are only two $\bf 10$
curves. The corresponding solutions for the matter curves in terms
of $\lambda_i$ after factorization are:
\begin{eqnarray}
\Sigma^{(1)}_{\bf 16}:&& \{\lambda_4\} \nonumber \\
\Sigma^{(3)}_{\bf 16}:&&  \{\lambda_1, \lambda_2, \lambda_3\} \nonumber\\
\Sigma^{(13)}_{\bf 10}: && \{ \lambda_1+\lambda_4,
\lambda_2+\lambda_4, \lambda_3+\lambda_4\}, \nonumber \\
\Sigma^{(33)}_{\bf 10}: && \{ \lambda_1+\lambda_2,
\lambda_2+\lambda_3, \lambda_3+\lambda_1 \}.
\end{eqnarray}

\begin{table}[h]
\begin{center}
\renewcommand{\arraystretch}{1}
\begin{tabular}{|c|c|c|c|} \hline
& $C^{(1),(1)}$ & $C^{(1),(3)}+C^{(3),(1)}$ & $C^{(3),(3)}$ \\
\hline

{\bf 16} & $\sigma\cdot\pi^{\ast}\xi$ & - & $\sigma\cdot\pi^{\ast}
(\eta-4c_1-\xi)$ \\ \hline

\multirow{2}{*}{\bf 10} & \multirow{2}{*}{$\pi^{\ast}
\xi\cdot\pi^{\ast} (c_1+\xi)$} & $2(\sigma+\pi^{\ast}(c_1+\xi))$ &
$(2\sigma+\pi^{\ast}(\eta-2c_1 -\xi))$\\
& & $\cdot~\pi^{\ast}(\eta-3c_1-\xi)+2\sigma\cdot \pi^{\ast}\xi$ &
$\cdot~\pi^{\ast}(\eta -3c_1-\xi)+2(\sigma+\pi^{\ast}c_1)\cdot
\pi^{\ast} \xi$ \\ \hline

\multirow{2}{*}{$\infty$} & \multirow{2}{*}{$\sigma_{c}
\cdot\pi^{\ast} (c_1+\xi)$} & \multirow{2}{*}{4$\sigma_{c}
\cdot \pi^{\ast} (c_1+\xi)$} & $\sigma_{c} \cdot
\pi^{\ast}(\eta-c_1 -\xi)$ \\
& & & $+2\sigma_{c} \cdot \pi^{\ast}(\eta -2c_1-2\xi)$
\\ \hline
\end{tabular}
\caption{Factorization $C_V=C^{(1)} + C^{(3)}$} \label{3-1}
\end{center}
\end{table}

The matter curves corresponding to $C^{(1)}$ and $C^{(3)}$ are
$\Sigma^{(1)}_a$ and $\Sigma^{(3)}_b$, where:
\begin{equation}
\Sigma^{(1)}_a=C^{(1)}\cdot \sigma,~~ \Sigma^{(3)}_b=C^{(3)}\cdot
\sigma.
\end{equation}
The details about the components of the curves are listed in Table
\ref{3-1} \footnote{In the table we abused notation. The covers
for the two $\bf 16$ curves should actually be denoted by $C^{(1)}$ and~$C^{(3)}$.}. In what follows, we summarize some properties of
this $(1,3)$ cover factorization. Our notation
follows~\cite{Marsano:2009gv, Marsano:2009wr}.

\subsection{Universal flux}

For an $SU(n)$ cover the first Chern class vanishes: $c_1(V)=0$.
Since the cover factorizes into $C^{(1)}$ and $C^{(3)}$, by
using $c_1(V_3\oplus L)=c_1(V_3)+c_1(L)$, the traceless condition
turns out to be:
\begin{equation}
c_1(p_{1\ast}\mathcal{L}^{(1)})+c_1(p_{3\ast}\mathcal{L}^{(3)})=0,
\end{equation}
where again $p_{i}: C^{(i)}\rightarrow S$. For the flux to be well
defined and supersymmetric, there are two more conditions
\cite{Marsano:2009wr}:
\begin{eqnarray}
&&\mathcal{L}^{(i)}\in H^2(C^{(i)},\mathbb{Z})\\
&&c_1(p_{1\ast}\mathcal{L}^{(1)})-c_1(p_{3\ast}\mathcal{L}^{(3)})~{\rm
is~SUSY~on~}S.
\end{eqnarray}
More details about these conditions will be discussed below.

The ramification divisor for the map $p_i$ on the cover $C_V$ is $
r= p^{\ast}c_1 -c_1(C_V)$. So for each component of the split
cover we get:
\begin{equation}
r_i=p_i^{\ast} c_1 -c_1(C^i),~~ c_1(C^i)=(c_1(T_{\bar
X})-C^i)\cdot C^i\,.
\end{equation}
The ramification
divisors for a $(1,3)$ factorization are:
\begin{eqnarray}
r^{(1)}_a=(-\sigma+\pi^{\ast}\xi)\cdot C^{(1)},~~
r^{(3)}_b=(\sigma+\pi^{\ast}(\eta-2c_1-\xi))\cdot C^{(3)}.
\end{eqnarray}
Using the formula $c_1(p_{i\ast}\mathcal{L}^{(i)})=p_{i\ast}
c_1(\mathcal{L}^{(i)})-\frac{1}{2}p_{i\ast}r_i$ we can then define
the flux on each cover:
\begin{equation}
\gamma_i=c_1(\mathcal{L}^{(i)})-\frac{1}{2}r_i~.
\end{equation}

\subsubsection{Splitting of the universal flux}

The universal flux encoded in the $n=4$ spectral cover has been
given in (\ref{universalflux}).  This flux is also separated due to the split spectral cover. The corresponding universal flux on each
factorized curve is:
\begin{eqnarray}
&&\gamma_a=\Sigma_a-p_a^{\ast} p_{a\ast} \Sigma_a =C^{(1)}\cdot
\left(  \sigma-\pi^{\ast}\xi\right), \nonumber \\
&&\gamma_b=3\Sigma_b-p_b^{\ast} p_{b\ast} \Sigma_b =C^{(3)}\cdot
\left( 3 \sigma-\pi^{\ast}(\eta-4c_1-\xi)\right).
\end{eqnarray}
Motivated by the expansion of $\gamma_u$ when splitting the cover,
there are two more choices of fluxes:
\begin{eqnarray}
&&\delta_a=3\Sigma_a-p_b^{\ast} p_{a\ast} \Sigma_a =C^{(1)}\cdot
3\sigma- C^{(3)}\cdot\pi^{\ast}\xi, \nonumber \\
&&\delta_b=\Sigma_b-p_a^{\ast} p_{b\ast} \Sigma_b =C^{(3)}\cdot
\sigma- C^{(1)}\cdot \pi^{\ast}(\eta-4c_1-\xi).
\end{eqnarray}
In addition, there is a third type of universal flux
$\tilde{\rho}$ that can be included~\cite{Marsano:2009wr}:
\begin{equation}
\tilde{\rho}=3\pi_a^{\ast}\rho-\pi_b^{\ast}\rho,~~ \rho\in
H_2(S,\mathbb{R}).
\end{equation}
$\rho$ does not have to be effective because we can build
$\tilde{\rho}$ with any real linear combination of
$\tilde{\rho}_i$ from effective $\rho_i$~\cite{Marsano:2009wr}.

We summarize the contributions of the above flux components to the chiralities of each factorized matter curve in the following two tables:
\begin{equation}
\begin{tabular}{l|c||c|c}
& class in $S$ & $\gamma_a$ & $\gamma_b$ \\\hline

${\bf 16}^{(1)}_a$ & $\xi$ & $-\xi(c_1+\xi)$ & 0 \\ \hline

${\bf 16}^{(3)}_b$ & $\eta-4c_1-\xi$ & 0 &
$-(\eta-c_1-\xi)\cdot(\eta-4c_1-\xi)$
\\\hline

${\bf 10}^{(1,3)}_{ab}$ & $\eta-3c_1$ & $-\xi(c_1+\xi)$ &
$-(\eta-3c_1-3\xi)\cdot(\eta-4c_1-\xi)$
\\\hline

${\bf 10}^{(3,3)}_{bb}$ & $\eta-3c_1$ & 0 &
$(\eta-3c_1-3\xi)\cdot(\eta-4c_1-\xi)$
\\\hline
\end{tabular} \label{splitn1}
\end{equation}

\begin{equation}
\begin{tabular}{l|c|c|c}
& $\delta_a$ & $\delta_b$ & $\tilde{\rho}$ \\\hline

${\bf 16}^{(1)}_a$ & $-3c_1\xi$ & $-\xi\cdot (\eta-4c_1-\xi)$ &
$3\rho\cdot \xi$ \\ \hline

${\bf 16}^{(3)}_b$ & $-\xi\cdot (\eta-4c_1-\xi)$ &
$-c_1\cdot(\eta-4c_1-\xi)$ & $-\rho\cdot (\eta-4c_1-\xi)$
\\\hline

${\bf 10}^{(1,3)}_{ab}$ & $\xi\cdot(2\eta-9c_1-3\xi)$ &
$-(\eta-3c_1-\xi)\cdot(\eta-4c_1-\xi)$& $2\cdot\rho(\eta-3c_1)$
\\\hline

${\bf 10}^{(3,3)}_{bb}$ & $-2\xi\cdot (\eta-3c_1)$ &
$(\eta-3c_1-\xi)\cdot(\eta-4c_1-\xi)$ & $-2\rho\cdot (\eta-3c_1)$
\\\hline
\end{tabular}  \label{splitn2}
\end{equation}

The total universal flux is then the linear combination of these
pieces:
\begin{equation}
\gamma_u=k_a \gamma_a+k_b\gamma_b +d_a \delta_a +d_b \delta_b +
\tilde{\rho}  \label{total flux}
\end{equation}
Thus, the fluxes on each component of the split cover are:
\begin{eqnarray}
&&\gamma_{ua}=C^{(1)}\cdot \left[(k_a+3d_a)\sigma
-\pi^{\ast}(k_a\xi+d_b(\eta-4c_1-\xi)-3\rho) \right], \nonumber \\
&&\gamma_{ub}=C^{(3)}\cdot \left[(3k_b+d_b)\sigma -\pi^{\ast}(k_b
(\eta-4c_1-\xi) +d_a\xi + \rho) \right],
\end{eqnarray}
where
\begin{eqnarray}
&&p_{a\ast} \gamma_{ua} = 3d_a\xi- d_b(\eta-4c_1-\xi)+3\rho \nonumber \\
&&p_{b\ast} \gamma_{ub} = -3d_a\xi +d_b(\eta-4c_1-\xi)-3\rho
\label{pullflux}
\end{eqnarray}
The coefficients have to obey the following quantization
conditions \cite{Marsano:2009wr}:
\begin{eqnarray}
(k_a+3d_a+\frac{1}{2})\sigma -\pi^{\ast}(k_a\xi
+d_b(\eta-4c_1-\xi)-3\rho -\frac{1}{2}\xi)
\in H_4(\bar X,\mathbb{Z}) , \nonumber \\
(3k_b+d_b-\frac{1}{2})\sigma -\pi^{\ast}(k_b(\eta-4c_1-\xi)+
d_a\xi + \rho-\frac{1}{2}(\eta-2c_1-\xi)) \in H_4(\bar X,\mathbb{Z}).
\end{eqnarray}

\subsubsection{Chirality on the curves}

From~(\ref{total flux}) we can summarize the number of generations on the matter curves as:
\begin{eqnarray}
n_{{\bf 16}^{(1)}_a} &=& (d_b-k_a)\xi^2
-d_b\xi\eta+(4d_b-k_a-3d_a)\xi c_1 + 3\rho\xi, \\
n_{{\bf 16}^{(3)}_b} &=&
-k_b(\eta^2+4c_1^2-5c_1\eta)+d_b(4c_1^2-c_1\eta)+(d_a-k_b)\xi^2+
(2k_b-d_a)\xi\eta  \nonumber \\&& +(4d_a-5k_b+d_b)\xi
c_1 -\rho\eta+4\rho c_1+\rho\xi,  \\
n_{{\bf 10}_{ab}} &=&
-(k_b+d_b)(\eta^2-7c_1\eta+12c_1^2)-(k_a+3k_b+d_b+3d_a)\xi^2
+(4k_b+2d_b+2d_a)\xi\eta \nonumber \\ &&-(k_a+15k_b+7d_b+9d_a)\xi
c_1 + 2\rho\eta-6\rho c_1, \\
n_{{\bf 10}_{bb}} &=&
(k_b+d_b)(\eta^2-7c_1\eta+12c_1^2)+(k_b+d_b)\xi^2
-(4k_b+2d_b+2d_a)\xi\eta \nonumber \\ &&+(15k_b+7d_b+6d_a)\xi c_1
- 2\rho\eta+6\rho c_1,
\end{eqnarray}

\subsubsection{Tadpole condition}

The tadpole condition for D3-branes is
\begin{equation}
N_{D3}=\frac{\chi(Y)}{24}-\frac{1}{2}\int_{Y} G\wedge G
\end{equation}
If there are non-abelian singularities in $Y$, they account for
additional contributions to the Euler characteristics, which are
\cite{Blumenhagen:2009yv}:
\begin{eqnarray}
\chi_{SU(n)}=\int_{S_{GUT}} \left[ c_1^2(n^3-n)+3n\eta(\eta-nc_1) \right] \label{chiSU}\\
\chi_{E_8} =\int_{S_{GUT}} 120(-27c_1\eta+62c_1^2+3\eta^2)~.~
\end{eqnarray}
For $n=4$,
\begin{equation}
\label{eulerformula}
\chi(X_4)=\chi(X^{\ast}_4)+\chi_{SU(4)}-\chi_{E_8}
\end{equation}
These equations were derived for models with a heterotic dual and conjectured to hold also for more general F-theory models in~\cite{Blumenhagen:2009yv}. We will put this formula to the test for the examples we will present in Section~\ref{sec-examples}. For three out of five examples find a discrepancy between the Euler numbers computed by~(\ref{eulerformula}) and those computed from the geometry of the Calabi-Yau fourfold. We will discuss possible causes for this discrepancy in the Conclusions.

For a $(1,3)$ factorization the bundle changes as follows: $SU(4)\rightarrow
S[U(3)\times U(1)]$. As a consequence the Euler number becomes
\cite{Blumenhagen:2009yv,Marsano:2009wr}:
\begin{equation}
\label{muceuler}
\chi(X_4)=\chi(X^{\ast}_4)+\chi_{SU(1)}+\chi_{SU(3)}-\chi_{E_8}
\end{equation}
The class $\eta$ in~(\ref{chiSU}) is decomposed into:
\begin{equation}
\eta^{(3)}=\eta-(c_1+\xi),~~~ \eta^{(1)}=c_1+\xi.
\end{equation}
Thus one gets:
\begin{equation}
\chi(X_4)=\chi(X^{\ast}_4)+\int_S 3\left(3\eta^2 +20c_1^2
-15c_1\eta +4\xi^2 -6 \xi\eta + 16\xi c_1 \right) -\chi_{E_8}
\end{equation}
On the other hand we have:
\begin{equation}
\frac{1}{2}\int_{X_4} G\wedge G =-\frac{1}{2}\gamma^2 =
-\frac{1}{2} (\gamma_a\cdot\gamma_a + \gamma_b \cdot \gamma_b ),
\end{equation}
and
\begin{eqnarray}
\gamma^2  &=& -(k_a+3d_a)^2\xi(c_1+\xi)
-\frac{1}{3}(3k_b+d_b)^2 (\eta-c_1-\xi)(\eta-4c_1-\xi) \nonumber \\
&& +\frac{4}{3}[d_b(\eta-4c_1-\xi)-3d_a\xi-3\rho]^2
\end{eqnarray}
The D3-branes are added to cancel the tadpoles. We want to avoid
a situation with anti-D3-branes, therefore the constraint is:
\begin{equation}
N_{D3}\geq 0.
\end{equation}

\subsubsection{Supersymmetry condition for the universal flux}

We have the condition that $c_1(p_{a\ast}\mathcal{L}_a)
-c_1(p_{b\ast}\mathcal{L}_b)$ is the Poincar\'{e} dual of a
supersymmetric cycle in $S$. Recall that $\mathcal{L}$ is the line bundle on
the cover and $p_i$ denotes the projection map from $C^{(i)}$ to
$S$. For the flux $\gamma$, the
supersymmetry constraint restricted on $S$ is:
\begin{equation}
\omega\cdot (p_{i\ast} \gamma_{ui})=0, \label{susyC}
\end{equation}
where $p_{i\ast}\gamma_{ui}$ is as in~(\ref{pullflux}) in the $(1,3)$
splitting. $\omega$ is an ample divisor dual to K\"ahler form $J$
of $S$:
\begin{equation}
\omega=\alpha_i \mathbf{C}_i, \label{susy}
\end{equation}
where the $\mathbf{C}_i$ are the generators of the Mori cone of $dP_n$,
and $\alpha_i>0$.

%%%%%%%%%%%%%%%%%%%%%%%%%%%%%%%%%%%%%%%%%%%%%%%%%%
%%%%%%%%%%%%%%%%%%%%%%%%%%%%%%%%%%%%%%%%%%%%%%%%%%
\section{Explicit examples}
\label{sec-examples}

\subsection{Geometric backgrounds} \label{sunsec-geo}

The GUT surface is a del Pezzo surface as well as a divisor in
the base $B$. Since we are interested in the physics on the GUT
surface, it is more convenient to extract the useful information
from the ambient space. We will present the geometry of the del
Pezzo surface $dP_k$ in terms of the hyperplane divisor $H$ of
$\mathbb{P}^2$ and the exceptional divisors $E_i$, $i=1,2,\dots,k$
with intersection numbers
\begin{equation}
H\cdot H=1,~~ H\cdot E_i=0,~~ E_i\cdot E_j=-\delta_{ij},~~\forall~
i,j~.
\end{equation}
The canonical divisor of $dP_k$ is then
\begin{equation}
K_S=-3H+\sum_i^k E_i.
\end{equation}
In what follows we will discuss five examples.

\subsubsection{Model $1$}
The starting point for this model is the Fano hypersurface $\mathbb{P}^4[4]$ where we perform two curve blowups. The weight matrix for the blowup of one curve is given in table~\ref{tab-p41curve1} in Appendix~\ref{app-models}. The second curve blowup is realized by adding line two of table~\ref{tab-p42curve} to the weight matrix in table~\ref{tab-p41curve1} whereby one adds a seventh column to table~\ref{tab-p41curve1} filled with zeros. For convenience reasons we rearrange this a bit and obtain the following table:
\begin{equation}
  \begin{array}{|c|c|c|c|c|c|c|c|}
\hline
 y_{1} & y_{2} & y_{3} & y_{4} & y_5 & y_6 & y_7 & \sum \tabularnewline
\hline\hline
1 & 1 & 0 & 0 & 1 & 1 & 1 & 5\\ \hline
1 & 0 & 0 & 1 & 0 & 0 & 1 & 3 \\ \hline
0 & 1 & 1 & 0 & 0 & 0 & 1 & 3 \\ \hline
  &   &   & J_1  & J_2 &   & J_3 & \\
\hline
\end{array}
\end{equation}
In the tables in Appendix~\ref{app-models}, models with this toric ambient space are labeled by $2C0P2$. There are at least two possible triangulations for the toric space. The relevant one here has the label $1$ in Appendix~\ref{app-models}. The hypersurface describing the base manifolds has degrees $(4,2,2)$.
The GUT divisor $S$ ($y_4=0$) is $dP_5$. To see this, we look at the submanifold of the toric ambient space after setting $y_4$ to zero. Since by the Stanley-Reisner ideal of the fourfold, $y_4$ and $y_1$ are not allowed to vanish simultaneously, we obtain the following equivalence relations:
\begin{equation}
 (1,\,y_2,\,y_3,\,0,\,y_5,\,y_6,\,\frac{y_7}{y_1})\sim(1,\,\rho\lambda y_2,\,\lambda y_3,\,0,\,\rho y_5,\,\rho y_6,\,\lambda \frac{y_7}{y_1})\qquad\forall \rho,\lambda\in \mathbb{C}^*\,,
\end{equation}
where $y_2,\ldots,\,y_6$ and $\frac{y_7}{y_1}$ are now the new homogeneous coordinates. Via a Segre like map we can embed this threefold $\mathcal{G}$ into $\mathbb{P}^4$, $y\mapsto[y_2:y_3 y_5:\frac{y_7}{y_1} y_6:y_3 y_6:\frac{y_7}{y_1} y_5]\in \mathbb{P}^4$. The Stanley-Reisner ideal guarantees that we do not map to $\vec 0$. Since the points of $\mathcal{G}$ fulfill a certain relation in terms of the homogeneous coordinates of  $\mathbb{P}^4$, $\mathcal{G}$ is realized as a hypersurface of degree two in $\mathbb{P}^4$. From the hypersurface equation, which is compatible with this map, one obtains a second degree two equation. Thus, we have a complete intersection of two degree two equations in $\mathbb{P}^4$, which is in fact a $dP_5$.

The relevant first Chern classes of the base manifold in terms of the ambient space divisor basis $\{J_1,J_2,J_3\}$ are:
\begin{eqnarray}
\label{tabmod1}
\begin{tabular}{l|c|c}
Chern class & in $B$  & on $S$ \\ \hline %
$c_1(B)$& $J_3$ & -  \\ \hline %
$c_1(N_{S|B})$& $J_1$ & $\displaystyle AH+\sum_i^5 B_i E_i$ \\ \hline %
$c_1(S)$ & $J_3-J_1$ & $\displaystyle 3H-\sum_i^5 E_i$ \\
\end{tabular}
\end{eqnarray}
where $A$ and $B_i$ are integers and will be determined. The
triple intersections are:
\begin{eqnarray}
J_1^2J_3=-2,~~ J_2J_3^2=4,~~ J_1J_2J_3=2,~~J_3^3=4.
\label{triple1}
\end{eqnarray}
Using equations~(\ref{gutvol}), we calculate the volumes of the base $B$ and the GUT brane $S$ in terms of the K\"ahler moduli $r_i>0$:
\begin{eqnarray}
\mathrm{Vol}(B)&=&16 r_1^3+36 r_1^2 r_2+24 r_1 r_2^2+4 r_2^3+24 r_1^2 r_3+36 r_1 r_2 r_3+12 r_2^2 r_3+6 r_1 r_3^2+6 r_2 r_3^2 \nonumber\\
\mathrm{Vol}(S)&=&4 r_1^2+4 r_1 r_2
\end{eqnarray}
Thus, for $r_3\rightarrow\infty$ the GUT divisor remains of finite size while the volume of the base becomes infinitely large. One can also check that in this limit all the other divisors inside $B$ also obtain infinite volume. For $r_1=0$ and at least $r_2\neq 0$ we can implement the mathematical decoupling limit. For $r_1=r_3=0$ a further divisor which is not del Pezzo will shrink to zero size.\\
The geometry for both pictures in~(\ref{tabmod1}) should be
consistent. We can use the triple intersections in (\ref{triple1})
to compute $c_1(N_{S|B})^2=J_1^3=0$ and $c_1(N_{S|B})\cdot
c_1(S)=J_1^2(J_3-J_1)=-2$ on $S$. There are two possible solutions
for $A$ and the $B_i$'s, which are
\begin{eqnarray}
c_1(N_{S|B}) &=& -H+E_i; \nonumber \\
&{\rm or}& -2H+E_i+E_j+E_k+E_l,~~ i\neq j\neq k\neq l.
\end{eqnarray}
These two solutions are actually equivalent via an
involution ($I_{\mathcal{B}_5}^{(5)}$) of the del Pezzo surface. Its explicit action can be found in Table 22 of \cite{Blumenhagen:2008zz}. For concreteness we will stick to $c_1(N_{S|B}) = -H+E_5$ in the following.

Finally we can collect the information we need for model building,
\begin{eqnarray}
\begin{array}{ccc}
c_1(S) & c_1 & 3H-E_1-E_2-E_3-E_4-E_5 \\
c_1(N_{S|B}) & -t & -H+E_5 \\
b_0=\eta & 6c_1-t &  17H -6(E_1+E_2+E_3+E_4+E_5)+E_5 \\
b_2 & \eta-2c_1 & 11H -4(E_1+E_2+E_3+E_4+E_5)+E_5 \\
b_3 & \eta-3c_1 & 8H -3(E_1+E_2+E_3+E_4+E_5)+E_5 \\
b_4 & \eta-4c_1 & 5H -2(E_1+E_2+E_3+E_4+E_5)+E_5 \\
\end{array}\,.   \label{dP5a}
\end{eqnarray}

%%%%%%%%%%%%%%%%%%%%%%%%%%%%%%%%%%%%%%%%%%%%%%%%%%%%%%%%%%%%%%%%%%%%%%%%%
\subsubsection{Model $2$}
Our second model is a point blowup in $\mathbb{P}^4[4]$. The corresponding weight matrix can be found in table~\ref{tab-p41point1} in Appendix~\ref{app-models}. We label the model by $0C1P1$. The base $B$ is a hypersurface of degree $(4,1)$.
The relevant first Chern classes of the base in terms of the ambient space divisor basis $\{J_1\sim [y_2],\,J_2\sim [y_1]\}$ are:
\begin{eqnarray}
\label{tabmod2}
\begin{tabular}{l|c|c}
Chern class & in $B$ & on $S$ \\ \hline %
$c_1(B)$& $J_2$ & -  \\ \hline %
$c_1(N_{S|B})$& $J_2-J_1$ & $\displaystyle AH+\sum_i^6 B_i E_i$ \\ \hline %
$c_1(S)$ & $J_1$ & $\displaystyle 3H-\sum_i^6 E_i$ \\
\end{tabular}
\end{eqnarray}
where $A$ and $B_i$ are integers and will be determined. The
triple intersections are:
\begin{eqnarray}
J_1^3=1,~~J_1^2J_2=4,~~ J_1J_2^2=4,~~J_2^3=4.
\end{eqnarray}
We can check the existence of a decoupling limit by computing the volumes of the base $B$ and the GUT divisor $S$ in terms of $r_i>0$:
\begin{eqnarray}
\mathrm{Vol}(B)&=&4 r_1^3+12 r_1^2 r_2+12 r_1 r_2^2+r_2^3\nonumber\\
\mathrm{Vol}(S)&=&3 r_2^2
\end{eqnarray}
For $r_1\rightarrow\infty$ we find the physical decoupling limit. In this limit no other divisor in $B$ remains of finite size. Setting $r_2=0$ while keeping $r_1\neq 0$ gives the mathematical decoupling limit. No divisors other than the GUT divisor will shrink to zero size in this limit.\\
The geometry for both pictures in~(\ref{tabmod2}) should be
consistent. We calculate $c_1(N_{S|B})^2=3$ and
$c_1(N_{S|B})\cdot c_1(S)=-3$ on $S$. Therefore one finds there is only
one possible solution in this case:
\begin{eqnarray}
c_1(N_{S|B}) &=& -3H+E_1+E_2+E_3+E_4+E_5+E_6.
\end{eqnarray}
Finally we can write down the information we need for model
building\footnote{Note that one can only construct an $SO(10)$
model in this geometry. In an $SU(5)$ model one would not obtain a
{\bf10} curve.}:
\begin{eqnarray}
\begin{array}{ccc}
c_1(S) & c_1 & 3H-E_1-E_2-E_3-E_4-E_5-E_6 \\
c_1(N_{S|B}) & -t & -3H+E_1+E_2+E_3+E_4+E_5+E_6 \\
b_0=\eta & 6c_1-t &  15H -5(E_1+E_2+E_3+E_4+E_5+E_6) \\
b_2 & \eta-2c_1 & 9H -3(E_1+E_2+E_3+E_4+E_5+E_6) \\
b_3 & \eta-3c_1 & 6H -2(E_1+E_2+E_3+E_4+E_5+E_6) \\
b_4 & \eta-4c_1 & 3H -(E_1+E_2+E_3+E_4+E_5+E_6) \\
\end{array}
\end{eqnarray}

%%%%%%%%%%%%%%%%%%%%%%%%%%%%%%%%%%%%%%%%%%%%%%%%%%%%%%%%%%%%%%%%%%%%%%%%
\subsubsection{Model \texorpdfstring{$3$}{3}}
The third model we would like to discuss is a blowup of one curve and one point in the Fano $\mathbb{P}^4[4]$. The weight matrix is written in table~\ref{tab-p41curve1point} in Appendix~\ref{app-models} where one has to replace the first '$\ast$' by a $0$ and the second '$\ast$' by a $1$. We give the model the identifier $1C1P2$. The base manifold is a hypersurface of degree $(4,2,1)$.
The GUT divisor $S$ ($y_6=0$) is $dP_4$. To make this explicit we again set the coordinate corresponding to the GUT divisor to zero and fix one scaling, in this case the second one ($y_5=1$). What we obtain from the new equivalence relations and the Stanley-Reisner ideal is $\mathbb{P}^2\times \mathbb{P}^1$. The hypersurface becomes a degree $(2,1)$ equation in the new degrees. Thus, we can bring it to the following form:
\begin{equation}\label{eq:dP4-3model}
 y_4\,g_1(y_1,y_2,y_3)=y_7\,g_2(y_1,y_2,y_3)\,.
\end{equation}
where $[y_4:y_7]$ and $[y_1:y_2:y_3]$ are the homogeneous coordinates of $\mathbb{P}^1$ and $\mathbb{P}^2$, respectively. Hence, the four points in $\mathbb{P}^2$ where $g_1=g_2=0$ vanish are blown up to four $\mathbb{P}^1$s, i.e.~$S$ is a $dP_4$.

The relevant first Chern classes of the base in terms of the ambient space divisor basis $\{J_1\sim [y_3],\,J_2\sim [y_4],\,J_3\sim [y_5]\}$ are:
\begin{eqnarray}
\label{tabmod3}
\begin{tabular}{l|c|c}
Chern class & in $B$ & on $S$ \\ \hline %
$c_1(B)$& $J_2$ & -  \\ \hline %
$c_1(N_{S|B})$& $J_3-J_1$ & $\displaystyle AH+\sum_i^4 B_i E_i$ \\ \hline %
$c_1(S)$ & $J_1+J_2-J_3$ & $\displaystyle 3H-\sum_i^4 E_i$ \\
\end{tabular}
\end{eqnarray}
where $A$ and $B_i$ are integers that will be determined below. The
triple intersections are:
\begin{eqnarray}
&& J_1^2J_2=2,~~J_1^2J_3=1,~~ J_2^3=4,~~
J_1J_2^2=4,~~J_2^2J_3=4,\nonumber \\&& J_3^3=1,~~ J_1J_3^2=1,~~
J_2J_3^2=4,~~J_1J_2J_3=4.
\end{eqnarray}
The volumes of the base $B$ and the GUT divisor $S$ can be determined from~(\ref{gutvol}):
\begin{eqnarray}
\mathrm{Vol}(B)&=&4 r_1^3+12 r_1^2 r_2+12 r_1 r_2^2+r_2^3+12 r_1^2 r_3+24 r_1 r_2 r_3+3 r_2^2 r_3+6 r_1 r_3^2+3 r_2 r_3^2\nonumber\\
\mathrm{Vol}(S)&=&4 r_1 r_3+r_3^2
\end{eqnarray}
The limit $r_2\rightarrow\infty$ keeps the volume of the GUT divisor finite while sending the volumes of the base and all other divisors to infinity. To obtain a mathematical decoupling limit we have to set $r_3=0$ while keeping $r_1$ or $r_2$ non-zero. Note that if we also set $r_2=0$ the volume of the base will still be finite but another divisor which does not have the topological characteristics of a del Pezzo will shrink to zero size.\\
The geometry for both pictures in~(\ref{tabmod3}) should be
consistent. We compute $c_1(N_{S|B})^2=1$ and $c_1(N_{S|B})\cdot
c_1(S)=-3$ on $S$. Then one finds two solutions for $A$ and the $B_i$'s:
\begin{eqnarray}\label{eq:cc-sol-3model}
c_1(N_{S|B}) &=& -2H+ E_i+E_j+E_k, ~~i\neq j\neq k; \nonumber \\
&{\rm or}& -H.
\end{eqnarray}
These solutions are related by automorphisms of the del Pezzo surface\footnote{To convince oneself that this is the case, one can have a look at the (dual) intersection graph of (-1)-curves of dP$_4$, see the left graph of Figure~9 in~\cite{Blumenhagen:2008zz} (Petersen graph). The graph has an obvious $\mathbb{Z}_5$ symmetry. If one assigns $H-E_4-E_2$, $H - E_3 - E_4$, $H - E_1 - E_3$, $E_4$ and $E_3$ to the inner points and performs a positive rotation one obtains the following transformations:
\begin{equation}
 H \mapsto 2H-E_1-E_2-E_4 \mapsto 2H-E_2-E_3-E_4 \mapsto 2H-E_1-E_3-E_4 \mapsto 2H-E_1-E_2-E_3
\end{equation}}.
Thus, the solutions are equivalent modulo automorphisms.
For concreteness we will only use the last solution of \eqref{eq:cc-sol-3model} in the following.

Finally we can collect the information we need for model
building:
% \begin{eqnarray}
% \begin{array}{ccc}
% c_1(S) & c_1 & 3H-E_1-E_2-E_3-E_4 \\
% c_1(N_{S|B}) & -t & -2H+E_i+E_j+E_k \\
% b_0=\eta & 6c_1-t &  16H -6(E_1+E_2+E_3+E_4)+E_i+E_j+E_k \\
% b_2 & \eta-2c_1 & 10H -4(E_1+E_2+E_3+E_4)+E_i+E_j+E_k \\
% b_3 & \eta-3c_1 & 7H -3(E_1+E_2+E_3+E_4)+E_i+E_j+E_k \\
% b_4 & \eta-4c_1 & 4H -2(E_1+E_2+E_3+E_4)+E_i+E_j+E_k \\
% \end{array} \label{dP4a}
% \end{eqnarray}
%
% The other case is:
\begin{eqnarray}
\begin{array}{ccc}
c_1(S) & c_1 & 3H-E_1-E_2-E_3-E_4 \\
c_1(N_{S|B}) & -t & -H \\
b_0=\eta & 6c_1-t &  17H -6(E_1+E_2+E_3+E_4) \\
b_2 & \eta-2c_1 & 11H -4(E_1+E_2+E_3+E_4) \\
b_3 & \eta-3c_1 & 8H -3(E_1+E_2+E_3+E_4) \\
b_4 & \eta-4c_1 & 5H -2(E_1+E_2+E_3+E_4) \\
\end{array} \label{dP4b}
\end{eqnarray}

%%%%%%%%%%%%%%%%%%%%%%%%%%%%%%%%%%%%%%%%%%%%%%%%%%%%%%%%%%%%%%%%%%%%%%%%%%
\subsubsection{Model $4$}
We consider a blowup of two curves and one point starting from $\mathbb{P}^4[4]$. The weight matrix for this model can be found in table~\ref{tab-p42curve1point2} in Appendix~\ref{app-models}, where one should replace the third '$\ast$' by $1$ and all the others by $0$. In our data tables this is labeled by $2C1P4$. The hypersurface we consider in this ambient space has degrees $(4,2,2,1)$.
The GUT divisor $S$ ($y_7=0$) is a $dP_5$~\footnote{Note that there is a second $dP_5$ in this geometry which also satisfies both the mathematical and the physical decoupling limit. However, there occurs a problem with the fourfold for this divisor: After removing points in the M-lattice in order to impose the $SO(10)$ gauge group the polyhedron describing the fourfold is no longer reflexive.}. To show that we set $y_7$ to zero and and 'scale away' the third row of table~\ref{tab-p42curve1point2}. We are allowed to do this because by the Stanley-Reisner ideal of this triangulation $y_7$ and $y_2$ must not vanish at the same time. The remaining weight matrix looks as follows:
\begin{equation}
 \begin{array}{|c|c|c|c|c|c|}
\hline
 y_{1} & y_{3} & \frac{y_{4}}{y_7} & y_{5} & y_6 & y_8\\\hline
\hline
1 & 1 & 0 & 1 & 0 & 0 \\ \hline
0 & 0 & 1 & 1 & 1 & 0  \\ \hline
0 & 0 & 1 & 0 & 0 & 1 \\ \hline
\end{array}\quad\textmd{with} \quad\textmd{SR-I}=\left\{\frac{y_{4}}{y_7}y_8,\,y_5 y_6,\,y_1y_3y_5,\,y_8 y_1 y_3\right\}.
\end{equation}
On this submanifold the hypersurface equation takes the form,
\begin{equation}
 y_8\,g_1^{(2,2,0)}=\frac{y_{4}}{y_7}\,g_2^{(2,1,0)}\,,\label{eq:hs-model4}
\end{equation}
where $g_1^{(2,2,0)}$ and $g_2^{(2,1,0)}$ are homogeneous functions of the indicated degree. Looking at the equivalence relations of the homogeneous coordinates appearing in $g_1$ and $g_2$ we observe that they are the ones of a $dP_1$. Also the scalings are correct, $y_1$ and $y_3$ are allowed to vanish simultaneously. However, $y_1=y_3=0$ is not a solution to the hypersurface equation so we can safely exclude it from the definition set. Equation~(\ref{eq:hs-model4}) tells us that the points of $dP_1$ where $g_1=g_2=0$ vanish are replaced by further point blowups ($\mathbb{P}^1$). Thus, we end up with a $dP_5$.

The relevant first Chern classes of the base in terms of the ambient space divisor basis $\{J_1\sim [y_1],\, J_2\sim [y_2]\,,J_3\sim [y_4],\,J_4\sim [y_5]\}$ are:
\begin{eqnarray}
\label{tabmod4}
\begin{tabular}{l|c|c}
Chern class & in $B$  & on $S$ \\ \hline %
$c_1(B)$& $J_3$ & -  \\ \hline %
$c_1(N_{S|B})$& $J_2-J_1$ & $\displaystyle AH+\sum_i^5 B_i E_i$ \\ \hline %
$c_1(S)$ & $J_1-J_2+J_3$ & $\displaystyle 3H-\sum_i^5 E_i$ \\
\end{tabular}
\end{eqnarray}
where $A$ and $B_i$ are integers and will be determined. The
triple intersections are:
\begin{eqnarray}
&& J_2^2J_3=2,~~J_2^2J_4=1,~~ J_1J_3^2=4,~~J_2J_3^2=4,~~
J_3^2J_4=4, ~~ J_3^3=4,~~ J_2J_4^2=1,\nonumber \\ && J_3J_4^2=2,~~
J_1J_2J_3=2,~~ J_1J_2J_4=1,~~ J_1J_3J_4=2,~~ J_2J_3J_4=4.
\end{eqnarray}
In order to study the decoupling limit we calculate the volumes of the base and the GUT brane in terms of $r_i>0$ using~(\ref{gutvol}):
\begin{eqnarray}
\mathrm{Vol}(B)&=&3 r_1^2 r_2+3 r_1 r_2^2+6 r_1^2 r_3+24 r_1 r_2 r_3+6 r_2^2 r_3+12 r_1 r_3^2+12 r_2 r_3^2+4 r_3^3+6 r_1^2 r_4+30 r_1 r_2 r_4\nonumber\\
&&+6 r_2^2 r_4+36 r_1 r_3 r_4+36 r_2 r_3 r_4+24 r_3^2 r_4+24 r_1 r_4^2+24 r_2 r_4^2+36 r_3 r_4^2+16 r_4^3\nonumber\\
\mathrm{Vol}(S)&=&r_1^2+4 r_1 r_3+6 r_1 r_4+4 r_3 r_4+4 r_4^2
\end{eqnarray}
For $r_2\rightarrow\infty$ the volume of the base becomes infinitely large while the volume of the GUT divisor remains finite. In this limit, also the volumes of all the other divisors become infinite. If we want to shrink the volume of the GUT divisor to zero size we have to set at least $r_1=r_4=0$ while keeping $r_3$ finite. If we also set $r_2=0$ $\mathrm{Vol}(B)$ will still be non-zero but two more $dP_5$--divisors will have zero volume.\\
The geometry for both pictures in~(\ref{tabmod4}) should be
consistent. We calculate $c_1(N_{S|B})^2=0$ and $c_1(N_{S|B})\cdot
c_1(S)=-2$ on $S$. Then one finds two possible
solutions in this case:
\begin{eqnarray}
c_1(N_{S|B}) &=& -H+E_i; \nonumber \\
&{\rm or}& -2H+E_i+E_j+E_k+E_l,~~ i\neq j\neq k\neq l.\nonumber
\end{eqnarray}
Again both solutions are related via an involution. To fix a basis we choose
\begin{equation}
 c_1(N_{S|B}) = -H+E_5
\end{equation}
for the first Chern class of the normal bundle.

Finally we can write down the information we need for model
building.  Since it is also a $dP_5$ space and the conditions are
the same as those in Model 1 on $S$, we again get:
\begin{eqnarray}
\begin{array}{ccc}
c_1(S) & c_1 & 3H-E_1-E_2-E_3-E_4-E_5 \\
c_1(N_{S|B}) & -t & -H+E_5 \\
b_0=\eta & 6c_1-t &  17H -6(E_1+E_2+E_3+E_4+E_5)+E_5 \\
b_2 & \eta-2c_1 & 11H -4(E_1+E_2+E_3+E_4+E_5)+E_5 \\
b_3 & \eta-3c_1 & 8H -3(E_1+E_2+E_3+E_4+E_5)+E_5 \\
b_4 & \eta-4c_1 & 5H -2(E_1+E_2+E_3+E_4+E_5)+E_5 \\
\end{array}
\end{eqnarray}
There is no difference to (\ref{dP5a}).
%Similarly, the other case has the same form as (\ref{dP5b}).

%%%%%%%%%%%%%%%%%%%%%%%%%%%%%%%%%%%%%%%%%%%%%%%%%%%%%%%%%%%%%%%%%%%%%%%%%
\subsubsection{Model $5$}
Our final example starts with the Fano $\mathbb{P}^4[3]$, and we blow up two curves and one point. The weight matrix can be found in Appendix~\ref{app-models} in table~\ref{tab-p42curve1point1}, where the '$\ast$' is to be replaced by $1$ and the '$\diamond$' by $0$. In the tables in Appendix~\ref{app-models} this model is labeled by $2C1P1$.  The hypersurface has degrees $(3,2,1,1)$. Due to singularities in the ambient space there are two triangulations. The particular triangulation we choose is denoted by $1$ in the tables in Appendix~\ref{app-models}.
The GUT divisor $S$ ($y_7=0$) is $dP_4$. To make this obvious, we use the same steps as in model 3 except that in this case the Stanley-Reisner ideal allows us to 'scale away' two rows, the second and the third. We end up again with $\mathbb{P}^2\times \mathbb{P}^1$. The modified hypersurface equation, we obtain an equivalent version of model 3,
\begin{equation}
 y_8\,g_1^{(2)}(\frac{y_5}{y_6},y_3,\frac{y_4}{y_6})=\frac{y_1}{y_2}\,g_2^{(2)}(\frac{y_5}{y_6},y_3,\frac{y_4}{y_6})\,,
\end{equation}
where $[\frac{y_5}{y_6}:y_3:\frac{y_4}{y_6}]$ and $[y_8:\frac{y_1}{y_2}]$ are the homogeneous coordinates of $\mathbb{P}^2$ and $\mathbb{P}^1$, respectively. Thus, $S$  is a $dP_4$.

The relevant first Chern classes of the base in terms of the ambient space divisor basis $\{J_1\sim [y_5],\,J_2\sim [y_1],\,J_3\sim [y_2],\,J_4\sim [y_6]\}$ are:
\begin{eqnarray}
\label{tabmod5}
\begin{tabular}{l|c|c}
Chern class & in $B$ & on $S$ \\ \hline %
$c_1(B)$& $J_2+J_3+J_4$ & -  \\ \hline %
$c_1(N_{S|B})$& $J_3+J_4-J_1$ & $\displaystyle AH+\sum_i^4 B_i E_i$ \\ \hline %
$c_1(S)$ & $J_1+J_2$ & $\displaystyle 3H-\sum_i^4 E_i$ \\
\end{tabular}
\end{eqnarray}
where $A$ and $B_i$ are integers and will be determined. The
triple intersections are:
\begin{eqnarray}
&& J_1^2J_2=1,~~J_1^2J_3=1,~~ J_1J_2^2=2,~~J_2^2J_4=2,~~
J_3^2J_4=2, ~~ J_3^3=-2,~~ J_1J_4^2=-1,\nonumber \\ &&
J_2J_4^2=-1,~~ J_3J_4^2=-1,~~ J_1J_2J_3=2,~~ J_1J_2J_4=1,~~
J_1J_3J_4=1,~~ J_2J_3J_4=2.
\end{eqnarray}
The volumes of the base and the GUT divisor can be computed using~(\ref{gutvol}):
\begin{eqnarray}
\mathrm{Vol}(B)&=&6 r_1^2 r_2+3 r_1 r_2^2+6 r_1^2 r_3+18 r_1 r_2 r_3+3 r_2^2 r_3+9 r_1 r_3^2+3 r_2 r_3^2+r_3^3+6 r_1^2 r_4+18 r_1 r_2 r_4\nonumber\\
&&+3 r_2^2 r_4+18 r_1 r_3 r_4+18 r_2 r_3 r_4+9 r_3^2 r_4+9 r_1 r_4^2+9 r_2 r_4^2+9 r_3 r_4^2+3 r_4^3\nonumber \\
\mathrm{Vol}(S)&=&4 r_1 r_2+r_2^2+4 r_2 r_4
\end{eqnarray}
The physical decoupling limit exists for $r_3\rightarrow\infty$. In this limit also the volumes of the other divisors in the base become infinitely large. In order to implement a mathematical decoupling limit we have to set at least $r_2=0$ while keeping finite values for $r_3$ or $r_4$. Setting also $r_3=0$ will make another non del Pezzo divisor shrink to zero size. For $r_1=0$ an extra $dP_1$ will get zero volume.\\
The geometry for both pictures in~(\ref{tabmod5}) should be
consistent, and we compute $c_1(N_{S|B})^2=1$ and
$c_1(N_{S|B})\cdot c_1(S)=-3$ on $S$.
One finds that there are two
possible solutions in this case:
\begin{eqnarray}
\label{mod5sol}
c_1(N_{S|B}) &=& -2H+ E_i+E_j+E_k, ~~i\neq j\neq k. \nonumber \\
&{\rm or}& -H\,,
\end{eqnarray}
which are again related by the automorphisms of model 3. For concreteness we chose the last solution in \eqref{mod5sol}.

Finally we can collect the information we need for model
building.  Since it is also a $dP_4$ space and the conditions are
the same as those in Model 3 on $S$, we can again write:
\begin{eqnarray}
\begin{array}{ccc}
c_1(S) & c_1 & 3H-E_1-E_2-E_3-E_4 \\
c_1(N_{S|B}) & -t & -H \\
b_0=\eta & 6c_1-t &  17H -6(E_1+E_2+E_3+E_4) \\
b_2 & \eta-2c_1 & 11H -4(E_1+E_2+E_3+E_4) \\
b_3 & \eta-3c_1 & 8H -3(E_1+E_2+E_3+E_4) \\
b_4 & \eta-4c_1 & 5H -2(E_1+E_2+E_3+E_4) \\
\end{array}
\end{eqnarray}
There is no difference to (\ref{dP4b}).
%Similarly, the other case has the same form as (\ref{dP4b}).

%%%%%%%%%%%%%%%%%%%%%%%%%%%%%%%%%%%%%%%%%%%%%%%%%%%%%%%%%%%%%%%%%%%%%%%%%%%%%%
%%%%%%%%%%%%%%%%%%%%%%%%%%%%%%%%%%%%%%%%%%%%%%%%%%%%%%%%%%%%%%%%%%%%%%%%%%%%%%
\subsection{Fourfolds}
Here we give the explicit data of the Calabi-Yau fourfold(s) constructed from the base manifold of the first Model. The Calabi-Yau fourfolds corresponding to the other base manifolds can be found in Appendix~\ref{fourfold-data}. For convenience we relabeled the vertices obtained from the base. The vertex corresponding to the GUT divisor is given by $\nu_4$ and we associate to it the coordinate $w$. The additional vertices/coordinates obtained after dualizing the reduced M-lattice polytope are denoted with a tilde. Furthermore we compute the Euler numbers for the $SO(10)$ model and compare with the formula~(\ref{eulerformula}).

The vertices in the N-lattice are:
\begin{equation}
\begin{tabular}{c|c||c|c}
nef-part.&vertices&weights&coordinates\\
\hline\hline
$\nabla_1$&$\nu_1=(\begin{array}{rrrrrrr}\nm 3&\nm 0&\nm 1&\nm 1&\nm 1&\nm 0\end{array})$&$\begin{array}{rrrr}2&2&2&2\end{array}$&$x$\\
&$\nu_2=(\begin{array}{rrrrrrr}-2&\nm 0&-1&-1&-1&\nm 0\end{array})$&$\begin{array}{rrrrrr}3&3&3&3\end{array}$&$y$\\
&$\nu_3=(\begin{array}{rrrrrrr}\nm 0&\nm 0&\nm 1&\nm 1&\nm 1&\nm 0\end{array})$&$\begin{array}{rrrr}0&0&0&1\end{array}$&$z$\\
&$\nu_4=(\begin{array}{rrrrrrr}\nm 0&\nm 0&\nm 0&\nm 0&\nm 0&\nm 1\end{array})$&$\begin{array}{rrrrrr}1&0&0&0\end{array}$&$w$\\
&$\nu_5=(\begin{array}{rrrrrrr}\nm 0&-1&\nm 0&-1&\nm 0&\nm 1\end{array})$&$\begin{array}{rrrr}0&1&0&0\end{array}$&$y_1$\\
&$\nu_6=(\begin{array}{rrrrrrr}\nm 0&\nm 0&\nm 0&\nm 0&\nm 1&\nm 0\end{array})$&$\begin{array}{rrrrrr}0&0&1&0\end{array}$&$y_2$\\
\hline
$\nabla_2$&$\nu_7=(\begin{array}{rrrrrrr}\nm 0&\nm 0&\nm 0&\nm 1&\nm 1&-1\end{array})$&$\begin{array}{rrrrrr}1&1&0&0\end{array}$&$y_3$\\
&$\nu_8=(\begin{array}{rrrrrrr}\nm 0&\nm 1&\nm 0&\nm 0&\nm 0&\nm 0\end{array})$&$\begin{array}{rrrrrr}0&1&0&0\end{array}$&$y_4$\\
&$\nu_9=(\begin{array}{rrrrrrr}\nm 0&\nm 0&\nm 1&\nm 0&\nm 0&\nm 0\end{array})$&$\begin{array}{rrrrrr}1&1&1&0\end{array}$&$y_5$\\
&$\nu_{10}=(\begin{array}{rrrrrrr}\nm 0&\nm 0&\nm 0&\nm 1&\nm 0&\nm 0\end{array})$&$\begin{array}{rrrrrr}0&1&1&0\end{array}$&$y_6$\\
\hline\hline
\end{tabular}
\end{equation}
After reducing the M-lattice polytope to the SO(10) case, we obtain for the dual N-lattice polytope:
\begin{equation}
\begin{tabular}{c|c||c|c}
nef-part.&vertices&weights&coordinates\\
\hline\hline
$\nabla_1$&$\nu_1=(\begin{array}{rrrrrrr}\nm 3&\nm 0&\nm 1&\nm 1&\nm 1&\nm 0\end{array})$&$\begin{array}{rrrrrr}1&2&2&0&2&2\end{array}$&$x$\\
&$\nu_2=(\begin{array}{rrrrrrr}-2&\nm 0&-1&-1&-1&\nm 0\end{array})$&$\begin{array}{rrrrrr}2&3&3&0&3&3\end{array}$&$y$\\
&$\nu_3=(\begin{array}{rrrrrrr}\nm 0&\nm 0&\nm 1&\nm 1&\nm 1&\nm 0\end{array})$&$\begin{array}{rrrrrr}0&0&0&0&0&1\end{array}$&$z$\\
%&$\nu_4=(\begin{array}{rrrrrrr}\nm 0&\nm 0&\nm 0&\nm 0&\nm 0&\nm 1\end{array})$&$\begin{array}{rrrr}1&0&0&0\end{array}$&$w$\\
&$\nu_5=(\begin{array}{rrrrrrr}\nm 0&-1&\nm 0&-1&\nm 0&\nm 1\end{array})$&$\begin{array}{rrrrrr}0&1&0&0&0&0\end{array}$&$y_1$\\
&$\nu_6=(\begin{array}{rrrrrrr}\nm 0&\nm 0&\nm 0&\nm 0&\nm 1&\nm 0\end{array})$&$\begin{array}{rrrrrr}0&0&0&0&1&0\end{array}$&$y_2$\\
\hline
$\nabla_2$&$\nu_7=(\begin{array}{rrrrrrr}\nm 0&\nm 0&\nm 0&\nm 1&\nm 1&-1\end{array})$&$\begin{array}{rrrrrr}2&1&2&1&0&0\end{array}$&$y_3$\\
&$\nu_8=(\begin{array}{rrrrrrr}\nm 0&\nm 1&\nm 0&\nm 0&\nm 0&\nm 0\end{array})$&$\begin{array}{rrrrrr}0&1&0&0&0&0\end{array}$&$y_4$\\
&$\nu_9=(\begin{array}{rrrrrrr}\nm 0&\nm 0&\nm 1&\nm 0&\nm 0&\nm 0\end{array})$&$\begin{array}{rrrrrr}2&1&2&1&1&0\end{array}$&$y_5$\\
&$\nu_{10}=(\begin{array}{rrrrrrr}\nm 0&\nm 0&\nm 0&\nm 1&\nm 0&\nm 0\end{array})$&$\begin{array}{rrrrrr}0&1&0&0&1&0\end{array}$&$y_6$\\
&$\tilde{\nu}_{11}=(\begin{array}{rrrrrrr}\nm 1&\nm 0&-1&-1&-1&\nm 2\end{array})$&$\begin{array}{rrrrrr}1&0&0&0&0&0\end{array}$&$\tilde{y}_7$\\
&$\tilde{\nu}_{12}=(\begin{array}{rrrrrrr}\nm 0&\nm 0&-1&-1&-1&\nm 1\end{array})$&$\begin{array}{rrrrrr}0&0&0&1&0&0\end{array}$&$\tilde{y}_8$\\
&$\tilde{\nu}_{13}=(\begin{array}{rrrrrrr}\nm 0&\nm 0&-1&-1&-1&\nm 2\end{array})$&$\begin{array}{rrrrrr}0&0&1&0&0&0\end{array}$&$\tilde{y}_9$\\
\hline\hline
\end{tabular}
\end{equation}
Note that the GUT divisor $\{w=0\}$ no longer corresponds to a vertex after this procedure. However it is still a point in the polytope $\nabla_2$. Furthermore the additional vertices appear in $\nabla_2$ and not in $\nabla_1$. The Euler number is $912$. Here we find a discrepancy with the Euler number computed via~(\ref{eulerformula}) where the result is $672$.

We also find a mismatch for  the Euler numbers computed in these two ways for the models 3 and 4. For the models 2 and 5 they agree.
%%%%%%%%%%%%%%%%%%%%%%%%%%%%%%%%%%%%%%%%%%%%%%%%%%%%%%%%%%%%%%%%%%%%%%%%%%%%%%%

\subsection{GUT models}

In this subsection we will use the geometric backgrounds
discussed in Section~\ref{sunsec-geo} to create examples of
$SO(10)$ models with a split spectral cover.  We demonstrate numerical
results for each $dP_n$ on $S$.

\subsubsection{Examples based on Model 1, $S=dP_5$}

From the normal bundle $c_1(N_{S|B})=-H+E_5$, we have:
\begin{equation}
\eta= 17H-6E_1-6E_2-6E_3-6E_4-5E_5.
\end{equation}

\paragraph{Model 1A}~\\

It is natural to set $\xi=\mathcal{O}$. For this case only the ${\bf
16}_b^{(3)}$ curve contributes to the Yukawa coupling, and there is no
contribution from the matter curve associated to $C^{(1)}$. By assuming
$\rho = xH-\sum y_i E_i$ it is possible to obtain a general
spectrum. Thus by (\ref{splitn1}) and (\ref{splitn2}) the
contributions from the components of the universal flux to the
curve chirality are:
\begin{equation}
\begin{tabular}{l@{}|@{}c@{}|c|@{}c@{}|c|@{}c@{}|@{}c@{}}
curve\, & $k_a\gamma_a$ & $k_b\gamma_b$ & $d_a\delta_a$ &
$d_b\delta_b$ & $\rho$ & chirality \\ \hline

${\bf 16}_a$ & 0 & 0 & 0 & 0 & 0 & 0  \\
${\bf 16}_b$ & 0 & -$26k_b$ & 0 & -6$d_b$ & $\displaystyle
-5x+2\sum_{i=1}^4y_i+y_5$ & -$(26k_b+6d_b)-\displaystyle
5x+2\sum_{i=1}^4y_i+y_5$ \\
${\bf 10}_{ab}$ & 0 & -14$k_b$ & 0 & -14$d_b$ & $\displaystyle
2(8x-3{\sum_{i=1}^4}y_i -2y_5)$ & -$14(k_b+k_d)+\displaystyle
2(8x-3{\sum_{i=1}^4}y_i -2y_5) $ \\
${\bf 10}_{bb}$ & 0 & 14$k_b$ & 0 & 14$d_b$ & -$\displaystyle
2(8x-3{\sum_{i=1}^4}y_i -2y_5)$ & $14(k_b+k_d)-\displaystyle
2(8x-3{\sum_{i=1}^4}y_i -2y_5) $
\end{tabular}
\end{equation}

We have sufficiently many degrees of freedom from the $dP_5$ surface that
we can tune the parameters to create a three generation $SO(10)$
model with the supersymmetry and tadpole conditions satisfied. The
spectrum of the model can be summarized as follows:
\begin{equation}
\begin{tabular}{c|c|c}
curve & class & generation   \\ \hline
${\bf 16}_a$ & $\mathcal{O}$ & 0      \\
${\bf 16}_b$ & $5H-2(E_1+E_2+E_3+E_4)-E_5$ & 3   \\
${\bf 10}_{ab}$ & $8H -3(E_1+E_2+E_3+E_4)-2E_5$ & $k$  \\
${\bf 10}_{bb}$ & $8H -3(E_1+E_2+E_3+E_4)-2E_5$ & -$k$
\end{tabular}
\end{equation}
In the table above we have indicated that the parameters can be
tuned such that we get three generations of fermions and $k$ Higgs
fields. The $\bf 10$ fields from the two different curves are
conjugate and have the same generation number. This implies that
the $SO(10)$ GUT spectrum has exotic ${\bf 10}$ fields. In what
follows we will present other examples with $\xi\neq\mathcal{O}$.

\paragraph{Model 1B}~\\

The structure for the models with $\xi\neq \mathcal{O}$ is
plentiful. We present an example of an $SO(10)$ model in
Table \ref{Model 1B} \footnote{In the following examples we impose
the condition $(c_1+\xi)\cdot\xi=0$ in order to keep the ramification of
the cover $C^{(1)}$ trivial.}:
\begin{table}[h]
\center
\begin{tabular}{c|c|c|c|c|c}
$k_a$ & $k_b$ & $d_a$ & $d_b$ & $\xi$ & $\rho$
\\ \hline -0.5 & 0.5 & 0 & -1 & $H-E_1+E_4+2E_5$ & $-H+E_4+E_5$
\end{tabular}
\caption{Parameters of Model 1B.} \label{Model 1B} %model 1
\end{table}

The matter spectrum and the corresponding classes are:
\begin{equation}
\begin{tabular}{c|c|c}
Matter & class with fixed $\xi$ & generation \\\hline

${\bf 16}_a$  &  $H-E_1+E_4+2E_5$ & 0  \\

${\bf 16}_b$ & $4H-E_1-2E_2-2E_3-3E_4-3E_5$ & 3   \\

${\bf 10}_{ab}$ & $8H-3E_1-3E_2-3E_3-3E_4-2E_5$ & 1  \\

${\bf 10}_{bb}$ & $8H-3E_1-3E_2-3E_3-3E_4-2E_5$ & -1 \\
\end{tabular}
\end{equation}
In this model the tadpole condition and (\ref{muceuler}) imply that $N_{D3}=8$.  The
supersymmetry condition is not very constrained, and for
simplicity we choose for $\omega$ in (\ref{susy}) the following
special form (for $dP_5$):
\begin{eqnarray}
\omega = \beta \left( (2H - \sum_{i=1}^5 E_i)+ \sum^5_{j\neq k}
(H-E_j-E_k) \right) + \alpha \sum_{l=1}^5 E_l.
\end{eqnarray}
The condition for $\omega$ being ample can be summarized as
$5\alpha>\beta>0$ and $5\beta>\alpha$. Then in this model we find
$\alpha/\beta=13/3$.

\subsubsection{Example based on Model 2}

In the previous section we obtained $c_1(N_{S|B})=-c_1(S)$, and
therefore $\eta=5c_1(S)$, so that the coefficients of the
exceptional divisors $E_i$ of $dP_6$ in $\eta$ are the same. This
implies that we need to choose a non-trivial $\xi$ if we want to
reserve the freedom of having restriction to the $\bf 16$ curves
by the flux $F_X=E_i-E_j$. Here we give an example of a
non-trivial $\xi$ and a three-generation $\bf 16$ curve:
\begin{table}[h] %611
\center
\begin{tabular}{c|c|c|c|c|c}
$k_a$ & $k_b$ & $d_a$ & $d_b$ & $\xi$ & $\rho$
\\ \hline -0.5 & -1.5 & 0 & -1 & $2H-2E_1-E_2-E_3$ & $-H+E_1+E_2+E_3$
\end{tabular}
\caption{Parameters of Model 2}
\end{table}

The matter spectrum and the corresponding classes are:
\begin{equation}
\begin{tabular}{c|c|c}
Matter & class with fixed $\xi$ & generation \\\hline

${\bf 16}^{(1)}_a$  &  $2H-2E_1-E_2-E_3$ & 0  \\

${\bf 16}^{(3)}_b$ & $H+E_1-E_4-E_5-E_6$ & 3   \\

${\bf 10}_{ab}$ & $6H-2E_1-2E_2-2E_3-2E_4-2E_5-2E_6$ & 3
\\

${\bf 10}_{bb}$ & $6H-2E_1-2E_2-2E_3-2E_4-2E_5-2E_6$ & -3
\\
\end{tabular}
\end{equation}
In this model we get $N_{D3}=27$ from the tadpole condition, using (\ref{muceuler}) to compute the Euler number.  We choose for the
supersymmetry condition (\ref{susy}) the following special
case:
\begin{eqnarray}
\omega &=& \beta \left( \sum^6_{m\neq n\neq p\neq q\neq r}(2H -
E_m-E_n-E_p-E_1-E_r)+ \sum^6_{j\neq k} (H-E_j-E_k) \right) +
\alpha \sum_{l=1}^6 E_l \nonumber \\
&=& 27\beta H -(10\beta-\alpha)\sum_{i=1}^6 E_i
\end{eqnarray}
Then we get $10\beta>\alpha>0$ and $\alpha/\beta=16/35$.

\subsubsection{Examples based on Model 5}

On this $dP_4$ surface we choose the first solution for
$c_1(N_{S|B})$ in (\ref{mod5sol}). With that we obtain:
\begin{equation}
\eta=17H-6E_1-6E_2-6E_3-6E_4.
\end{equation}
Again in this model it is possible to choose $\xi=\mathcal{O}$,
analogous to what we discussed in Model 1. Here we present two
examples with $\xi$ non-trivial.

\paragraph{Example 5A}~\\

In this example we present a three-generation model. The
parameters in this example are:
\begin{table}[h]
\center
\begin{tabular}{c|c|c|c|c|c}
$k_a$ & $k_b$ & $d_a$ & $d_b$ & $\xi$ & $\rho$
\\ \hline % -1.5 & 0.5 & 1 & 0 & $H-E_3+2E_4$ & $-H+E_1+E_2+2E_3+2E_4$  %%1
-1.5 & -1.5 & 0 & 0 & $H-E_1-E_2+E_3+E_4$ &
$-H+2E_1+2E_2+E_3+2E_4$
\end{tabular}
\caption{Parameters of Example 5A.}  %3-5xn  18
\end{table}

The matter spectrum and the corresponding classes are:
\begin{equation}
\begin{tabular}{c|c|c}
Matter & class with fixed $\xi$ & generation \\\hline

${\bf 16}^{(1)}_a$  &  $H-E_1-E_2+E_3+E_4$ & 0   \\

${\bf 16}^{(3)}_b$ & $4H-E_1-E_2-3E_3-3E_4$ & 3  \\

${\bf 10}_{ab}$ & $8H-3E_1-3E_2-3E_3-3E_4$ & 2 \\

${\bf 10}_{bb}$ & $8H-3E_1-3E_2-3E_3-3E_4$ & -2 \\
\end{tabular}
\end{equation}
In this model the tadpole cancellation condition and
(\ref{muceuler}) imply that $N_{D3}=81$. For simplicity the
supersymmetry condition in (\ref{susy}) is chosen to have the
following form:
\begin{eqnarray}
\omega = \beta \left(  \sum^4_{j\neq k} (H-E_j-E_k) \right) +
\alpha \sum_{l=1}^4 E_l,
\end{eqnarray}
where $3\beta>\alpha>0$. For this model we find
$\alpha/\beta=15/7$.

\paragraph{Example 5B}~\\

In this example we present a four-generation model in $SO(10)$.
The reason to consider this situation is that if the flux $F_X$ has
non-zero restriction on the $\bf 16$ curve,
the chirality of the matter representations from this
$\bf 16$ curve will be modified such that the model may no longer
have a three-generation interpretation in the flipped $SU(5)$
picture. The parameters in this example are listed in Table
\ref{M5B}.
\begin{table}[h]
\center
\begin{tabular}{c|c|c|c|c|c}
$k_a$ & $k_b$ & $d_a$ & $d_b$ & $\xi$ & $\rho$
\\ \hline %-1.5 & -0.5 & 0 & 1 & $2H-2E_1-E_2-E_4$ & $E_1-2E_2-2E_3+2E_4$   %%%61
-2& -1 & -0.5 & 1.5 & $H-E_1-E_2+E_4$ & $\frac{1}{2}H+E_2+2E_3$
\end{tabular}
\caption{Parameters of Example 5B.}  \label{M5B} %5x4n  930 new-11
\end{table}

The matter content and the corresponding classes are:
\begin{equation}
\begin{tabular}{c|c|c}
Matter & class with fixed $\xi$ & generation \\\hline

${\bf 16}^{(1)}_a$  &  $H-E_1-E_2+E_4$ & 0   \\

${\bf 16}^{(3)}_b$ & $4H-E_1-E_2-2E_3-3E_4$ & 4  \\

${\bf 10}_{ab}$ & $8H-3E_1-3E_2-3E_3-3E_4$ & 8
\\

${\bf 10}_{bb}$ & $8H-3E_1-3E_2-3E_3-3E_4$ & -8
\\
\end{tabular}
\end{equation}
In this model the tadpole cancellation condition and
(\ref{muceuler}) imply that $N_{D3}=111$. The supersymmetry
condition implies $\alpha/\beta=2/3$.

\section{Phenomenology}
\label{sec-pheno}

In this section we will give a detailed interpretation of the
numerical results we obtained in the previous section.  First we
discuss an $SO(10)$ GUT models with a minimal spectrum. After
turning on the flux $F_X$, the $SO(10)$ gauge group is broken to
$SU(5)\times U(1)_X$, which will be interpreted as a flipped
$SU(5)$ GUT from which further gauge breaking to the MSSM is
possible.

\subsection{$SO(10)$ GUT}

The examples we presented in the previous section have the
following general spectrum:
\begin{equation}
\begin{tabular}{c|c|c|c}
Matter & Rep. & generation & $U(1)_C$ \\  \hline %
${\bf 16}_M$ & ${\bf 16}^{(3)}$ & 3 & 1 \\
${\bf 10}_H$ & ${\bf 10}^{(1,3)}$ & 1 & -2\\ \hline\hline
${\bf 10}_H$ & ${\bf 10}^{(1,3)}$ & $k-1$ & -2\\
- & ${\bf 10}^{(3,3)}$ & $-k$ & 2
\end{tabular}
\end{equation}
Here $k$ is the number of generations on the ${\bf 10}$ curve, as
given in the above examples. The $U(1)_C$ is of the Cartan
subalgebra of $SU(4)_{\bot}$ that is not removed by the monodromy,
as discussed in \cite{Marsano:2009gv}. The Yukawa coupling is
filtered by the conservation of this $U(1)_C$.
%We may interpret the matter content as one set of $(k-1)$ ${\bf
%10} + \overline{\bf 10}$ pairs and one exotic $\overline{\bf 10}$.
The superpotential is:
\begin{equation}
W \supset {\bf 16}_1 {\bf 16}_1 {\bf 10}_{-2} + \dots
\end{equation}
This model only satisfies the minimum requirement of a three
generation $SO(10)$ GUT model.  Some Higgs fields, such as $\bf
210$, $\bf 120$, and ${\bf 126}+\overline{\bf 126}$ that break the
$SO(10)$ gauge group to the $SU(5)$ and MSSM gauge groups, are
absent in the F-theory construction.  There are two possible ways
to break the gauge group.  One is to break to MSSM-like models by
introducing a non-abelian instanton (flux) which can be further
broken into a product of $U(1)$'s~\cite{Beasley:2008kw,
Chung:2009ib}. The other possibility is to introduce abelian flux
of the form \cite{Beasley:2008kw, Donagi:2008kj} $[F_X]=E_i-E_j$
to break $SO(10)$ to $SU(5)\times U(1)_X$\footnote{In the global
$U(1)$-restricted Tate model the GUT gauge group is $SO(10)\times
U(1)_C$, therefore it breaks into $SU(5)\times U(1)_X \times
U(1)_C$ after the flux is turned on. Since $U(1)_C$ is used to
confine the Yukawa couplings of the $SO(10)$ curves, in what
follows we focus the discussion on the phenomenology of the
$SU(5)\times U(1)_X$ gauge group.}. This is discussed in detail
below.

\subsection{A flipped $SU(5)$ interpretation}

\subsubsection{Restriction of $F_X$}

When the abelian flux $F_X$ is turned on, the breaking pattern of
the $SO(10)$ gauge group is:
\begin{eqnarray}
SO(10) &\supset& SU(5)\times U(1)_X \nonumber \\
{\bf 45} &\rightarrow& {\bf 24}_0 + {\bf 1}_0 + {\bf 10}_4
+\overline{\bf 10}_{-4} \nonumber \\
{\bf 16} &\rightarrow& {\bf 10}_{-1} + \bar{\bf 5}_3 + {\bf
1}_{-5} \nonumber \\
{\bf 10} &\rightarrow& {\bf 5}_{2} + \bar{\bf 5}_{-2}
\end{eqnarray}
Since we do not need ${\bf 10}_4$ and $\overline{\bf 10}_{-4}$
from the adjoint representation in the spectrum, we require their
chirality to be equal to zero. Hence if $L_X$ is the line bundle
associated with $F_X$, we set $\chi(S,L_X^4)=0$ and
$\chi(S,L_X^{-4})=0$. Then~\cite{Beasley:2008kw}:
\begin{equation}
L_X = \mathcal{O}_S(E_i-E_j)^{1/4},
\end{equation}
where $E_i$ are the exceptional divisors of $dP_k$. $L_X$ will
change the chirality of the matter fields on each curve if $F_X$
restricts non-trivially to that curve, i.e. if
$F_X|_{\Sigma}=c_1(F_X)\cdot_S \Sigma\neq 0$.  The net matter
chirality on the curves in terms of the bundles can be summarized as:
\begin{eqnarray}
n_{{\bf 10}_{-1}} &=& -\chi(\Sigma_{\bf 16}, K_{\Sigma_{\bf
16}}^{1/2}\otimes V \otimes L^{-1}_X|_{\Sigma_{\bf 16}}), \\
n_{\bar{\bf 5}_{3}} &=& -\chi(\Sigma_{\bf 16}, K_{\Sigma_{\bf
16}}^{1/2}\otimes V \otimes L^{3}_X|_{\Sigma_{\bf 16}}), \\
n_{{\bf 1}_{-5}} &=& -\chi(\Sigma_{\bf 16}, K_{\Sigma_{\bf
16}}^{1/2}\otimes V \otimes L^{-5}_X|_{\Sigma_{\bf 16}}), \\
n_{{\bf 5}_2} &=& -\chi(\Sigma_{\bf 10}, K_{\Sigma_{\bf
10}}^{1/2}\otimes \wedge^2V \otimes L^{2}_X|_{\Sigma_{\bf 10}}).
\end{eqnarray}
We can calculate the net chiralities $n_{\Sigma}$ by using $\chi(\Sigma,
K_{\Sigma}^{1/2}\otimes V \otimes L|_{\Sigma})= \deg(V\otimes
L|_{\Sigma})$, and the formula
\begin{equation}
\deg(V\otimes L)=c_1(V\otimes L)=c_1(V)+c_1(L^r),
\end{equation}
where $r$ is the rank of $V$. Since $L_X$ is fractional and only
$L_X^4$ makes sense physically, we require that $L_X^{1/4}\otimes
V_{16}$ takes the value $n_{\Sigma}$ on $\Sigma$, i.e.
$(L_X^{1/4}\otimes V)|_{\Sigma} = n_{\Sigma}$, while we set
$L_X^4|_{\Sigma}= N_{\Sigma}$
\cite{Blumenhagen:2006wj,Hayashi:2009bt,Marsano:2009wr}. We
summarize the modified chiralities on the matter curves with
non-trivial restrictions of $F_X$ in Table \ref{3-1FX}.

\begin{table}[h]
\begin{center}
\renewcommand{\arraystretch}{1.25}
\begin{tabular}{|c|c|c||c|c|} \hline
Curve & Matter & Chirality & Model 5B & rest. $F_X$ \\ \hline

\multirow{3}{*}{${\bf 16}^{(1)}_{-3}$} & ${\bf 10}_{-3,-1}$
& $n_{\bf 16}^{(1)}$ & $0$ & \multirow{3}{*}{1}\\

& $\bar{\bf 5}_{-3,3}$  & $n_{\bf 16}^{(1)}+N_{\bf 16}^{(1)}$ & $0+1$&\\

& ${\bf 1}_{-3,-5}$ & $n_{\bf 16}^{(1)}-N_{\bf 16}^{(1)}$ & $0-1$&
\\ \hline

\multirow{3}{*}{${\bf 16}^{(3)}_1$} & ${\bf 10}_{1,-1}$ &
$n_{\bf 16}^{(3)}$ & $4$ & \multirow{3}{*}{-1}\\

& $\bar{\bf 5}_{1,3}$ & $n_{\bf 16}^{(3)}+N_{\bf 16}^{(3)}$ & $4-1$ &\\

& ${\bf 1}_{1,-5}$ & $n_{\bf 16}^{(3)}-N_{\bf 16}^{(3)}$ & $4+1$ &\\
\hline

\multirow{2}{*}{${\bf 10}^{(1,3)}_{-2}$} & ${\bf 5}_{-2,2}$
& $n_{\bf 10}^{(1,3)}+N_{\bf 10}^{(1,3)}$ & $8+0$ & \multirow{2}{*}{0}\\

& $\bar{\bf 5}_{-2,-2}$  & $n_{\bf 10}^{(1,3)}$ &$8$ &\\ \hline

\multirow{2}{*}{${\bf 10}^{(3,3)}_{2}$} & ${\bf 5}_{2,2}$
& $n_{\bf 10}^{(3,3)}+N_{\bf 10}^{(3,3)}$ & $-8+0$ & \multirow{2}{*}{0}\\

& $\bar{\bf 5}_{2,-2}$ & $n_{\bf 10}^{(3,3)}$ & $-8$ & \\ \hline
\end{tabular}
\caption{The spectrum after $[F_X]=E_2-E_3$ is turned on and
restricted to the matter curves, with the numerical results from
the example of Model 5B. The first number of the subscription of
the matter representation is the $U(1)_C$ charge and the second is
the $U(1)_X$ charge.} \label{3-1FX}
\end{center}
\end{table}

\subsubsection{An example of a flipped $SU(5)$ GUT}

In Example 5A we choose for $F_X$ the class $F_X=E_1-E_2$ such
that there are trivial restrictions to all the curves. Thus the
chirality on each curve remains unchanged. After the flux is
turned on, we can interpret the spectrum as that of the flipped
$SU(5)$ model:
\begin{equation}
\begin{tabular}{c|c|c}
Matter & rep. & generation \\ \hline

${\bf 10}_M$ & ${\bf 10}_{1,-1}$ & 3 \\
$\bar{\bf 5}_M$ & $\bar{\bf 5}_{1,3}$ & 3 \\
${\bf 1}_M$ & ${\bf 1}_{1,-5}$ & 3 \\ \hline

${\bf 5}_h$ & ${\bf 5}_{-2,2}$ & 1 \\
$\bar{\bf 5}_h$ & $\bar{\bf 5}_{-2,-2}$ & 1 \\\hline

${\bf 10}_H+\overline{\bf 10}_H$ & ${\bf 10}_{1,-1}+\overline{\bf 10}_{1,1}$ & 1 \\
\hline \hline

\multirow{2}{*}{${\bf 5}+\bar{\bf 5}$} & ${\bf 5}_{-2,2}+\bar{\bf 5}_{-2,-2}$ & 1 \\
& ${\bf 5}_{2,2}+\bar{\bf 5}_{2,-2}$ & -2

\end{tabular}
\end{equation}

Although the advantage of this choice is that there is no exotic
fermion and the quantum numbers of the matter are typical, the superheavy Higgses ${\bf 10}_H$ and $\overline{\bf 10}_H$ which are needed for breaking the gauge group to the MSSM are not obvious
from the spectrum. We may claim that they are a vector-like pair
from the ${\bf 16}^{(3)}$ curve, but we are not able to fix the
number such of pairs. Therefore, we propose a configuration where
the flux $F_X$ restricts non-trivially to both the $16$ curves.

Consider $[F_X]=E_2-E_3$ in Example 5B. The flux then takes the
value $1$ on ${\bf 16}^{(1)}$ and $-1$ on ${\bf 16}^{(3)}$.
Therefore it will reduce the generation number of $\bar{\bf 5}$
representation from curve ${\bf 16}^{(3)}$ by one. That is the
reason we consider Example 5B as a four-generation model. The
detailed effect of this flux can be also found in
Table~\ref{3-1FX}. We conclude that the flipped $SU(5)$ spectrum
of Example 5B is:
\begin{equation}
\begin{tabular}{c|c|c}
Matter & rep. & generation \\ \hline

${\bf 10}_M$ & ${\bf 10}_{1,-1}$ & 3 \\
$\bar{\bf 5}_M$ & $\bar{\bf 5}_{1,3}$ & 3 \\
${\bf 1}_M$ & ${\bf 1}_{1,-5}$ & 3 \\ \hline

${\bf 10}_H+\overline{\bf 10}_H$ & ${\bf 10}_{1,-1}+\overline{\bf
10}_{1,1}$ & 1 \\ \hline

${\bf 5}_h$ & ${\bf 5}_{-2,2}$ & 1 \\
$\bar{\bf 5}_h$ & ${\bf 5}_{-2,-2}$ & 1 \\   \hline \hline

${\bf 10}$ & ${\bf 10}_{1,-1}$ & 1 \\
$\bar{\bf 5}$ & $\bar{\bf 5}_{-3,3}$ & 1 \\

${\bf 1}$ &  ${\bf 1}_{3,5}$ & 1 \\
${\bf 1}$ &  ${\bf 1}_{1,-5}$ & 2 \\ \hline

\multirow{2}{*}{${\bf 5}+\bar{\bf 5}$} & ${\bf 5}_{-2,2}+\bar{\bf 5}_{-2,-2}$ & 7 \\
& ${\bf 5}_{2,2}+\bar{\bf 5}_{2,-2}$ & -8
\end{tabular}
\end{equation}
The Yukawa couplings are standard:
\begin{eqnarray}
W&\supset&{\bf 10}_{1,-1M} {\bf 10}_{1,-1M} {\bf 5}_{-2,2h} + {\bf
10}_{1,-1M} \bar{\bf 5}_{1,3M}\bar{\bf 5}_{-2,-2h} + \bar{\bf
5}_{1,3M} {\bf 1}_{1,-5M} {\bf 5}_{-2,2h} \nonumber \\ &+&  {\bf
10}_{1,-1H} {\bf 10}_{1,-1H} {\bf 5}_{-2,2h} + \overline{\bf
10}_{1,1H} \overline{\bf 10}_{1,1H} \bar{\bf 5}_{-2,-2h}+\dots
\end{eqnarray}
Again the $\bf 10_H$ and $\overline{\bf 10}_H$ Higgs fields have
to come from ${\bf 16}^{(3)}$, which is not obvious. Although the
flux can reduce the chirality of one field, it can increase the
chirality of another field.  The total effect shows that some
exotic fields from the $\bf 16$ curves are unavoidable at the
current stage.

\subsubsection{The Singlet Higgs}

In the $SU(5)$ spectral cover the singlet matter is not obvious. A
semi-local approach to this problem suggests that the singlet
fields localize on $\lambda_i=\lambda_j$ for $i\neq j$
\cite{Marsano:2009wr}. In the Georgi-Glashow $SU(5)$ the singlet
is taken as the right-handed neutrino, while in the flipped
$SU(5)$ model this singlet is interpreted as the right-handed
electron which must clearly be included. Since in our discussion
the model starts from an $SO(10)$ gauge group and is then broken
to $SU(5)$, the matter singlet is naturally embedded into the $\bf
16$ representation. Therefore we may avoid the singlet in the
$SU(4)$ spectral cover setup.

On the other hand, in order to explain the neutrino mass problem
by a seesaw mechanism, there is a Yukawa coupling term in the
superpotential that completes the flipped $SU(5)$ model. This
Yukawa coupling term is \cite{Nanopoulos:2002qk, Georgi:1979dq}:
\begin{equation}
{\bf 10}_{-1M}\overline{\bf 10}_{1H} {\bf 1}_{0\phi}.
\end{equation}
This singlet ${\bf 1}_{0}$ can be found neither on the $\bf 16$
nor on the $\bf 10$ curves we have discussed.  It is an $SO(10)$
object, which can be identified in the $SU(4)$ cover with the
locus given by $\prod_{i<j}(\lambda_i-\lambda_j)=0$. Since it is
antisymmetric, we can square it to make it
symmetric~\cite{Marsano:2009wr}.  To calculate its matter
chirality we need to compute the genus of the curve and the degree
of the line bundle.  However the mechanism is still not clear, and
we hope to obtain a more global understanding of this singlet
curve along the lines of~\cite{Marsano:2009wr} in the future.
Therefore, here we just assume that this singlet exists and
provides the above Yukawa coupling.

\section{Conclusions and outlook}
\label{sec-conclusion}
In this paper we have explicitly constructed a class of global $SO(10)$ F-theory models. We have shown that toric geometry makes it possible to obtain a large number of Calabi-Yau fourfolds which are elliptic fibrations over non-Fano base manifolds. Inside the base manifolds we could identify a large number of divisors where one can construct GUT models on. These divisors are del Pezzo and satisfy a (mathematical and/or physical) decoupling limit. We also found that many of the base manifolds we have constructed satisfy the definitions of almost Fano manifolds. Constructing the elliptically fibered fourfolds further reduces the number of possible global models since not all base geometries can be extended to geometries of fourfolds that are torically realized as reflexive polyhedra with the right nef partition. Despite these issues we have managed to construct a significant number of geometries which are suitable for non-trivial global F-theory GUT models.

The second goal of our paper was to construct $SO(10)$ GUT models.
We worked out several examples in different geometric setups in
detail. We factorized the spectral cover in order to obtain
non-zero generation number on the $\bf 10$ matter curve. This
gives us a Higgs field in the $SO(10)$ GUT, as well as further
degrees of freedom which can be tuned to get more realistic
models.  By turning on the massless gauge flux $F_X$ we break the
$SO(10)$ gauge group to $SU(5)\times U(1)_X$. This can be
interpreted as a flipped $SU(5)$ GUT model. Superheavy Higgs
fields can be identified in the spectrum which implies that this GUT
model can be broken to the MSSM by the associated Higgs
mechanism.

There are several directions for further research. Firstly, it would be interesting to construct more general fourfolds. Our approach was to first construct a base manifold and then the elliptic fibrations. All the base manifolds were obtained from point and curve blowups in Fano threefolds which are hypersurfaces in $\mathbb{P}^4$. It can be expected that the models one gets from this rather restricted class may have very similar properties. It would be interesting to investigate the F-theory models one gets from more general base manifolds. Such a task will require an extensive computer search for models. This may be useful for the discussion of an ``F-theory landscape``.

A second issue which we have not touched at all in this paper is the problem of moduli stabilization. We have shown in explicit examples that it is possible to find three generation models in our geometries. However, saying that one can tune the moduli to specific values in order to get interesting physical properties does not imply that this necessarily happens. Finding a way to stabilize the moduli in explicit examples is a crucial requirement for the success of global F-theory GUTs. Related to this issue is the calculation of superpotentials and instanton corrections in F-theory. This has been recently discussed in~\cite{Heckman:2008es,Marsano:2008py,Marchesano:2009rz,Cvetic:2009ah,Blumenhagen:2010ja,Cvetic:2010rq} from various viewpoints. In a different context the problem of computing the superpotential has been recently addressed in~\cite{Alim:2009bx,Grimm:2009ef,Grimm:2009sy,Jockers:2009ti} where the calculation of F-theory superpotentials has been related to open string mirror symmetry. It would be interesting to investigate whether this approach can also be applied to the F-theory models we have considered here.

In our discussion of the $SO(10)$ models we relied on a split spectral cover in order to produce chiral matter on the ${\bf 10}$ curves. Recently, it has been pointed out in~\cite{Hayashi:2010zp} that the split spectral cover may not be well-defined globally. In $SU(5)$ models this leads to a possible generation of degree $4$ proton decay operators despite the introduction of a split spectral cover. A similar problem may leave us without chiral matter on the ${\bf 10}$ curves in the $SO(10)$ models. Therefore an analysis along the lines of~\cite{Hayashi:2010zp} should be done also for $SO(10)$ F-theory GUTs.

In order to discuss tadpole cancellation we have calculated the Euler number for the Calabi-Yau fourfolds after a crepant resolution of the $SO(10)$ singularities. We have compared our results with a conjectured formula for the Euler number given in~\cite{Blumenhagen:2009yv}. For three out of the five examples we have discussed we found a mismatch between the two ways of calculation. So far we have not been able to pin down where the mismatch in the results is rooted. An obvious explanation is that some of the assumptions under which the formula in~\cite{Blumenhagen:2009yv} was claimed to hold were violated. One possibility would be that there are more non-abelian enhancements in the Weierstrass model than just the $SO(10)$. Evidence for this can be collected by calculating $h^{1,1}$ of the fourfold. In the examples with the discrepancy in the Euler numbers it can be shown that after the resolution of the $SO(10)$ singularities $h^{1,1}$ changes by more than the rank of $SO(10)$\footnote{We thank T. Grimm for pointing this out to us.}. However, there can also be further reasons which can contribute to a discrepancy in the Euler numbers. One further possibility may be that the fourfold geometries exhibit terminal singularities. Given the recent results of~\cite{Hayashi:2010zp} one may also speculate whether the discrepancy in the Euler numbers is due to a globally ill-defined spectral cover, at least for models without a heterotic dual. One possibility to collect evidence for this is to check whether the models for which the Euler numbers match actually have a heterotic dual. For this to hold the fourfold $Y$ has to admit a $K3$ fibration over $S_\textmd{GUT}$  \cite{Friedman:1997yq}.  For the model with the GUT brane on $dP_7$  in \cite{Blumenhagen:2009yv} and the $dP_5$ model discussed in \cite{Grimm:2009yu} as well as our model $5$ we indeed found such a fibration structure, which indicates that these models do have heterotic duals. We intend to explain and resolve this Euler number discrepancy in the future.

While we were preparing this paper for publication \cite{Heckman:2010pv} appeared where it was discussed that the decoupling of gravity in the four-dimensional theory implies that the GUT brane wraps a non-commutative four cycle. It would be interesting to refine our discussion of decoupling limit by taking into account non-commutativity.

\appendix

%%%%%%%%%%%%%%%%%%%%%%%%%%%%%%%%%%%%%%%%%%%%%%%%%%%%%%%%%%%%%%%%%%%%%%%%%%%%
\section{List of geometries}
\label{app-models}
In this appendix we provide a class of base manifolds which come from up to three blowups of curves and points in $\mathbb{P}^4[d]$ with $d=2,3,4$, and have at least one del Pezzo divisor with a physical or mathematical decoupling limit. In the specification of the base manifold we restrict ourselves to those models where the degrees of the hypersurface equation which determine the base are strictly smaller than the sums of the weights. This class includes all examples with up to three blowups which have been discussed previously in the literature. For technical reasons we will only allow for weight matrices which do not lead to further weight vectors in the lattice polytope they create. For the study of the decoupling limit it is convenient to look only at models where the number of generators of the Mori cone is the same as the number of K\"ahler moduli. This technical requirement does not significantly reduce the number of models.

We have examined $241$ base geometries in total. $208$ of these geometries have at least one divisor which is del Pezzo and is subject to a mathematical and/or physical decoupling limit. For all the models we have checked the 'almost Fano' property and whether it is possible to construct an elliptically fibered Calabi-Yau fourfold which is characterized by a reflexive polyhedron in toric geometry. For $86$ of the models we find a reflexive polyhedron for the fourfold. We will explicitly give the data of the geometries which satisfy the requirements of an 'almost Fano' base and/or reflexivity of the fourfold polyhedron.
%%%%%%%%%%%%%%%%%%%%%%%%%%%%%%%%%%%%%%%%%%%%%%%%%%%%%%%%%%%%%%%%%%%%%%%%%%%%%%%
\subsection{Weight matrices}
We will now discuss the weight matrices for the models we would like to construct. These weight matrices together with the specification of the hypersurface divisor of the base $B$ encode all the data we need for our calculations. Since the ambient space of the three examples of Fano threefolds is always $\mathbb{P}^{4}$ the weight matrices will be the same for each Fano. Only the degrees of the hypersurface equations specifying the base will change. Therefore we can discuss the weight matrices for all three Fanos at once.

The weight vector of $\mathbb{P}^4$ is given in table~\ref{tab-p4ambient1}.
\begin{table}
\begin{center}
\begin{tabular}{c|ccccc|c}
&$y_1$&$y_2$&$y_3$&$y_4$&$y_5$&$\sum$\\
\hline
$w_1$&$1$&$1$&$1$&$1$&$1$&$5$\\
\end{tabular}
\end{center}
\caption{Weight vector for $\mathbb{P}^4[d]$.}\label{tab-p4ambient1}
\end{table}
Let us first consider curve blowups.  Since all variables have the same weight the choice is unique up to permutation of variables. The weight vector is given in table~\ref{tab-p41curve1}.
 \begin{table}
\begin{center}
\begin{tabular}{c|cccccc|c}
&$y_1$&$y_2$&$y_3$&$y_4$&$y_5$&$y_6$&$\sum$\\
\hline
$w_1$&$1$&$1$&$1$&$1$&$1$&$0$&$5$\\
$w_2$&$0$&$0$&$0$&$1$&$1$&$1$&$3$\\
\end{tabular}
\end{center}
\caption{Blowup of the first curve.}\label{tab-p41curve1}
\end{table}
For $\mathbb{P}^4[3]$ and $\mathbb{P}^4[4]$ it is possible to tune the complex structure moduli in such a way that there is a singularity at $(0,0,0,y_4,y_5)$~\cite{Blumenhagen:2009yv}.

For our models to have a decoupling limit it is usually not enough to blow up just one curve. Taking into account the symmetries, there are four possibilities to blow up a second curve, as shown in table~\ref{tab-p42curve}.
 \begin{table}
\begin{center}
\begin{tabular}{c|ccccccc|c}
&$y_1$&$y_2$&$y_3$&$y_4$&$y_5$&$y_6$&$y_7$&$\sum$\\
\hline
$w_{3,1}$&$1$&$1$&$0$&$0$&$0$&$0$&$1$&$3$\\
$w_{3,2}$&$0$&$1$&$0$&$1$&$0$&$0$&$1$&$3$\\
$w_{3,3}$&$0$&$1$&$0$&$0$&$0$&$1$&$1$&$3$\\
$w_{3,4}$&$0$&$0$&$0$&$1$&$0$&$1$&$1$&$3$\\
\end{tabular}
\end{center}
\caption{Possibilities to blow up a second curve.}\label{tab-p42curve}
\end{table}
Let us discuss these four weight vectors in turn. It looks like $w_{3,1}$ comes from a singular transition at $(y_1,y_2,0,0,0)$. However, looking at table  \ref{tab-p41curve} we find that $y_4=y_5=y_6=0$ is in the Stanley-Reisner ideal, so this point actually has to be excluded. The weight vector $w_{3,2}$ describes the second curve blowup in~\cite{Grimm:2009yu}. It comes from the blowup of the singularity at $(0,y_2,0,y_4,0,0)$. Note however that we do not insist on singular transitions in our models. In our further discussion we will not consider the weight vectors $w_{3,3}$ and $w_{3,4}$ because the polytope generated by these weight vectors has more (for $w_{3,3}$) or fewer (for $w_{3,4}$, where the same curve is blown up twice) weight vectors than those given by the weight matrix.

We also consider blowups of three curves. The weight matrix for this setup consists of $w_1,w_2,w_{3,1},w_{3,2}$, as shown in table~\ref{tab-p43curve}.
\begin{table}
\begin{center}
\begin{tabular}{c|cccccccc|c}
&$y_1$&$y_2$&$y_3$&$y_4$&$y_5$&$y_6$&$y_7$&$y_8$&$\sum$\\
\hline
$w_1$&$1$&$1$&$1$&$1$&$1$&$0$&$0$&$0$&$5$\\
$w_2$&$0$&$0$&$0$&$1$&$1$&$1$&$0$&$0$&$3$\\
$w_3$&$1$&$1$&$0$&$0$&$0$&$0$&$1$&$0$&$3$\\
$w_4$&$0$&$1$&$0$&$1$&$0$&$0$&$0$&$1$&$3$\\
\end{tabular}
\end{center}
\caption{Blowup of three curves.}\label{tab-p43curve}
\end{table}

Apart from blowing up curves, we can also make point blowups. If we blow up one curve and one point there are two inequivalent possibilities summarized in table~\ref{tab-p41curve1point}. Every '$\ast$'-- or '$\diamond$'--entry in the tables stands for one possible position of a $1$. Since we want a point blowup only one of the'$\ast$'-- or '$\diamond$'--entries can be set to one while the others must be set to zero. The difference between the '$\ast$'-- and '$\diamond$'--entries will be explained below. Setups which lead to additional weight vectors are again excluded.
\begin{table}
\begin{center}
\begin{tabular}{c|ccccccc|c}
&$y_1$&$y_2$&$y_3$&$y_4$&$y_5$&$y_6$&$y_7$&$\sum$\\
\hline
$w_1$&$1$&$1$&$1$&$1$&$1$&$0$&$0$&$5$\\
$w_2$&$0$&$0$&$0$&$1$&$1$&$1$&$0$&$3$\\
$w_3$&$\ast$&$0$&$0$&$\ast$&$0$&$0$&$1$&$2$\\
\end{tabular}
\end{center}
\caption{Blowups of one curve and one point.}\label{tab-p41curve1point}
\end{table}
Blowing up one curve and two points, we can distinguish between two cases, coming from the two possibilities of blowing up one curve and one point. Altogether we find four possible structures given in tables \ref{tab-p41curve2point1} and \ref{tab-p41curve2point2}.
\begin{table}
\begin{center}
\begin{tabular}{c|cccccccc|c}
&$y_1$&$y_2$&$y_3$&$y_4$&$y_5$&$y_6$&$y_7$&$y_8$&$\sum$\\
\hline
$w_1$&$1$&$1$&$1$&$1$&$1$&$0$&$0$&$0$&$5$\\
$w_2$&$0$&$0$&$0$&$1$&$1$&$1$&$0$&$0$&$3$\\
$w_3$&$1$&$0$&$0$&$0$&$0$&$0$&$1$&$0$&$2$\\
$w_4$&$0$&$\ast$&$0$&$\ast$&$0$&$0$&$0$&$1$&$2$\\
\end{tabular}
\end{center}
\caption{Blowups of one curve and two points, first case.}\label{tab-p41curve2point1}
\end{table}
\begin{table}
\begin{center}
\begin{tabular}{c|cccccccc|c}
&$y_1$&$y_2$&$y_3$&$y_4$&$y_5$&$y_6$&$y_7$&$y_8$&$\sum$\\
\hline
$w_1$&$1$&$1$&$1$&$1$&$1$&$0$&$0$&$0$&$5$\\
$w_2$&$0$&$0$&$0$&$1$&$1$&$1$&$0$&$0$&$3$\\
$w_3$&$0$&$0$&$0$&$1$&$0$&$0$&$1$&$0$&$2$\\
$w_4$&$\diamond$&$0$&$0$&$0$&$\ast$&$0$&$0$&$1$&$2$
\end{tabular}
\end{center}
\caption{Blowups of one curve and two points, second case.}\label{tab-p41curve2point2}
\end{table}
If we blow up two curves and one point we can distinguish between the cases where we choose $w_{3,1}$ or $w_{3,2}$ from table~\ref{tab-p42curve} as our second curve. The inequivalent possibilities we have are listed in tables \ref{tab-p42curve1point1} and \ref{tab-p42curve1point2}.
\begin{table}
\begin{center}
\begin{tabular}{c|cccccccc|c}
&$y_1$&$y_2$&$y_3$&$y_4$&$y_5$&$y_6$&$y_7$&$y_8$&$\sum$\\
\hline
$w_1$&$1$&$1$&$1$&$1$&$1$&$0$&$0$&$0$&$5$\\
$w_2$&$0$&$0$&$0$&$1$&$1$&$1$&$0$&$0$&$3$\\
$w_3$&$1$&$1$&$0$&$0$&$0$&$0$&$1$&$0$&$3$\\
$w_4$&$\ast$&$0$&$0$&$\diamond$&$0$&$0$&$0$&$1$&$2$
\end{tabular}
\end{center}
\caption{Blowups of two curves and one point, first case.}\label{tab-p42curve1point1}
\end{table}
\begin{table}
\begin{center}
\begin{tabular}{c|cccccccc|c}
&$y_1$&$y_2$&$y_3$&$y_4$&$y_5$&$y_6$&$y_7$&$y_8$&$\sum$\\
\hline
$w_1$&$1$&$1$&$1$&$1$&$1$&$0$&$0$&$0$&$5$\\
$w_2$&$0$&$0$&$0$&$1$&$1$&$1$&$0$&$0$&$3$\\
$w_3$&$0$&$1$&$0$&$1$&$0$&$0$&$1$&$0$&$3$\\
$w_4$&$\ast$&$\ast$&$0$&$\ast$&$\diamond$&$0$&$0$&$1$&$2$
\end{tabular}
\end{center}
\caption{Blowups of two curves and one point, second case.}\label{tab-p42curve1point2}
\end{table}
Finally, we can also consider cases where we do not blow up any curves but up to three points. Given the symmetry of our setup and the restrictions we imposed there is only one possibility to blow up a single point. The corresponding weight matrix is given in table~\ref{tab-p41point1}.
 \begin{table}
\begin{center}
\begin{tabular}{c|cccccc|c}
&$y_1$&$y_2$&$y_3$&$y_4$&$y_5$&$y_6$&$\sum$\\
\hline
$w_1$&$1$&$1$&$1$&$1$&$1$&$0$&$5$\\
$w_2$&$1$&$0$&$0$&$0$&$0$&$1$&$2$\\
\end{tabular}
\end{center}
\caption{Blowup of one point.}\label{tab-p41point1}
\end{table}
If we blow up two points there is only one weight matrix which meets our requirements. This is shown in table~\ref{tab-p42point}.
\begin{table}
\begin{center}
\begin{tabular}{c|ccccccc|c}
&$y_1$&$y_2$&$y_3$&$y_4$&$y_5$&$y_6$&$y_7$&$\sum$\\
\hline
$w_1$&$1$&$1$&$1$&$1$&$1$&$0$&$0$&$5$\\
$w_2$&$1$&$0$&$0$&$0$&$0$&$1$&$0$&$2$\\
$w_3$&$0$&$1$&$0$&$0$&$0$&$0$&$1$&$2$\\
\end{tabular}
\end{center}
\caption{Blowup of two points.}\label{tab-p42point}
\end{table}
If we blow up three points it turns out that there is also only one possibility, which is given in table~\ref{tab-p43point}.\\
\begin{table}
\begin{center}
\begin{tabular}{c|cccccccc|c}
&$y_1$&$y_2$&$y_3$&$y_4$&$y_5$&$y_6$&$y_7$&$y_8$&$\sum$\\
\hline
$w_1$&$1$&$1$&$1$&$1$&$1$&$0$&$0$&$0$&$5$\\
$w_2$&$1$&$0$&$0$&$0$&$0$&$1$&$0$&$0$&$2$\\
$w_3$&$0$&$1$&$0$&$0$&$0$&$0$&$1$&$0$&$2$\\
$w_4$&$0$&$0$&$1$&$0$&$0$&$0$&$0$&$1$&$2$\\
\end{tabular}
\end{center}
\caption{Blowup of three points.}\label{tab-p43point}
\end{table}
So far, we have built up our weight matrices line by line and have taken into account symmetries which come from the exchange of columns in the weight matrices because this only amounts to a permutation of coordinates. However, there are further symmetries which come from a combined exchange of rows and columns in the weight matrices. This leads to redundancies in the weight matrices given in the tables. We have marked the redundant choices by a '$\diamond$'. Altogether there are three pairs of weight matrices that are equivalent: The second choice of weights in table~\ref{tab-p41curve2point1} can be transformed into the first choice of weights in table~\ref{tab-p41curve2point2} by exchanging the last two rows and permuting the columns. Similar manipulations transform the first choice of weights in table~\ref{tab-p42curve1point1} into the second choice in this table. Furthermore in table~\ref{tab-p42curve1point2} the choice with the $1$ in the last line positioned in the second column can be transformed into the last configuration where the $1$ is in the fifth position, by exchanging the second and the third row and permuting the columns. Due to these extra symmetries of the weight matrices, one can always place a '$0$' at every position where a '$\diamond$' entry appears.

There may also arise additional symmetries from the choice of degrees in the hypersurface equations. Furthermore different triangulations of the ambient space may also lead to the same results. We have not removed these redundancies in the tables below.
%%%%%%%%%%%%%%%%%%%%%%%%%%%%%%%%%%%%%%%%%%%%%%%%%%%%%%%%%%%%%%%%%%%%%%%%%%%%%%
\subsection{Base manifolds and GUT divisors}
We will now discuss specific base geometries. Since the number of base manifolds we have constructed is quite large we will only list those geometries which satisfy the 'almost Fano' property and/or where the associated fourfold is characterized by a reflexive polytope. We will organize the data into tables with the following entries.
\begin{enumerate}
\item {\bf Weights}: This specifies one of the weight matrices listed above, formatted as $n$C$m$P$k$, where $n$ and $m$ are the number of blown up curves and points, respectively. $k$ indicates the position in the list of weight matrices.
\item {\bf Triangulation}: For some weight matrices the N-lattice polytope does not define the ambient space uniquely. The one-cones given by all the points of the N-lattice polytope may realized by different fans. This entry in the tables labels different triangulations of the polytopes, that is, the different fans.
\item {\bf Base}: We give a list of degrees which specify the base manifold $B$. The ordering is given by the ordering of the weight vectors. Whenever the 'almost Fano' criterion is satisfied we will add a $()^{\ast}$ to the vector of degrees. If the polytope of the associated Calabi-Yau fourfold is reflexive we will add a $()^{\circ}$.
\item {\bf GUT divisor}: This lists the divisors in a given model which are del Pezzo surfaces satisfying a physical and/or mathematical decoupling limit. If a particular model has been discussed in detail in Section~\ref{sec-examples} we mark this by (M${x}$) ($x$ stands for the number of the model) next to the corresponding GUT divisor.
\item {\bf Genus}: Here we give the genera of the matter curves for $SO(10)$ models\footnote{We give $SO(10)$ specific data here, but one can of course also construct $SU(5)$ models on these geometries.}, ordered as $(g_{\textrm{\bf 10}},g_{\textrm{\bf 16}})$.
\item {\bf Yukawa}: This gives the number of Yukawa couplings for $SO(10)$ models ordered as  $(n_{E_7},n_{SO(14)})$.
\item {\bf Decoupling}: This entry indicates whether there is a mathematical ('m') or physical ('p') decoupling limit.
\end{enumerate}
%%%%%%%%%%%%%%%%%%%%%%%%%%%%%%%%%%%%%%%%%%%%%%%%%%%%%%%%%%%%%%%%%%%%%%%%%%%%%%
\subsubsection{$\mathbb{P}^4[4]$}
\subsubsection*{Blowup of $1$ curve}
%\begin{equation}
\begin{longtable}{c|c|c|c|c|c|c}
{\bf Weights}&{\bf Triang.}&{\bf Base}&{\bf dP}&{\bf Genus}&{\bf Yukawa}&{\bf Decoupling}\\
\hline
$1$C$0$P$1$&1&$(4,2)^{\ast\circ}$&$dP_7$&$(2,6)$&$(10,44)$&m\\
\hline\hline
\end{longtable}
%\end{equation}
%%%%%%%%%%%%%%%%%%%%%%%%%%%%%%%%%%%%%%%%%%%%%%%%%%%%%%%%%%%%%%%%%%%%%%%%%%%%%%%%
\subsubsection*{Blowup of $2$ curves}
%\begin{equation}
\begin{longtable}{c|c|c|c|c|c|c}
{\bf Weights}&{\bf Triang.}&{\bf Base}&{\bf dP}&{\bf Genus}&{\bf Yukawa}&{\bf Decoupling}\\
\hline
$2$C$0$P$1$&$1$&$(4,2,2)^{\ast\circ}$&$dP_7$&$(2,6)$&$(10,44)$&m\\
&&&$dP_7$&$(2,6)$&$(10,44)$&m\\
\hline
$2$C$0$P$2$&$1$&$(4,2,2)^{\ast\circ}$&$dP_5$ (M$1$)&$(2,8)$&$(14,68)$&mp\\
\hline
$2$C$0$P$2$&$2$&$(4,2,2)^{\ast\circ}$&$dP_5$&$(2,8)$&$(14,68)$&mp\\
\hline\hline
\end{longtable}
%\end{equation}
%%%%%%%%%%%%%%%%%%%%%%%%%%%%%%%%%%%%%%%%%%%%%%%%%%%%%%%%%%%%%%%%%%%%%%%%%%%%%%%
\subsubsection*{Blowup of $3$ curves}
%\begin{equation}
\begin{longtable}{c|c|c|c|c|c|c}
{\bf Weights}&{\bf Triang.}&{\bf Base}&{\bf dP}&{\bf Genus}&{\bf Yukawa}&{\bf Decoupling}\\
\hline
$3$C$0$P$1$&$1$&$(4,2,2,2)^{\ast\circ}$&$dP_3$&$(2,10)$&$(18,92)$&mp\\
&&&$dP_1$&$(6,20)$&$(38,164)$&m\\
\hline
$3$C$0$P$1$&$2$&$(4,2,2,2)^{\ast}$&$dP_1$&$(6,20)$&$(38,164)$&m\\
&&&$dP_5$&$(2,8)$&$(14,68)$&m\\
\hline
$3$C$0$P$1$&$3$&$(4,2,2,2)^{\ast}$&$dP_5$&$(2,8)$&$(14,68)$&m\\
&&&$dP_1$&$(6,20)$&$(38,164)$&m\\
\hline\hline
\end{longtable}
%\end{equation}
%%%%%%%%%%%%%%%%%%%%%%%%%%%%%%%%%%%%%%%%%%%%%%%%%%%%%%%%%%%%%%%%%%%%%%%%%%%%%%%
\subsubsection*{Blowup of $1$ curve and $1$ point}
%\begin{equation}
\begin{longtable}{c|c|c|c|c|c|c}
{\bf Weights}&{\bf Triang.}&{\bf Base}&{\bf dP}&{\bf Genus}&{\bf Yukawa}&{\bf Decoupling}\\
\hline
$1$C$1$P$1$&$2$&$(4,2,1)^{\ast}$&$dP_7$&$(2,6)$&$(10,44)$&m\\
&&&$dP_6$&$(1,4)$&$(16,36)$&mp\\
\hline
$1$C$1$P$2$&$1$&$(4,2,1)^{\ast\circ}$&$dP_4$ (M$3$)&$(2,9)$&$(16,80)$&mp\\
\hline
\hline
\end{longtable}
%\end{equation}
%%%%%%%%%%%%%%%%%%%%%%%%%%%%%%%%%%%%%%%%%%%%%%%%%%%%%%%%%%%%%%%%%%%%%%%%%%%%%%%
\subsubsection*{Blowup of $1$ curve and $2$ points}
%\begin{equation}
\begin{longtable}{c|c|c|c|c|c|c}
{\bf Weights}&{\bf Triang.}&{\bf Base}&{\bf dP}&{\bf Genus}&{\bf Yukawa}&{\bf Decoupling}\\
\hline
$1$C$2$P$3$&$1$&$(4,2,1,1)^{\ast\circ}$&$dP_1$&$(2,12)$&$(22,116)$&mp\\
\hline
\hline
\end{longtable}
%\end{equation}
%%%%%%%%%%%%%%%%%%%%%%%%%%%%%%%%%%%%%%%%%%%%%%%%%%%%%%%%%%%%%%%%%%%%%%%%%%%%%%%
\subsubsection*{Blowup of $2$ curves and $1$ point}
%\begin{equation}
\begin{longtable}{c|c|c|c|c|c|c}
{\bf Weights}&{\bf Triang.}&{\bf Base}&{\bf dP}&{\bf Genus}&{\bf Yukawa}&{\bf Decoupling}\\
\hline
$2$C$1$P$1$&$1$&$(4,2,2,1)^{\ast}$&$dP_4$&$(2,9)$&$(16,80)$&mp\\
&&&$dP_7$&$(2,6)$&$(10,44)$&m\\
\hline
$2$C$1$P$3$&$1$&$(4,2,2,1)^{\ast}$&$dP_2$&$(2,11)$&$(20,104)$&mp \\
\hline
$2$C$1$P$3$&$2$&$(4,2,2,1)^{\ast}$&$dP_5$&$(2,8)$&$(14,68)$&mp \\
\hline
$2$C$1$P$4$&$1$&$(4,2,2,1)^{\ast\circ}$&$dP_5$&$(2,8)$&$(14,68)$&m \\
&&&$dP_5$&$(1,5)$&$(8,48)$&mp \\
&&&$dP_5$ (M$4$)&$(2,8)$&$(14,68)$&mp \\
\hline
\hline
\end{longtable}
%\end{equation}
%%%%%%%%%%%%%%%%%%%%%%%%%%%%%%%%%%%%%%%%%%%%%%%%%%%%%%%%%%%%%%%%%%%%%%%%%%%%%%
\subsubsection*{Blowup of $1$ point}
\begin{longtable}{c|c|c|c|c|c|c}
{\bf Weights}&{\bf Triang.}&{\bf Base}&{\bf dP}&{\bf Genus}&{\bf Yukawa}&{\bf Decoupling}\\
\hline
$0$C$1$P$1$&$1$&$(4,1)^{\ast\circ}$&$dP_6$ (M$2$)&$(1,4)$&$(6,36)$&mp\\
\hline\hline
\end{longtable}
%%%%%%%%%%%%%%%%%%%%%%%%%%%%%%%%%%%%%%%%%%%%%%%%%%%%%%%%%%%%%%%%%%%%%%%%%%%%%%
\subsubsection{$\mathbb{P}^4[3]$}
\subsubsection*{Blowup of $1$ curve}
%\begin{equation}
\begin{longtable}{c|c|c|c|c|c|c}
{\bf Weights}&{\bf Triang.}&{\bf Base}&{\bf dP}&{\bf Genus}&{\bf Yukawa}&{\bf Decoupling}\\
\hline
$1$C$0$P$1$&$1$&$(3,2)^{\ast\circ}$&$dP_1$&$(6,20)$&$(38,164)$&m\\
\hline
$1$C$0$P$1$&$1$&$(3,1)^{\ast\circ}$&$dP_4$&$(3,12)$&$(22,100)$&m\\
\hline
\hline
\end{longtable}
%\end{equation}
%%%%%%%%%%%%%%%%%%%%%%%%%%%%%%%%%%%%%%%%%%%%%%%%%%%%%%%%%%%%%%%%%%%%%%%%%%%%%%
\subsubsection*{$2$ curves}
%\begin{equation}
\begin{longtable}{c|c|c|c|c|c|c}
{\bf Weights}&{\bf Triang.}&{\bf Base}&{\bf dP}&{\bf Genus}&{\bf Yukawa}&{\bf Decoupling}\\
\hline
$2$C$0$P$1$&$1$&$(3,1,2)^{\ast\circ}$&$dP_4$&$(3,12)$&$(22,100)$&m\\
&&&$dP_1$&$(9,26)$&$(50,200)$&m\\
&&&$dP_1$&$(6,20)$&$(38,164)$&m\\
\hline
$2$C$0$P$1$&$1$&$(3,2,2)^{\ast\circ}$&$dP_1$&$(6,20)$&$(38,164)$&m\\
&&&$dP_1$&$(6,20)$&$(38,164)$&m\\
\hline
$2$C$0$P$1$&$1$&$(3,2,1)^{\ast\circ}$&$dP_1$&$(6,20)$&$(38,164)$&m\\
&&&$dP_1$&$(9,26)$&$(50,200)$&m\\
&&&$dP_4$&$(3,12)$&$(22,100)$&m\\
\hline
$2$C$0$P$1$&$1$&$(3,1,1)^{\ast\circ}$&$dP_4$&$(3,12)$&$(22,100)$&m\\
&&&$dP_4$&$(3,12)$&$(22,100)$&m\\
\hline
$2$C$0$P$2$&$1$&$(3,1,2)^{\ast}$&$dP_5$&$(2,8)$&$(14,68)$&mp\\
\hline
$2$C$0$P$2$&$1$&$(3,2,2)^{\ast\circ}$&$dP_2$&$(5,17)$&$(32,140)$&m\\
&&&$dP_1$&$(5,18)$&$(34,152)$&mp\\
\hline
$2$C$0$P$2$&$1$&$(3,1,1)^{\ast}$&$dP_2$&$(3,14)$&$(26,124)$&mp\\
\hline
$2$C$0$P$2$&$2$&$(3,2,2)^{\ast\circ}$&$dP_1$&$(5,18)$&$(34,152)$&mp\\
&&&$dP_2$&$(5,17)$&$(32,140)$&m\\
\hline
$2$C$0$P$2$&$2$&$(3,2,1)^{\ast}$&$dP_5$&$(2,8)$&$(14,68)$&mp\\
\hline
$2$C$0$P$2$&$2$&$(3,1,1)^{\ast}$&$dP_2$&$(3,14)$&$(26,124)$&mp\\
\hline
\hline
\end{longtable}
%\end{equation}
%%%%%%%%%%%%%%%%%%%%%%%%%%%%%%%%%%%%%%%%%%%%%%%%%%%%%%%%%%%%%%%%%%%%%%%%%%%%%%
\subsubsection*{Blowup of $3$ curves}
%\begin{equation}
\begin{longtable}{c|c|c|c|c|c|c}
{\bf Weights}&{\bf Triang.}&{\bf Base}&{\bf dP}&{\bf Genus}&{\bf Yukawa}&{\bf Decoupling}\\
\hline
$3$C$0$P$1$&$1$&$(3,2,2,2)^{\ast\circ}$&$dP_2$&$(5,17)$&$(32,140)$&m\\
&&&$dP_1$&$(4,16)$&$(30,140)$&mp\\
&&&$dP_2$&$(5,17)$&$(32,140)$&m\\
\hline
$3$C$0$P$1$&$1$&$(3,2,2,1)^{\ast}$&$dP_6$&$(1,4)$&$(6,36)$&mp\\
&&&$dP_3$&$(4,14)$&$(26,116)$&m\\
\hline
$3$C$0$P$1$&$1$&$(3,1,2,1)^{\ast}$&$dP_3$&$(2,10)$&$(18,92)$&mp\\
\hline
$3$C$0$P$1$&$1$&$(3,2,1,1)^{\ast}$&$dP_3$&$(2,10)$&$(18,92)$&mp\\
\hline
$3$C$0$P$1$&$1$&$(3,1,1,1)^{\ast\circ}$&$dP_0$&$(3,16)$&$(30,48)$&mp\\
\hline
$3$C$0$P$1$&$2$&$(3,2,2,2)^{\ast\circ}$&$dP_2$&$(5,17)$&$(32,140)$&m\\
&&&$dP_2$&$(4,15)$&$(28,128)$&mp\\
\hline
$3$C$0$P$1$&$2$&$(3,1,2,2)^{\ast}$&$dP_2$&$(5,17)$&$(32,140)$&m\\
&&&$dP_1$&$(6,19)$&$(36,152)$&m\\
&&&$dP_5$&$(2,8)$&$(14,68)$&m\\
\hline
$3$C$0$P$1$&$2$&$(3,1,2,1)^{\ast}$&$dP_2$&$(3,14)$&$(26,124)$&m\\
\hline
$3$C$0$P$1$&$3$&$(3,2,2,2)^{\ast\circ}$&$dP_1$&$(5,18)$&$(34,152)$&m\\
&&&$dP_2$&$(4,15)$&$(28,128)$&mp\\
&&&$dP_2$&$(5,17)$&$(32,140)$&m\\
\hline
$3$C$0$P$1$&$3$&$(3,2,1,2)^{\ast}$&$dP_5$&$(2,8)$&$(14,68)$&m\\
&&&$dP_2$&$(6,19)$&$(36,152)$&m\\
&&&$dP_2$&$(5,17)$&$(32,140)$&m\\
\hline
$3$C$0$P$1$&$3$&$(3,2,1,1)^{\ast}$&$dP_2$&$(3,14)$&$(26,124)$&m\\
\hline
\hline
\end{longtable}
%\end{equation}
%%%%%%%%%%%%%%%%%%%%%%%%%%%%%%%%%%%%%%%%%%%%%%%%%%%%%%%%%%%%%%%%%%%%%%%%%%%%%%
\subsubsection*{Blowup of $1$ curve and $1$ point}
\begin{longtable}{c|c|c|c|c|c|c}
{\bf Weights}&{\bf Triang.}&{\bf Base}&{\bf dP}&{\bf Genus}&{\bf Yukawa}&{\bf Decoupling}\\
\hline
$1$C$1$P$1$&$1$&$(3,2,1)^{\ast\circ}$&$dP_3$&$(6,18)$&$(34,140)$&m\\
&&&$dP_4$&$(6,16)$&$(30,120)$&p\\
&&&$dP_3$&$(4,14)$&$(26,116)$&p\\
\hline
$1$C$1$P$1$&$2$&$(3,2,1)^{\ast\circ}$&$dP_1$&$(6,20)$&$(38,164)$&m\\
&&&$dP_1$&$(4,16)$&$(30,140)$&mp\\
\hline
$1$C$1$P$1$&$2$&$(3,1,1)^{\ast}$&$dP_4$&$(3,12)$&$(22,100)$&m\\
&&&$dP_1$&$(4,16)$&$(30,140)$&mp\\
\hline
$1$C$1$P$2$&$1$&$(3,2,1)^{\ast\circ}$&$dP_1$&$(5,18)$&$(34,152)$&mp\\
&&&$dP_2$&$(3,13)$&$(24,116)$&mp\\
\hline
$1$C$1$P$2$&$1$&$(3,1,1)^{\ast\circ}$&$dP_4$&$(2,9)$&$(16,80)$&mp\\
\hline
\hline
\end{longtable}
%%%%%%%%%%%%%%%%%%%%%%%%%%%%%%%%%%%%%%%%%%%%%%%%%%%%%%%%%%%%%%%%%%%%%%%%%%%%%%
\subsubsection*{Blowup of $1$ curve and $2$ points}
\begin{longtable}{c|c|c|c|c|c|c}
{\bf Weights}&{\bf Triang.}&{\bf Base}&{\bf dP}&{\bf Genus}&{\bf Yukawa}&{\bf Decoupling}\\
\hline
$1$C$2$P$1$&$1$&$(3,2,1,1)^{\ast\circ}$&$dP_2$&$(4,15)$&$(28,128)$&mp\\
&&&$dP_2$&$(4,15)$&$(28,128)$&mp\\
&&&$dP_1$&$(6,20)$&$(39,164)$&m\\
\hline
$1$C$2$P$1$&$1$&$(3,1,1,1)^{\ast}$&$dP_2$&$(4,15)$&$(28,128)$&mp\\
&&&$dP_2$&$(4,15)$&$(28,128)$&mp\\
&&&$dP_4$&$(3,12)$&$(22,100)$&m\\
\hline
$1$C$2$P$1$&$2$&$(3,2,1,1)^{\ast\circ}$&$dP_2$&$(4,15)$&$(28,128)$&mp\\
&&&$dP_3$&$(5,16)$&$(30,128)$&mp\\
&&&$dP_4$&$(6,16)$&$(30,120)$&p\\
&&&$dP_3$&$(6,18)$&$(34,140)$&m\\
\hline
$1$C$2$P$1$&$3$&$(3,2,1,1)^{\ast\circ}$&$dP_6$&$(7,15)$&$(28,100)$&m\\
&&&$dP_2$&$(4,15)$&$(28,128)$&mp\\
&&&$dP_3$&$(5,16)$&$(30,128)$&mp\\
&&&$dP_4$&$(6,16)$&$(30,120)$&p\\
&&&$dP_3$&$(6,18)$&$(34,140)$&m\\
\hline
$1$C$2$P$1$&$4$&$(3,2,1,1)^{\ast\circ}$&$dP_4$&$(4,13)$&$(24,104)$&p\\
&&&$dP_4$&$(4,13)$&$(24,104)$&p\\
&&&$dP_1$&$(5,18)$&$(34,152)$&mp\\
&&&$dP_4$&$(6,16)$&$(30,120)$&p\\
&&&$dP_4$&$(6,16)$&$(30,120)$&p\\
\hline
$1$C$2$P$2$&$1$&$(3,2,1,1)^{\ast\circ}$&$dP_4$&$(6,16)$&$(30,120)$&p\\
&&&$dP_2$&$(5,17)$&$(32,140)$&p\\
&&&$dP_4$&$(6,16)$&$(30,120)$&p\\
&&&$dP_3$&$(4,14)$&$(26,116)$&p\\
&&&$dP_3$&$(3,12)$&$(22,104)$&p\\
\hline
$1$C$2$P$2$&$2$&$(3,2,1,1)^{\ast\circ}$&$dP_1$&$(5,18)$&$(34,152)$&mp\\
&&&$dP_2$&$(4,15)$&$(28,128)$&mp\\
&&&$dP_3$&$(3,12)$&$(22,104)$&mp\\
\hline
$1$C$2$P$2$&$2$&$(3,1,1,1)^{\ast}$&$dP_4$&$(2,9)$&$(16,80)$&p\\
&&&$dP_2$&$(4,15)$&$(28,128)$&mp\\
\hline
$1$C$2$P$3$&$1$&$(3,2,1,1)^{\ast\circ}$&$dP_2$&$(3,13)$&$(24,116)$&mp\\
&&&$dP_1$&$(4,16)$&$(30,140)$&mp\\
&&&$dP_2$&$(3,13)$&$(24,116)$&mp\\
\hline
$1$C$2$P$3$&$1$&$(3,1,1,1)^{\ast\circ}$&$dP_4$&$(1,6)$&$(10,60)$&mp\\
\hline
\hline
\end{longtable}
%%%%%%%%%%%%%%%%%%%%%%%%%%%%%%%%%%%%%%%%%%%%%%%%%%%%%%%%%%%%%%%%%%%%%%%%%%%%%%
\subsubsection*{Blowup of $2$ curves and $1$ point}
\begin{longtable}{c|c|c|c|c|c|c}
{\bf Weights}&{\bf Triang.}&{\bf Base}&{\bf dP}&{\bf Genus}&{\bf Yukawa}&{\bf Decoupling}\\
\hline
$2$C$1$P$1$&$1$&$(3,2,2,1)^{\ast\circ}$&$dP_1$&$(5,18)$&$(34,152)$&mp\\
&&&$dP_2$&$(3,13)$&$(24,116)$&mp\\
&&&$dP_1$&$(6,20)$&$(38,164)$&m\\
\hline
$2$C$1$P$1$&$1$&$(3,1,2,1)^{\ast}$&$dP_1$&$(5,18)$&$(34,152)$&mp\\
&&&$dP_2$&$(3,13)$&$(24,116)$&mp\\
&&&$dP_4$&$(3,12)$&$(22,100)$&m\\
\hline
$2$C$1$P$1$&$1$&$(3,2,1,1)^{\ast\circ}$&$dP_4$ (M$5$)&$(2,9)$&$(16,80)$&mp\\
&&&$dP_1$&$(6,20)$&$(38,164)$&m\\
\hline
$2$C$1$P$1$&$1$&$(3,1,1,1)^{\ast\circ}$&$dP_4$&$(2,9)$&$(16,80)$&mp\\
&&&$dP_4$&$(3,12)$&$(22,100)$&m\\
\hline
$2$C$1$P$1$&$2$&$(3,2,2,1)^{\ast\circ}$&$dP_1$&$(5,18)$&$(34,152)$&mp\\
&&&$dP_4$&$(3,11)$&$(20,92)$&p\\
&&&$dP_4$&$(4,13)$&$(24,104)$&p\\
&&&$dP_4$&$(6,16)$&$(30,120)$&p\\
&&&$dP_3$&$(6,18)$&$(34,140)$&m\\
\hline
$2$C$1$P$1$&$2$&$(3,2,1,1)^{\ast\circ}$&$dP_4$&$(2,9)$&$(16,80)$&mp\\
&&&$dP_1$&$(7,22)$&$(42,176)$&mp\\
&&&$dP_4$&$(6,16)$&$(30,120)$&p\\
&&&$dP_3$&$(6,18)$&$(34,140)$&m\\
\hline
$2$C$1$P$1$&$2$&$(3,1,1,1)^{\circ}$&$dP_4$&$(2,9)$&$(16,80)$&mp\\
&&&$dP_1$&$(9,25)$&$(48,192)$&p\\
&&&$dP_4$&$(5,15)$&$(28,116)$&m\\
\hline
$2$C$1$P$2$&$1$&$(3,2,2,1)^{\ast\circ}$&$dP_1$&$(6,20)$&$(38,164)$&m\\
&&&$dP_2$&$(3,13)$&$(24,116)$&mp\\
&&&$dP_1$&$(5,18)$&$(34,152)$&mp\\
\hline
$2$C$1$P$2$&$1$&$(3,1,2,1)^{\ast\circ}$&$dP_1$&$(6,20)$&$(38,164)$&m\\
&&&$dP_4$&$(2,9)$&$(16,80)$&mp\\
\hline
$2$C$1$P$2$&$1$&$(3,2,2,1)^{\ast\circ}$&$dP_4$&$(6,16)$&$(30,120)$&p\\
&&&$dP_4$&$(5,15)$&$(28,116)$&m\\
&&&$dP_5$&$(5,16)$&$(30,128)$&mp\\
&&&$dP_2$&$(4,15)$&$(28,128)$&mp\\
\hline
$2$C$1$P$2$&$2$&$(3,2,2,1)^{\ast\circ}$&$dP_4$&$(6,16)$&$(30,120)$&p\\
&&&$dP_3$&$(5,16)$&$(30,128)$&mp\\
&&&$dP_4$&$(5,15)$&$(28,116)$&m\\
&&&$dP_2$&$(4,15)$&$(28,128)$&mp\\
\hline
$2$C$1$P$2$&$3$&$(3,2,2,1)^{\ast\circ}$&$dP_4$&$(6,16)$&$(30,120)$&p\\
&&&$dP_1$&$(5,18)$&$(34,152)$&mp\\
&&&$dP_4$&$(5,15)$&$(28,116)$&m\\
&&&$dP_3$&$(4,14)$&$(26,116)$&p\\
&&&$dP_4$&$(4,13)$&$(24,104)$&p\\
\hline
$2$C$1$P$2$&$3$&$(3,2,1,1)^{\ast}$&$dP_4$&$(6,16)$&$(30,120)$&p\\
&&&$dP_5$&$(2,8)$&$(14,68)$&mp\\
&&&$dP_3$&$(4,14)$&$(26,116)$&p\\
&&&$dP_0$&$(7,23)$&$(44,188)$&p\\
\hline
$2$C$1$P$2$&$4$&$(3,2,2,1)^{\ast\circ}$&$dP_4$&$(6,16)$&$(30,120)$&p\\
&&&$dP_4$&$(5,15)$&$(28,116)$&m\\
&&&$dP_1$&$(5,18)$&$(34,152)$&mp\\
&&&$dP_3$&$(4,14)$&$(26,116)$&p\\
&&&$dP_4$&$(4,13)$&$(24,104)$&p\\
\hline
$2$C$1$P$2$&$4$&$(3,1,2,1)^{\ast}$&$dP_4$&$(6,16)$&$(30,120)$&p\\
&&&$dP_5$&$(2,8)$&$(14,68)$&mp\\
&&&$dP_3$&$(4,14)$&$(26,116)$&p\\
&&&$dP_0$&$(7,23)$&$(44,188)$&p\\
\hline
$2$C$1$P$3$&$1$&$(3,2,2,1)^{\ast\circ}$&$dP_2$&$(5,17)$&$(32,140)$&m\\
&&&$dP_1$&$(4,16)$&$(30,140)$&mp\\
&&&$dP_2$&$(3,13)$&$(24,116)$&mp\\
\hline
$2$C$1$P$3$&$1$&$(3,2,1,1)^{\ast}$&$dP_5$&$(1,5)$&$(8,48)$&mp\\
\hline
$2$C$1$P$3$&$1$&$(3,1,1,1)^{\ast}$&$dP_2$&$(2,11)$&$(20,104)$&mp\\
\hline
$2$C$1$P$3$&$2$&$(3,2,2,1)^{\ast\circ}$&$dP_1$&$(5,18)$&$(34,151)$&mp\\
&&&$dP_2$&$(4,15)$&$(28,128)$&mp\\
&&&$dP_2$&$(3,13)$&$(24,116)$&mp\\
\hline
$2$C$1$P$3$&$2$&$(3,1,2,1)^{\ast}$&$dP_5$&$(2,8)$&$(14,68)$&mp\\
&&&$dP_2$&$(3,13)$&$(24,116)$&mp\\
\hline
$2$C$1$P$3$&$2$&$(3,1,1,1)^{\ast}$&$dP_2$&$(3,14)$&$(26,124)$&mp\\
\hline
$2$C$1$P$3$&$3$&$(3,2,2,1)^{\ast\circ}$&$dP_5$&$(5,13)$&$(24,96)$&p\\
&&&$dP_3$&$(5,16)$&$(30,128)$&m\\
&&&$dP_1$&$(4,16)$&$(30,140)$&mp\\
&&&$dP_5$&$(5,13)$&$(24,96)$&p\\
&&&$dP_3$&$(3,12)$&$(22,104)$&p\\
\hline
$2$C$1$P$3$&$3$&$(3,2,1,1)^{\ast}$&$dP_5$&$(1,5)$&$(8,48)$&mp\\
\hline
$2$C$1$P$3$&$4$&$(3,2,2,1)^{\ast\circ}$&$dP_5$&$(5,13)$&$(24,96)$&p\\
&&&$dP_5$&$(5,17)$&$(32,140)$&mp\\
&&&$dP_2$&$(4,15)$&$(28,128)$&mp\\
&&&$dP_5$&$(5,13)$&$(24,96)$&p\\
\hline
$2$C$1$P$4$&$1$&$(3,2,2,1)^{\ast\circ}$&$dP_1$&$(5,18)$&$(34,152)$&mp\\
&&&$dP_3$&$(2,10)$&$(18,92)$&mp\\
&&&$dP_1$&$(5,18)$&$(34,152)$&mp\\
\hline
$2$C$1$P$4$&$1$&$(3,2,1,1)^{\ast\circ}$&$dP_5$&$(2,8)$&$(14,68)$&mp\\
&&&$dP_1$&$(3,14)$&$(26,128)$&mp\\
&&&$dP_2$&$(5,17)$&$(32,140)$&mp\\
\hline
$2$C$1$P$4$&$1$&$(3,1,2,1)^{\ast\circ}$&$dP_2$&$(5,17)$&$(32,140)$&mp\\
&&&$dP_1$&$(3,14)$&$(26,128)$&mp\\
&&&$dP_5$&$(2,8)$&$(14,68)$&mp\\
\hline
\hline
\end{longtable}
%%%%%%%%%%%%%%%%%%%%%%%%%%%%%%%%%%%%%%%%%%%%%%%%%%%%%%%%%%%%%%%%%%%%%%%%%%%%%%
\subsubsection*{Blowup of $1$ point}
\begin{longtable}{c|c|c|c|c|c|c}
{\bf Weights}&{\bf Triang.}&{\bf Base}&{\bf dP}&{\bf Genus}&{\bf Yukawa}&{\bf Decoupling}\\
\hline
$0$C$1$P$0$&$1$&$(3,1)^{\ast\circ}$&$dP_1$&$(4,16)$&$(30,140)$&mp\\
\hline\hline
\end{longtable}
%%%%%%%%%%%%%%%%%%%%%%%%%%%%%%%%%%%%%%%%%%%%%%%%%%%%%%%%%%%%%%%%%%%%%%%%%%%%%%
\subsubsection*{Blowup of $2$ points}
\begin{longtable}{c|c|c|c|c|c|c}
{\bf Weights}&{\bf Triang.}&{\bf Base}&{\bf dP}&{\bf Genus}&{\bf Yukawa}&{\bf Decoupling}\\
\hline
$0$C$2$P$0$&$1$&$(3,1,1)^{\ast\circ}$&$dP_2$&$(4,15)$&$(28,128)$&mp\\
&&&$dP_2$&$(4,15)$&$(28,128)$&mp\\
\hline\hline
\end{longtable}
%%%%%%%%%%%%%%%%%%%%%%%%%%%%%%%%%%%%%%%%%%%%%%%%%%%%%%%%%%%%%%%%%%%%%%%%%%%%%%
\subsubsection*{Blowup of $3$ points}
\begin{longtable}{c|c|c|c|c|c|c}
{\bf Weights}&{\bf Triang.}&{\bf Base}&{\bf dP}&{\bf Genus}&{\bf Yukawa}&{\bf Decoupling}\\
\hline
$0$C$3$P$0$&$1$&$(3,1,1,1)^{\ast\circ}$&$dP_3$&$(4,14)$&$(26,116)$&p\\
&&&$dP_3$&$(4,14)$&$(26,116)$&p\\
&&&$dP_3$&$(4,14)$&$(26,116)$&p\\
\hline\hline
\end{longtable}
%%%%%%%%%%%%%%%%%%%%%%%%%%%%%%%%%%%%%%%%%%%%%%%%%%%%%%%%%%%%%%%%%%%%%%%%%%%%%%
\subsubsection{$\mathbb{P}^4[2]$}
\subsubsection*{Blowup of $1$ curve}
\begin{longtable}{c|c|c|c|c|c|c}
{\bf Weights}&{\bf Triang.}&{\bf Base}&{\bf dP}&{\bf Genus}&{\bf Yukawa}&{\bf Decoupling}\\
\hline
$1$C$0$P$1$&$1$&$(2,1)^{\ast\circ}$&$dP_1$&$(7,22)$&$(42,176)$&m\\
\hline\hline
\end{longtable}
%%%%%%%%%%%%%%%%%%%%%%%%%%%%%%%%%%%%%%%%%%%%%%%%%%%%%%%%%%%%%%%%%%%%%%%%%%%%%%%
\subsubsection*{Blowup of $2$ curves}
\begin{longtable}{c|c|c|c|c|c|c}
{\bf Weights}&{\bf Triang.}&{\bf Base}&{\bf dP}&{\bf Genus}&{\bf Yukawa}&{\bf Decoupling}\\
\hline
$2$C$0$P$1$&$1$&$(2,1,1)^{\ast\circ}$&$dP_1$&$(7,22)$&$(42,176)$&m\\
&&&$dP_1$&$(8,24)$&$(76,188)$&m\\
&&&$dP_1$&$(7,22)$&$(42,176)$&m\\
\hline
$2$C$0$P$2$&$1$&$(2,1,1)^{\ast\circ}$&$dP_2$&$(6,19)$&$(36,152)$&m\\
&&&$dP_1$&$(6,20)$&$(38,164)$&mp\\
\hline
$2$C$0$P$2$&$2$&$(2,1,1)^{\ast\circ}$&$dP_1$&$(12,30)$&$(58,220)$&p\\
&&&$dP_1$&$(6,20)$&$(38,164)$&mp\\
&&&$dP_2$&$(6,19)$&$(36,152)$&m\\
\hline
\hline
\end{longtable}
%%%%%%%%%%%%%%%%%%%%%%%%%%%%%%%%%%%%%%%%%%%%%%%%%%%%%%%%%%%%%%%%%%%%%%%%%%%%%%%
\subsubsection*{Blowup of $3$ curves}
\begin{longtable}{c|c|c|c|c|c|c}
{\bf Weights}&{\bf Triang.}&{\bf Base}&{\bf dP}&{\bf Genus}&{\bf Yukawa}&{\bf Decoupling}\\
\hline
$3$C$0$P$1$&$1$&$(2,1,1,1)^{\ast\circ}$&$dP_2$&$(6,19)$&$(36,152)$&m\\
&&&$dP_1$&$(5,18)$&$(34,152)$&mp\\
&&&$dP_1$&$(6,20)$&$(38,164)$&m\\
\hline
$3$C$0$P$1$&$2$&$(2,1,1,1)^{\ast\circ}$&$dP_2$&$(6,19)$&$(36,152)$&m\\
&&&$dP_2$&$(5,17)$&$(32,140)$&mp\\
&&&$dP_1$&$(6,20)$&$(38,164)$&m\\
&&&$dP_1$&$(6,20)$&$(38,164)$&m\\
\hline
$3$C$0$P$1$&$3$&$(2,1,1,1)^{\ast\circ}$&$dP_1$&$(6,20)$&$(38,164)$&m\\
&&&$dP_2$&$(5,17)$&$(32,140)$&mp\\
&&&$dP_1$&$(6,20)$&$(38,164)$&m\\
&&&$dP_2$&$(6,19)$&$(36,152)$&m\\
\hline
\hline
\end{longtable}
%%%%%%%%%%%%%%%%%%%%%%%%%%%%%%%%%%%%%%%%%%%%%%%%%%%%%%%%%%%%%%%%%%%%%%%%%%%%%%
\subsubsection*{Blowup of $1$ curve and $1$ point}
\begin{longtable}{c|c|c|c|c|c|c}
{\bf Weights}&{\bf Triang.}&{\bf Base}&{\bf dP}&{\bf Genus}&{\bf Yukawa}&{\bf Decoupling}\\
\hline
$1$C$1$P$1$&$1$&$(2,1,1)^{\ast\circ}$&$dP_1$&$(9,25)$&$(48,192)$&m\\
&&&$dP_1$&$(6,20)$&$(38,164)$&p\\
\hline
$1$C$1$P$1$&$2$&$(2,1,1)^{\ast\circ}$&$dP_1$&$(7,22)$&$(42,176)$&m\\
&&&$dP_0$&$(6,21)$&$(40,176)$&mp\\
\hline
$1$C$1$P$2$&$1$&$(2,1,1)^{\ast\circ}$&$dP_1$&$(5,18)$&$(34,152)$&mp\\
&&&$dP_1$&$(5,18)$&$(34,152)$&mp\\
\hline
\hline
\end{longtable}
%%%%%%%%%%%%%%%%%%%%%%%%%%%%%%%%%%%%%%%%%%%%%%%%%%%%%%%%%%%%%%%%%%%%%%%%%%%%%%
\subsubsection*{Blowup of $1$ curve and $2$ points}
\begin{longtable}{c|c|c|c|c|c|c}
{\bf Weights}&{\bf Triang.}&{\bf Base}&{\bf dP}&{\bf Genus}&{\bf Yukawa}&{\bf Decoupling}\\
\hline
$1$C$2$P$1$&$1$&$(2,1,1,1)^{\ast\circ}$&$dP_0$&$(6,21)$&$(40,176)$&mp\\
&&&$dP_0$&$(6,21)$&$(40,176)$&mp\\
&&&$dP_1$&$(7,22)$&$(42,176)$&m\\
\hline
$1$C$2$P$1$&$2$&$(2,1,1,1)^{\ast\circ}$&$dP_0$&$(6,21)$&$(40,176)$&mp\\
&&&$dP_2$&$(6,19)$&$(36,152)$&mp\\
&&&$dP_1$&$(9,25)$&$(48,192)$&mp\\
&&&$dP_2$&$(7,21)$&$(40,164)$&m\\
\hline
$1$C$2$P$1$&$3$&$(2,1,1,1)^{\ast\circ}$&$dP_0$&$(6,21)$&$(40,176)$&mp\\
&&&$dP_2$&$(6,19)$&$(36,152)$&mp\\
&&&$dP_1$&$(9,25)$&$(48,192)$&p\\
&&&$dP_2$&$(9,24)$&$(46,180)$&m\\
&&&$dP_2$&$(7,21)$&$(40,164)$&m\\
\hline
$1$C$2$P$1$&$4$&$(2,1,1,1)^{\ast\circ}$&$dP_1$&$(6,20)$&$(38,164)$&p\\
&&&$dP_1$&$(6,20)$&$(38,164)$&p\\
&&&$dP_1$&$(6,20)$&$(38,164)$&mp\\
&&&$dP_1$&$(9,25)$&$(48,192)$&p\\
&&&$dP_1$&$(9,25)$&$(48,192)$&p\\
\hline
$1$C$2$P$2$&$1$&$(2,1,1,1)^{\ast\circ}$&$dP_2$&$(8,22)$&$(42,168)$&p\\
&&&$dP_2$&$(5,17)$&$(32,140)$&mp\\
&&&$dP_2$&$(5,17)$&$(32,140)$&p\\
&&&$dP_1$&$(6,20)$&$(38,164)$&p\\
&&&$dP_1$&$(5,18)$&$(34,152)$&mp\\
\hline
$1$C$2$P$2$&$2$&$(2,1,1,1)^{\ast\circ}$&$dP_1$&$(5,18)$&$(34,152)$&mp\\
&&&$dP_0$&$(6,21)$&$(40,176)$&mp\\
&&&$dP_1$&$(5,18)$&$(34,152)$&mp\\
\hline
$1$C$2$P$3$&$1$&$(2,1,1,1)^{\ast\circ}$&$dP_1$&$(5,18)$&$(34,152)$&mp\\
&&&$dP_1$&$(3,14)$&$(26,128)$&mp\\
&&&$dP_1$&$(5,18)$&$(34,152)$&mp\\
\hline
\hline
\end{longtable}
%%%%%%%%%%%%%%%%%%%%%%%%%%%%%%%%%%%%%%%%%%%%%%%%%%%%%%%%%%%%%%%%%%%%%%%%%%%%%%
\subsubsection*{$2$ curves, $1$ point}
\begin{longtable}{c|c|c|c|c|c|c}
{\bf Weights}&{\bf Triang.}&{\bf Base}&{\bf dP}&{\bf Genus}&{\bf Yukawa}&{\bf Couplings}\\
\hline
$2$C$1$P$1$&$1$&$(2,1,1,1)^{\ast\circ}$&$dP_1$&$(5,18)$&$(34,152)$&mp\\
&&&$dP_1$&$(5,18)$&$(34,152)$&mp\\
&&&$dP_2$&$(7,21)$&$(40,164)$&m\\
&&&$dP_1$&$(7,22)$&$(42,176)$&m\\
\hline
$2$C$1$P$1$&$2$&$(2,1,1,1)^{\ast\circ}$&$dP_1$&$(5,18)$&$(34,152)$&mp\\
&&&$dP_2$&$(5,17)$&$(32,140)$&mp\\
&&&$dP_1$&$(7,22)$&$(42,176)$&mp\\
&&&$dP_2$&$(9,25)$&$(48,192)$&p\\
&&&$dP_2$&$(7,21)$&$(40,164)$&m\\
\hline
$2$C$1$P$2$&$1$&$(2,1,1,1)^{\ast\circ}$&$dP_1$&$(9,25)$&$(48,192)$&p\\
&&&$dP_3$&$(6,18)$&$(34,140)$&m\\
&&&$dP_2$&$(6,19)$&$(36,152)$&mp\\
&&&$dP_0$&$(6,21)$&$(40,176)$&mp\\
\hline
$2$C$1$P$2$&$2$&$(2,1,1,1)^{\ast\circ}$&$dP_1$&$(9,25)$&$(48,192)$&p\\
&&&$dP_2$&$(6,19)$&$(36,152)$&mp\\
&&&$dP_3$&$(6,18)$&$(34,140)$&m\\
&&&$dP_0$&$(6,21)$&$(40,176)$&mp\\
\hline
$2$C$1$P$2$&$3$&$(2,1,1,1)^{\ast\circ}$&$dP_1$&$(6,20)$&$(38,164)$&mp\\
&&&$dP_3$&$(6,18)$&$(34,140)$&m\\
&&&$dP_1$&$(6,20)$&$(38,164)$&p\\
&&&$dP_1$&$(6,20)$&$(38,164)$&p\\
\hline
$2$C$1$P$2$&$4$&$(2,1,1,1)^{\ast\circ}$&$dP_1$&$(9,25)$&$(48,192)$&p\\
&&&$dP_3$&$(6,18)$&$(34,140)$&m\\
&&&$dP_1$&$(6,20)$&$(38,164)$&mp\\
&&&$dP_1$&$(6,20)$&$(38,164)$&p\\
&&&$dP_1$&$(6,20)$&$(38,164)$&p\\
\hline
$2$C$1$P$3$&$1$&$(2,1,1,1)^{\ast\circ}$&$dP_2$&$(6,19)$&$(36,152)$&m\\
&&&$dP_1$&$(4,16)$&$(30,140)$&mp\\
&&&$dP_1$&$(5,18)$&$(34,152)$&mp\\
\hline
$2$C$1$P$3$&$2$&$(2,1,1,1)^{\ast\circ}$&$dP_1$&$(6,20)$&$(38,164)$&mp\\
&&&$dP_2$&$(4,15)$&$(28,128)$&mp\\
&&&$dP_1$&$(5,18)$&$(34,152)$&mp\\
\hline
$2$C$1$P$3$&$3$&$(2,1,1,1)^{\ast\circ}$&$dP_2$&$(8,22)$&$(42,168)$&p\\
&&&$dP_2$&$(6,19)$&$(36,152)$&m\\
&&&$dP_1$&$(4,16)$&$(30,140)$&mp\\
&&&$dP_2$&$(8,22)$&$(42,168)$&p\\
&&&$dP_1$&$(5,18)$&$(34,152)$&mp\\
\hline
$2$C$1$P$3$&$4$&$(2,1,1,1)^{\ast\circ}$&$dP_2$&$(8,22)$&$(42,168)$&p\\
&&&$dP_1$&$(6,20)$&$(38,164)$&mp\\
&&&$dP_2$&$(4,15)$&$(28,128)$&mp\\
&&&$dP_2$&$(8,22)$&$(42,168)$&p\\
&&&$dP_1$&$(5,18)$&$(34,152)$&m\\
\hline
$2$C$1$P$4$&$1$&$(2,1,1,1)^{\ast\circ}$&$dP_2$&$(5,17)$&$(32,140)$&mp\\
&&&$dP_2$&$(4,16)$&$(30,140)$&mp\\
&&&$dP_2$&$(5,17)$&$(32,140)$&mp\\
\hline
\hline
\end{longtable}
%%%%%%%%%%%%%%%%%%%%%%%%%%%%%%%%%%%%%%%%%%%%%%%%%%%%%%%%%%%%%%%%%%%%%%%%%%%%%%
\subsubsection*{Blowup of $1$ point}
\begin{longtable}{c|c|c|c|c|c|c}
{\bf Weights}&{\bf Triang.}&{\bf Base}&{\bf dP}&{\bf Genus}&{\bf Yukawa}&{\bf Decoupling}\\
\hline
$0$C$1$P$1$&$1$&$(2,1)^{\ast\circ}$&$dP_0$&$(6,21)$&$(40,176)$&mp\\
\hline\hline
\end{longtable}
%%%%%%%%%%%%%%%%%%%%%%%%%%%%%%%%%%%%%%%%%%%%%%%%%%%%%%%%%%%%%%%%%%%%%%%%%%%%%%
\subsubsection*{Blowup of $2$ points}
\begin{longtable}{c|c|c|c|c|c|c}
{\bf Weights}&{\bf Triang.}&{\bf Base}&{\bf dP}&{\bf Genus}&{\bf Yukawa}&{\bf Decoupling}\\
\hline
$0$C$2$P$1$&$1$&$(2,1,1)^{\ast\circ}$&$dP_0$&$(6,21)$&$(40,176)$&mp\\
&&&$dP_0$&$(6,21)$&$(40,176)$&mp\\
\hline
\hline
\end{longtable}
%%%%%%%%%%%%%%%%%%%%%%%%%%%%%%%%%%%%%%%%%%%%%%%%%%%%%%%%%%%%%%%%%%%%%%%%%%%%%%%
\subsubsection*{Blowup of $3$ points}
\begin{longtable}{c|c|c|c|c|c|c}
{\bf Weights}&{\bf Triang.}&{\bf Base}&{\bf dP}&{\bf Genus}&{\bf Yukawa}&{\bf Decoupling}\\
\hline
$0$C$2$P$1$&$1$&$(2,1,1,1)^{\ast\circ}$&$dP_0$&$(6,21)$&$(40,176)$&p\\
&&&$dP_0$&$(6,21)$&$(40,176)$&p\\
&&&$dP_0$&$(6,21)$&$(40,176)$&p\\
\hline
\hline
\end{longtable}

\section{Fourfold data}\label{fourfold-data}
Here we give the explicit data of the Calabi-Yau fourfolds constructed from the base manifolds of models 2, 3, 4 and 5. For convenience we relabeled the vertices obtained from the base. The vertex corresponding to the GUT divisor is given the coordinate $w$. The additional vertices/coordinates obtained after dualizing the reduced M-lattice polytope are denoted with a tilde. Furthermore we compute the Euler numbers for the $SO(10)$ model and compare with the formula~(\ref{eulerformula}).

\subsection{Model $2$}
The vertices in the N-lattice are:
\begin{equation}
\begin{tabular}{c|c||c|c}
nef-part.&vertices&weights&coordinates\\
\hline\hline
$\nabla_1$&$\nu_1=(\begin{array}{rrrrrrr}\nm 3&\nm 1&\nm 0&\nm 0&\nm 0&\nm 1\end{array})$&$\begin{array}{rrr}2&2&2\end{array}$&$x$\\
&$\nu_2=(\begin{array}{rrrrrrr}-2&-1&\nm 0&\nm 0&\nm 0&\nm 1\end{array})$&$\begin{array}{rrr}3&3&3\end{array}$&$y$\\
&$\nu_3=(\begin{array}{rrrrrrr}\nm 0&\nm 1&\nm 0&\nm 0&\nm 0&\nm 1\end{array})$&$\begin{array}{rrr}1&0&0\end{array}$&$z$\\
&$\nu_4=(\begin{array}{rrrrrrr}\nm 0&\nm 0&\nm 0&\nm 0&\nm 0&\nm 1\end{array})$&$\begin{array}{rrr}0&0&1\end{array}$&$w$\\
&$\nu_5=(\begin{array}{rrrrrrr}\nm 0&\nm 0&\nm 0&\nm 1&\nm 0&\nm 0\end{array})$&$\begin{array}{rrr}0&1&0\end{array}$&$y_1$\\
\hline
$\nabla_2$&$\nu_6=(\begin{array}{rrrrrrr}\nm 0&\nm 0&\nm 0&\nm 1&\nm 0&\nm 0\end{array})$&$\begin{array}{rrr}0&1&0\end{array}$&$y_2$\\
&$\nu_7=(\begin{array}{rrrrrrr}\nm 0&\nm 0&\nm 0&\nm 0&\nm 1&\nm 0\end{array})$&$\begin{array}{rrr}0&1&0\end{array}$&$y_3$\\
&$\nu_8=(\begin{array}{rrrrrrr}\nm 0&\nm 1&\nm 0&\nm 0&\nm 0&\nm 0\end{array})$&$\begin{array}{rrr}0&1&1\end{array}$&$y_4$\\
&$\nu_9=(\begin{array}{rrrrrrr}\nm 0&\nm 0&-1&-1&-1&\nm 1\end{array})$&$\begin{array}{rrr}0&1&0\end{array}$&$y_5$\\
\hline\hline
\end{tabular}
\end{equation}
%%%%%
After reducing the M-lattice polytope to the SO(10) case, we obtain for the dual N-lattice polytope:
\begin{equation}
\begin{tabular}{c|c||c|c}
nef-part.&vertices&weights&coordinates\\
\hline\hline
$\nabla_1$&$\nu_1=(\begin{array}{rrrrrrr}\nm 3&\nm 1&\nm 0&\nm 0&\nm 0&\nm 1\end{array})$&$\begin{array}{rrrrr}2&2&1&2&0\end{array}$&$x$\\
&$\nu_2=(\begin{array}{rrrrrrr}-2&-1&\nm 0&\nm 0&\nm 0&\nm 1\end{array})$&$\begin{array}{rrrrr}3&3&2&3&0\end{array}$&$y$\\
&$\nu_3=(\begin{array}{rrrrrrr}\nm 0&\nm 1&\nm 0&\nm 0&\nm 0&\nm 1\end{array})$&$\begin{array}{rrrrr}1&0&0&0&0\end{array}$&$z$\\
%&$\nu_4=(\begin{array}{rrrrrrr}\nm 0&\nm 0&\nm 0&\nm 0&\nm 0&\nm 1\end{array})$&$\begin{array}{rrrrr}0&0&1\end{array}$&$w$\\
&$\nu_5=(\begin{array}{rrrrrrr}\nm 0&\nm 0&\nm 0&\nm 1&\nm 0&\nm 0\end{array})$&$\begin{array}{rrrrr}0&1&0&0&0\end{array}$&$y_1$\\
&$\tilde{\nu}_{10}=(\begin{array}{rrrrrrr}\nm 1&-1&\nm 0&\nm 0&\nm 0&\nm 1\end{array})$&$\begin{array}{rrrrr}0&0&1&0&0\end{array}$&$\tilde{y}_6$\\
&$\tilde{\nu}_{11}=(\begin{array}{rrrrrrr}\nm 0&-1&\nm 0&\nm 0&\nm 0&\nm 0\end{array})$&$\begin{array}{rrrrr}0&0&0&0&1\end{array}$&$\tilde{y}_7$\\
&$\tilde{\nu}_{12}=(\begin{array}{rrrrrrr}\nm 0&-1&\nm 0&\nm 0&\nm 0&\nm 1\end{array})$&$\begin{array}{rrrrr}0&0&0&1&0\end{array}$&$\tilde{y}_8$\\
\hline
$\nabla_2$&$\nu_6=(\begin{array}{rrrrrrr}\nm 0&\nm 0&\nm 0&\nm 1&\nm 0&\nm 0\end{array})$&$\begin{array}{rrrrr}0&1&0&0&0\end{array}$&$y_2$\\
&$\nu_7=(\begin{array}{rrrrrrr}\nm 0&\nm 0&\nm 0&\nm 0&\nm 1&\nm 0\end{array})$&$\begin{array}{rrrrr}0&1&0&0&0\end{array}$&$y_3$\\
&$\nu_8=(\begin{array}{rrrrrrr}\nm 0&\nm 1&\nm 0&\nm 0&\nm 0&\nm 0\end{array})$&$\begin{array}{rrrrr}0&1&2&2&1\end{array}$&$y_4$\\
&$\nu_9=(\begin{array}{rrrrrrr}\nm 0&\nm 0&-1&-1&-1&\nm 1\end{array})$&$\begin{array}{rrrrr}0&1&0&0&0\end{array}$&$y_5$\\
\hline\hline
\end{tabular}
\end{equation}
The GUT vertex is again no longer a vertex but a point in $\nabla_1$. The Euler number is $1368$ which matches with the result of the calculation using~(\ref{eulerformula}).
%%%%%%%%%%%%%%%%%%%%%%%%%%%%%%%%%%%%%%%%%%%%%%%%%%%%%%%%%%%%%%%%%%%%%%%%%%%%%%%
\subsection{Model $3$}
We choose the following nef partition:
\begin{equation}
\begin{tabular}{c|c||c|c}
nef-part.&vertices&weights&coordinates\\
\hline\hline
$\nabla_1$&$\nu_1=(\begin{array}{rrrrrrr}\nm 3&\nm 0&\nm 0&\nm 1&\nm 0&\nm 1\end{array})$&$\begin{array}{rrrr}2&2&2&2\end{array}$&$x$\\
&$\nu_2=(\begin{array}{rrrrrrr}-2&\nm 0&\nm 0&-1&\nm 0&-1\end{array})$&$\begin{array}{rrrr}3&3&3&3\end{array}$&$y$\\
&$\nu_3=(\begin{array}{rrrrrrr}\nm 0&\nm 0&\nm 0&\nm 1&\nm 0&\nm 1\end{array})$&$\begin{array}{rrrr}1&0&0&0\end{array}$&$z$\\
&$\nu_4=(\begin{array}{rrrrrrr}\nm 0&\nm 0&\nm 0&\nm 0&\nm 1&\nm 0\end{array})$&$\begin{array}{rrrr}0&0&1&0\end{array}$&$w$\\
&$\nu_5=(\begin{array}{rrrrrrr}\nm 0&\nm 1&\nm 0&\nm 0&\nm 0&\nm 0\end{array})$&$\begin{array}{rrrr}1&0&0&0\end{array}$&$y_1$\\
&$\nu_6=(\begin{array}{rrrrrrr}\nm 0&\nm 0&\nm 0&\nm 0&\nm 0&\nm 1\end{array})$&$\begin{array}{rrrr}0&0&0&1\end{array}$&$y_2$\\
\hline
$\nabla_2$&$\nu_7=(\begin{array}{rrrrrrr}\nm 0&\nm 0&\nm 1&\nm 0&\nm 0&\nm 0\end{array})$&$\begin{array}{rrrr}1&0&0&0\end{array}$&$y_3$\\
&$\nu_8=(\begin{array}{rrrrrrr}\nm 0&\nm 0&\nm 0&\nm 1&\nm 0&\nm 0\end{array})$&$\begin{array}{rrrr}1&0&1&1\end{array}$&$y_4$\\
&$\nu_9=(\begin{array}{rrrrrrr}\nm 0&\nm 0&\nm 0&\nm 0&-1&\nm 1\end{array})$&$\begin{array}{rrrr}1&0&1&0\end{array}$&$y_5$\\
&$\nu_{10}=(\begin{array}{rrrrrrr}\nm 0&-1&-1&\nm 0&\nm 1&\nm 0\end{array})$&$\begin{array}{rrrr}1&0&0&0\end{array}$&$y_6$\\
\hline\hline
\end{tabular}
\end{equation}
%%%%
After reducing the M-lattice polytope to the SO(10) case, we obtain for the dual N-lattice polytope:
\begin{equation}
\begin{tabular}{c|c||c|c}
nef-part.&vertices&weights&coordinates\\
\hline\hline
$\nabla_1$&$\nu_1=(\begin{array}{rrrrrrr}\nm 3&\nm 0&\nm 0&\nm 1&\nm 0&\nm 1\end{array})$&$\begin{array}{rrrrrrrr}2&2&1&2&1&2&0&0\end{array}$&$x$\\
&$\nu_2=(\begin{array}{rrrrrrr}-2&\nm 0&\nm 0&-1&\nm 0&-1\end{array})$&$\begin{array}{rrrrrrrr}3&3&2&3&2&3&0&0\end{array}$&$y$\\
&$\nu_3=(\begin{array}{rrrrrrr}\nm 0&\nm 0&\nm 0&\nm 1&\nm 0&\nm 1\end{array})$&$\begin{array}{rrrrrrrr}0&1&0&0&0&0&0&0\end{array}$&$z$\\
%&$\nu_4=(\begin{array}{rrrrrrr}\nm 0&\nm 0&\nm 0&\nm 0&\nm 1&\nm 0\end{array})$&$\begin{array}{rrrrrrrr}0&0&1&0\end{array}$&$w$\\
&$\nu_5=(\begin{array}{rrrrrrr}\nm 0&\nm 1&\nm 0&\nm 0&\nm 0&\nm 0\end{array})$&$\begin{array}{rrrrrrrr}1&0&0&0&0&0&0&0\end{array}$&$y_1$\\
%&$\nu_6=(\begin{array}{rrrrrrr}\nm 0&\nm 0&\nm 0&\nm 0&\nm 0&\nm 1\end{array})$&$\begin{array}{rrrrrrrr}0&0&0&1\end{array}$&$y_2$\\
&$\tilde{\nu}_{11}=(\begin{array}{rrrrrrr}\nm 0&\nm 0&\nm 0&-1&\nm 0&\nm 1\end{array})$&$\begin{array}{rrrrrrrr}0&0&0&0&1&0&0&0\end{array}$&$\tilde{y}_7$\\
&$\tilde{\nu}_{12}=(\begin{array}{rrrrrrr}\nm 1&\nm 0&\nm 0&-1&\nm 2&\nm 1\end{array})$&$\begin{array}{rrrrrrrr}0&0&1&0&0&0&0&0\end{array}$&$\tilde{y}_8$\\
&$\tilde{\nu}_{12}=(\begin{array}{rrrrrrr}\nm 1&\nm 0&\nm 0&-1&\nm 0&\nm 0\end{array})$&$\begin{array}{rrrrrrrr}0&0&0&0&0&0&0&1\end{array}$&$\tilde{y}_9$\\
&$\tilde{\nu}_{13}=(\begin{array}{rrrrrrr}\nm 0&\nm 0&\nm 0&-1&\nm 1&-1\end{array})$&$\begin{array}{rrrrrrrr}0&0&0&0&0&0&1&0\end{array}$&$\tilde{y}_{10}$\\
&$\tilde{\nu}_{14}=(\begin{array}{rrrrrrr}\nm 0&\nm 0&\nm 0&-1&\nm 0&\nm 1\end{array})$&$\begin{array}{rrrrrrrr}0&0&0&0&0&1&0&0\end{array}$&$\tilde{y}_{11}$\\
&$\tilde{\nu}_{15}=(\begin{array}{rrrrrrr}\nm 0&\nm 0&\nm 0&-1&\nm 2&-1\end{array})$&$\begin{array}{rrrrrrrr}0&0&0&1&0&0&0&0\end{array}$&$\tilde{y}_{12}$\\
\hline
$\nabla_2$&$\nu_7=(\begin{array}{rrrrrrr}\nm 0&\nm 0&\nm 1&\nm 0&\nm 0&\nm 0\end{array})$&$\begin{array}{rrrrrrrr}1&0&0&0&0&0&0&0\end{array}$&$y_3$\\
&$\nu_8=(\begin{array}{rrrrrrr}\nm 0&\nm 0&\nm 0&\nm 1&\nm 0&\nm 0\end{array})$&$\begin{array}{rrrrrrrr}1&0&2&2&2&2&1&1\end{array}$&$y_4$\\
&$\nu_9=(\begin{array}{rrrrrrr}\nm 0&\nm 0&\nm 0&\nm 0&-1&\nm 1\end{array})$&$\begin{array}{rrrrrrrr}1&0&2&2&0&0&1&0\end{array}$&$y_5$\\
&$\nu_{10}=(\begin{array}{rrrrrrr}\nm 0&-1&-1&\nm 0&\nm 1&\nm 0\end{array})$&$\begin{array}{rrrrrrrr}1&0&0&0&0&0&0&0\end{array}$&$y_6$\\
\hline\hline
\end{tabular}
\end{equation}
Note that the divisors corresponding to the coordinates $w$ (GUT divisor) and $y_2$ no longer appear as vertices of $\nabla_1$. However, they are still points. The Euler number is $960$. Here we find a mismatch with the calculation using~(\ref{eulerformula}) where one obtains $552$ as a result for the Euler number.
%%%%%%%%%%%%%%%%%%%%%%%%%%%%%%%%%%%%%%%%%%%%%%%%%%%%%%%%%%%%%%%%%%%%%%%%%%%%%%%
\subsection{Model $4$}
The vertices in the $N$-lattice are:
\begin{equation}
\begin{tabular}{c|c||c|c}
nef-part.&vertices&weights&coordinates\\
\hline\hline
$\nabla_1$&$\nu_1=(\begin{array}{rrrrrrr}\nm 3&\nm 0&\nm 1&\nm 0&\nm 0&\nm 1\end{array})$&$\begin{array}{rrrrr}2&2&2&2&2\end{array}$&$x$\\
&$\nu_2=(\begin{array}{rrrrrrr}-2&\nm 0&-1&\nm 0&\nm 0&-1\end{array})$&$\begin{array}{rrrrr}3&3&3&3&3\end{array}$&$y$\\
&$\nu_3=(\begin{array}{rrrrrrr}\nm 0&\nm 0&\nm 1&\nm 0&\nm 0&\nm 1\end{array})$&$\begin{array}{rrrrr}1&0&0&0&0\end{array}$&$z$\\
&$\nu_4=(\begin{array}{rrrrrrr}\nm 0&\nm 0&\nm 0&\nm 0&\nm 1&\nm 0\end{array})$&$\begin{array}{rrrrr}0&0&1&0&0\end{array}$&$w$\\
&$\nu_5=(\begin{array}{rrrrrrr}\nm 0&\nm 1&\nm 0&\nm 0&\nm 0&\nm 0\end{array})$&$\begin{array}{rrrrr}1&0&0&0&0\end{array}$&$y_1$\\
&$\nu_6=(\begin{array}{rrrrrrr}\nm 0&\nm 0&\nm 0&\nm 1&\nm 0&\nm 0\end{array})$&$\begin{array}{rrrrr}0&0&1&0&0\end{array}$&$y_2$\\
&$\nu_7=(\begin{array}{rrrrrrr}\nm 0&\nm 0&\nm 0&\nm 0&\nm 0&\nm 1\end{array})$&$\begin{array}{rrrrr}0&0&0&0&1\end{array}$&$y_3$\\
\hline
$\nabla_2$&$\nu_8=(\begin{array}{rrrrrrr}\nm 0&\nm 0&\nm 1&\nm 0&\nm 0&\nm 0\end{array})$&$\begin{array}{rrrrr}1&0&1&1&1\end{array}$&$y_4$\\
&$\nu_9=(\begin{array}{rrrrrrr}\nm 0&\nm 0&\nm 0&\nm 0&-1&\nm 1\end{array})$&$\begin{array}{rrrrr}1&0&0&1&0\end{array}$&$y_5$\\
&$\nu_{10}=(\begin{array}{rrrrrrr}\nm 0&\nm 0&\nm 0&-1&\nm 0&\nm 1\end{array})$&$\begin{array}{rrrrr}1&0&1&0&0\end{array}$&$y_6$\\
&$\nu_{11}=(\begin{array}{rrrrrrr}\nm 0&-1&\nm 0&\nm 1&\nm 1&-1\end{array})$&$\begin{array}{rrrrr}1&0&0&0&0\end{array}$&$y_7$\\
\hline\hline
\end{tabular}
\end{equation}
%%%%%
After reducing the M-lattice polytope to the SO(10) case, we obtain for the dual N-lattice polytope:
\begin{equation}
\begin{tabular}{c|c||c|c}
nef-part.&vertices&weights&coordinates\\
\hline\hline
$\nabla_1$&$\nu_1=(\begin{array}{rrrrrrr}\nm 3&\nm 0&\nm 1&\nm 0&\nm 0&\nm 1\end{array})$&$\begin{array}{rrrrrrrrr}2&2&2&1&2&1&2&0&0\end{array}$&$x$\\
&$\nu_2=(\begin{array}{rrrrrrr}-2&\nm 0&-1&\nm 0&\nm 0&-1\end{array})$&$\begin{array}{rrrrrrrrr}3&3&3&2&3&2&3&0&0\end{array}$&$y$\\
&$\nu_3=(\begin{array}{rrrrrrr}\nm 0&\nm 0&\nm 1&\nm 0&\nm 0&\nm 1\end{array})$&$\begin{array}{rrrrrrrrr}0&1&0&0&0&0&0&0&0\end{array}$&$z$\\
%&$\nu_4=(\begin{array}{rrrrrrr}\nm 0&\nm 0&\nm 0&\nm 0&\nm 1&\nm 0\end{array})$&$\begin{array}{rrrrrrrrr}0&0&1&0&0\end{array}$&$w$\\
&$\nu_5=(\begin{array}{rrrrrrr}\nm 0&\nm 1&\nm 0&\nm 0&\nm 0&\nm 0\end{array})$&$\begin{array}{rrrrrrrrr}1&0&0&0&0&0&0&0&0\end{array}$&$y_1$\\
&$\nu_6=(\begin{array}{rrrrrrr}\nm 0&\nm 0&\nm 0&\nm 1&\nm 0&\nm 0\end{array})$&$\begin{array}{rrrrrrrrr}0&0&1&0&0&0&0&0&0\end{array}$&$y_2$\\
%&$\nu_7=(\begin{array}{rrrrrrr}\nm 0&\nm 0&\nm 0&\nm 0&\nm 0&\nm 1\end{array})$&$\begin{array}{rrrrrrrrr}0&0&0&0&1\end{array}$&$y_3$\\
&$\tilde{\nu}_{12}=(\begin{array}{rrrrrrr}\nm 1&\nm 0&-1&\nm 0&\nm 0&\nm 1\end{array})$&$\begin{array}{rrrrrrrrr}0&0&0&0&0&1&0&0&0\end{array}$&$\tilde{y}_8$\\
&$\tilde{\nu}_{13}=(\begin{array}{rrrrrrr}\nm 1&\nm 0&-1&\nm 0&\nm 2&-1\end{array})$&$\begin{array}{rrrrrrrrr}0&0&0&0&1&0&0&0&0\end{array}$&$\tilde{y}_9$\\
&$\tilde{\nu}_{14}=(\begin{array}{rrrrrrr}\nm 0&\nm 0&-1&\nm 0&\nm 0&\nm 0\end{array})$&$\begin{array}{rrrrrrrrr}0&0&0&0&0&0&0&0&1\end{array}$&$\tilde{y}_{10}$\\
&$\tilde{\nu}_{15}=(\begin{array}{rrrrrrr}\nm 0&\nm 0&-1&\nm 0&\nm 1&-1\end{array})$&$\begin{array}{rrrrrrrrr}0&0&0&0&0&0&0&1&0\end{array}$&$\tilde{y}_{11}$\\
&$\tilde{\nu}_{16}=(\begin{array}{rrrrrrr}\nm 0&\nm 0&-1&\nm 0&\nm 0&\nm 1\end{array})$&$\begin{array}{rrrrrrrrr}0&0&0&0&0&0&1&0&0\end{array}$&$\tilde{y}_{12}$\\
&$\tilde{\nu}_{17}=(\begin{array}{rrrrrrr}\nm 0&\nm 0&-1&\nm 0&\nm 2&-1\end{array})$&$\begin{array}{rrrrrrrrr}0&0&0&0&1&0&0&0&0\end{array}$&$\tilde{y}_{13}$\\
\hline
$\nabla_2$&$\nu_8=(\begin{array}{rrrrrrr}\nm 0&\nm 0&\nm 1&\nm 0&\nm 0&\nm 0\end{array})$&$\begin{array}{rrrrrrrrr}1&0&1&2&2&2&2&1&1\end{array}$&$y_4$\\
&$\nu_9=(\begin{array}{rrrrrrr}\nm 0&\nm 0&\nm 0&\nm 0&-1&\nm 1\end{array})$&$\begin{array}{rrrrrrrrr}1&0&0&2&2&0&0&1&0\end{array}$&$y_5$\\
&$\nu_{10}=(\begin{array}{rrrrrrr}\nm 0&\nm 0&\nm 0&-1&\nm 0&\nm 1\end{array})$&$\begin{array}{rrrrrrrrr}1&0&1&0&0&0&0&0&0\end{array}$&$y_6$\\
&$\nu_{11}=(\begin{array}{rrrrrrr}\nm 0&-1&\nm 0&\nm 1&\nm 1&-1\end{array})$&$\begin{array}{rrrrrrrrr}1&0&0&0&0&0&0&0&0\end{array}$&$y_7$\\
\hline\hline
\end{tabular}
\end{equation}
The vertices corresponding to the GUT divisor and to the coordinate $y_3$ are no longer vertices but points in $\nabla_1$. The Euler number is $960$. Using~(\ref{eulerformula}) to compute the Euler number we obtain $672$ -- so again, we find a mismatch.
%%%%%%%%%%%%%%%%%%%%%%%%%%%%%%%%%%%%%%%%%%%%%%%%%%%%%%%%%%%%%%%%%%%%%%%%%%%%%%
\subsection{Model $5$}
The vertices in the N-lattice are:
\begin{equation}
\begin{tabular}{c|c||c|c}
nef-part.&vertices&weights&coordinates\\
\hline\hline
$\nabla_1$&$\nu_1=(\begin{array}{rrrrrrr}\nm 3&\nm 1&\nm 1&\nm 1&\nm 0&\nm 0\end{array})$&$\begin{array}{rrrrr}2&4&4&2&2\end{array}$&$x$\\
&$\nu_2=(\begin{array}{rrrrrrr}-2&-1&-1&-1&\nm 0&\nm 0\end{array})$&$\begin{array}{rrrrr}3&6&6&3&3\end{array}$&$y$\\
&$\nu_3=(\begin{array}{rrrrrrr}\nm 0&\nm 1&\nm 1&\nm 1&\nm 0&\nm 0\end{array})$&$\begin{array}{rrrrr}0&0&0&0&1\end{array}$&$z$\\
&$\nu_4=(\begin{array}{rrrrrrr}\nm 0&\nm 0&\nm 0&\nm 0&\nm 1&\nm 0\end{array})$&$\begin{array}{rrrrr}0&0&1&0&0\end{array}$&$w$\\
&$\nu_5=(\begin{array}{rrrrrrr}\nm 0&\nm 0&\nm 0&\nm 0&\nm 0&\nm 1\end{array})$&$\begin{array}{rrrrr}1&0&0&0&0\end{array}$&$y_1$\\
&$\nu_6=(\begin{array}{rrrrrrr}\nm 0&-1&-1&\nm 0&\nm 1&\nm 0\end{array})$&$\begin{array}{rrrrr}0&1&0&0&0\end{array}$&$y_2$\\
&$\nu_7=(\begin{array}{rrrrrrr}\nm 0&\nm 1&\nm 1&\nm 1&-1&\nm 1\end{array})$&$\begin{array}{rrrrr}0&1&1&0&0\end{array}$&$y_3$\\
&$\nu_8=(\begin{array}{rrrrrrr}\nm 0&\nm 0&\nm 0&\nm 1&\nm 0&\nm 0\end{array})$&$\begin{array}{rrrrr}0&0&0&1&0\end{array}$&$y_4$\\
\hline
$\nabla_2$&$\nu_9=(\begin{array}{rrrrrrr}\nm 0&\nm 1&\nm 1&\nm 1&\nm 0&-1\end{array})$&$\begin{array}{rrrrr}1&1&1&0&0\end{array}$&$y_5$\\
&$\nu_{10}=(\begin{array}{rrrrrrr}\nm 0&\nm 0&\nm 1&\nm 0&\nm 0&\nm 0\end{array})$&$\begin{array}{rrrrr}0&1&0&1&0\end{array}$&$y_6$\\
&$\nu_{11}=(\begin{array}{rrrrrrr}\nm 0&\nm 1&\nm 0&\nm 0&\nm 0&\nm 0\end{array})$&$\begin{array}{rrrrr}0&1&0&1&0\end{array}$&$y_7$\\
\hline\hline
\end{tabular}
\end{equation}
%%%%
After reducing the M-lattice polytope to the SO(10) case, we obtain for the dual N-lattice polytope:
\begin{equation}
\begin{tabular}{c|c||c|c}
nef-part.&vertices&weights&coordinates\\
\hline\hline
$\nabla_1$&$\nu_1=(\begin{array}{rrrrrrr}\nm 3&\nm 1&\nm 1&\nm 1&\nm 0&\nm 0\end{array})$&$\begin{array}{rrrrrrr}2&4&6&5&2&2&2\end{array}$&$x$\\
&$\nu_2=(\begin{array}{rrrrrrr}-2&-1&-1&-1&\nm 0&\nm 0\end{array})$&$\begin{array}{rrrrrrr}3&6&9&8&3&3&3\end{array}$&$y$\\
&$\nu_3=(\begin{array}{rrrrrrr}\nm 0&\nm 1&\nm 1&\nm 1&\nm 0&\nm 0\end{array})$&$\begin{array}{rrrrrrr}0&0&0&0&0&0&1\end{array}$&$z$\\
%&$\nu_4=(\begin{array}{rrrrrrr}\nm 0&\nm 0&\nm 0&\nm 0&\nm 1&\nm 0\end{array})$&$\begin{array}{rrrrrrr}0&0&1&0&0\end{array}$&$w$\\
&$\nu_5=(\begin{array}{rrrrrrr}\nm 0&\nm 0&\nm 0&\nm 0&\nm 0&\nm 1\end{array})$&$\begin{array}{rrrrrrr}1&0&0&0&0&0&0\end{array}$&$y_1$\\
&$\nu_6=(\begin{array}{rrrrrrr}\nm 0&-1&-1&\nm 0&\nm 1&\nm 0\end{array})$&$\begin{array}{rrrrrrr}0&1&0&0&0&0&0\end{array}$&$y_2$\\
&$\nu_7=(\begin{array}{rrrrrrr}\nm 0&\nm 1&\nm 1&\nm 1&-1&\nm 1\end{array})$&$\begin{array}{rrrrrrr}0&1&2&2&1&0&0\end{array}$&$y_3$\\
&$\nu_8=(\begin{array}{rrrrrrr}\nm 0&\nm 0&\nm 0&\nm 1&\nm 0&\nm 0\end{array})$&$\begin{array}{rrrrrrr}0&0&0&0&0&1&0\end{array}$&$y_4$\\
&$\tilde{\nu}_{12}=(\begin{array}{rrrrrrr}\nm 0&-1&-1&-1&\nm 2&\nm 0\end{array})$&$\begin{array}{rrrrrrr}0&0&1&0&0&0&0\end{array}$&$\tilde{y}_8$\\
&$\tilde{\nu}_{13}=(\begin{array}{rrrrrrr}\nm 1&-1&-1&-1&\nm 2&\nm 0\end{array})$&$\begin{array}{rrrrrrr}0&0&0&1&0&0&0\end{array}$&$\tilde{y}_9$\\
&$\tilde{\nu}_{14}=(\begin{array}{rrrrrrr}\nm 0&-1&-1&-1&\nm 1&\nm 0\end{array})$&$\begin{array}{rrrrrrr}0&0&0&0&1&0&0\end{array}$&$\tilde{y}_{10}$\\
\hline
$\nabla_2$&$\nu_9=(\begin{array}{rrrrrrr}\nm 0&\nm 1&\nm 1&\nm 1&\nm 0&-1\end{array})$&$\begin{array}{rrrrrrr}1&1&2&2&1&0&0\end{array}$&$y_5$\\
&$\nu_{10}=(\begin{array}{rrrrrrr}\nm 0&\nm 0&\nm 1&\nm 0&\nm 0&\nm 0\end{array})$&$\begin{array}{rrrrrrr}0&1&0&0&0&1&0\end{array}$&$y_6$\\
&$\nu_{11}=(\begin{array}{rrrrrrr}\nm 0&\nm 1&\nm 0&\nm 0&\nm 0&\nm 0\end{array})$&$\begin{array}{rrrrrrr}0&1&0&0&0&1&0\end{array}$&$y_7$\\
\hline\hline
\end{tabular}
\end{equation}
The GUT divisor is no longer a vertex but a point in $\nabla_1$. The Euler number is $4872$. This coincides with the result one gets from~(\ref{eulerformula}).

\bibliographystyle{fullsort}
\addcontentsline{toc}{section}{References}
\bibliography{papers_f-theory}

\end{document}